\documentclass{aa}  

\usepackage{graphicx}
\usepackage{txfonts}
\usepackage{xcolor} 
\usepackage{hyperref}
\newcommand{\ind}[1]{_{\mathrm{#1}}}

\newcommand\numax{\nu\ind{max}}

\newcommand\Teff{T\ind{eff}}

\newcommand\cahk{Ca\,\textsc{ii}\,H\&K\xspace}
\newcommand\cair{Ca\,\textsc{ii}\,IR\xspace}
\newcommand\caii{Ca\,\textsc{ii}\xspace}
\newcommand\mgi{Mg\,\textsc{i}\xspace}
\newcommand\ha{H$\alpha$\xspace}
\newcommand\fei{Fe\,\textsc{i}\xspace}

\begin{document}

   \title{Magnetic activity of red giants: correlation between the amplitude of solar-like oscillations and chromospheric indicators}

\author{%
C. Gehan\inst{1}
\and
D. Godoy-Rivera\inst{2,3}
\and
P. Gaulme\inst{4}
}

\institute{Max-Planck-Institut für Sonnensystemforschung, Justus-von-Liebig-Weg 3, 37077 Göttingen, Germany ; \texttt{gehan@mps.mpg.de}
\and Instituto de Astrofísica de Canarias, E-38205 La Laguna, Tenerife, Spain
\and Universidad de La Laguna, Departamento de Astrofísica, E-38206 La Laguna, Tenerife, Spain
\and Thüringer Landessternwarte, Sternwarte 5, 07778 Tautenburg, Germany
}



   \date{Received September 15, 1996; accepted March 16, 1997}

 
\abstract{
Previous studies have found that red giants (RGs) in close binary systems undergoing spin-orbit resonance exhibit an enhanced level of magnetic activity with respect to single RGs rotating at the same rate, from measurements of photometric variability $S'\ind{ph}$ and chromospheric emission S-index $S\ind{\caii}$. Here, we consider a sample of 4465 RGs observed by the NASA \textit{Kepler} mission studied by \citet{Gaulme_2020} and \citet{Gehan_2022} to measure additional activity indicators that probe different heights in the chromosphere: the near-ultraviolet (NUV) excess from NASA \textit{GALEX} photometric data, and chromospheric indices based on the depth of H$\alpha$, Mg\,\textsc{i} and infared Ca\,\textsc{ii} absorption lines from LAMOST spectroscopic data. Firstly, as for \cahk, we observe that RGs belonging to close binaries in a state of spin-orbit resonance display larger chromospheric emission than the cohort of RGs, as illustrated by an NUV excess and shallower \ha and infrared \caii lines. We report no excess of \mgi emission. This result reinforces previous claims that tidal locking leads to enhanced magnetic fields, and allows us to provide criteria to classify active red giants -- single or binary --, based on their rotation periods and magnetic activity indices. 
Secondly, we strikingly observe that the depths of the \mgi and \ha lines are anti-correlated and correlated, respectively, with the amplitude of solar-like oscillations for a given surface gravity $\log g$, regardless of the presence of photometric rotational modulation. Such a correlation opens up future possibilities of estimating the value of magnetic fields at the surface of RG stars, whether quiet or active, by combining spectroscopic and asteroseismic measurements with three-dimensional atmospheric models including radiative transfer.}


\authorrunning{C. Gehan, D. Godoy-Rivera, P. Gaulme}
\titlerunning{Chromospheric activity of red giants versus solar-like oscillation amplitude}

\maketitle
%

\section{Introduction}\label{introduction}

A complex dynamo mechanism can be established in the external convective envelope of low-mass stars, where surface magnetic fields are efficiently generated when the rotation period is shorter than the convective turnover time according to dynamo theory \citep[e.g.,][]{Skumanich_1972, Charbonneau_2014}. Magnetic fields play an important role in stellar evolution, as they are believed to regulate rotation throughout the life of low-mass stars \citep{Vidotto_2014}. They can also induce losses of angular momentum through a coupling with the magnetized stellar wind \citep[e.g.,][]{Weber_1967, Gallet_2013}. Magnetic fields are also likely to drive the important mass loss observed among red giants (RGs), although the specific mechanisms are still poorly understood \citep{Harper_2018}.

Magnetic fields generate various phenomena when they emerge from the outer convective envelope, which are grouped under the name of stellar magnetic activity \citep{Hall_2008}.
Characterizing stellar activity is crucial for the fields of stellar physics and evolution, as well as exoplanetary science. Indeed, it is an important probe of the magnetic field configuration that can provide clues on the physical dynamo process at work \citep{Reinhold_2019, Metcalfe_2022}. Stellar activity also hampers the detection of exoplanetary signals \citep{Borgniet_2015} and affects space weather and planets' habitability \citep{Airapetian_2020}.

In general, as stars age and evolve off the main sequence, they expand, spin down, and become less active \citep[e.g.,][]{Wilson_1964, Skumanich_1972}. Consequently, the rotational modulation signal caused by stellar spots in co-rotation with their photosphere tends to become smaller with longer periods, making it difficult to measure the rotation periods of evolved stars \citep{Santos_2021}. However, \cite{Gaulme_2020} showed that $\sim$8\% of the RGs observed by the NASA \textit{Kepler} mission \citep{Borucki_2010} display photometric rotational modulation, of which approximately 15\% belong to close binary systems. Fast rotation is not the only parameter explaining such enhanced magnetic activity for RGs in close binaries, as \cite{Gaulme_2020} also noticed that at a given rotation period, a RG that belongs to a close binary system displays a photometric modulation about one order of magnitude larger than that of a single RG with similar physical properties (mass and radius). This difference in the rotational modulation of binary and single RGs could be due to either a different spot distribution on their photosphere, or a stronger surface magnetic field.

\begin{table*}
\centering
\caption{Characteristics of our sample.}
\begin{tabular}{l r}
\hline
Sample & Number of stars\\
\hline
Initial \citep{Gaulme_2020} & 4465\\
With S-index \citep{Gehan_2022} & 3130\\
With $S\ind{\cahk}$ & 3105\\
With $S\ind{\mgi}$ & 3155\\
With $S\ind{H\alpha}$ & 3260\\
With $S\ind{\fei}$ & 3260\\
With $S\ind{\cair}$ & 3265\\
With $\Delta m\ind{NUV}$ & 842\\
\hline
\end{tabular}
\label{table:0}
\end{table*}


To discriminate between these two possibilities, \cite{Gehan_2022} used spectroscopic data from the Large Sky Area Multi-Object Fiber Spectroscopic Telescope survey (LAMOST) to measure the S-index from the H and K lines of \caii for 3130 RGs analyzed by \cite{Gaulme_2014, Gaulme_2020} and \cite{Benbakoura_2021}.  The S-index quantifies the level of chromospheric activity \citep{Wilson_1978, Duncan_1991, Karoff_2016, Zhang_2020, Gomes_2021}, and is proportional to the strength of surface magnetic fields \citep{Babcock_1961, Petit_2008, Auriere_2015, Brown_2022}. \cite{Gehan_2022} found that RGs in close binaries show a significantly larger S-index than single RGs and RGs in wide binaries, indicating that close binaries do display enhanced surface magnetic fields. More specifically, \cite{Gehan_2022} showed that the RGs in close binaries with enhanced magnetic activity are either synchronized (with their rotation period being the same as their orbital period), or undergoing spin-orbit resonance (with their rotation period being an integer ratio of their orbital period). This result highlights that tidal locking, rather than close binarity itself, somehow leads to larger magnetic field strengths.
Additionally, \cite{Gehan_2022} examined the relationship between different stellar activity indicators, and found a correlation between the photometric index, the chromospheric S-index, and the detection of rotational modulation due to stellar spots. These investigations provide important information for understanding the mechanisms behind the formation and evolution of stellar magnetic fields \citep{Reinhold_2019}.

To fully probe the dynamos and magnetic fields of evolved stars, it is fundamental to be able to classify the active evolved stars, whether they belong to close binary systems or not. Unfortunately, $S'\ind{ph}$ and $S\ind{\caii}$ are not always available and additional indicators from other surveys could help classifying the active red giants. 
On the one hand, the NUV excess is a relatively under-explored activity proxy that quantifies the NUV flux as measured above the photospheric level, thus providing an indication about chromospheric activity \citep{Findeisen_2010, Olmedo_2015, Godoy-Rivera_2021}. 
On the other hand, the \ha \citep{Cincunegui_2007, Newton_2017},  \mgi \citep{Sasso_2017} and \caii infrared \citep{Busa_2007, Martinez-Arnaiz_2011, Martin_2017} lines have been considered as possible indicators of chromospheric activity for late-type stars. Since the aforementioned spectral lines are located at longer wavelengths compared to the \caii lines used to measure the S-index, they appear as more suitable activity indicators for cool late-type stars that have an intrinsically low signal-to-noise ratio towards the blue part of their optical spectrum \citep{Cincunegui_2007}.

In this paper, we aim at extending the work done by \cite{Gaulme_2020} and \cite{Gehan_2022} on 4465 single and binary RGs by investigating the possibility of using the following indicators of chromospheric activity: the near ultraviolet (NUV) excess, measured using data from the \textit{Galaxy Evolution Explorer} (\textit{GALEX}) that provided UV photometry for about 90 millions stars \citep{Bianchi_2005}, as well as the \ha, \mgi and infrared \caii spectral lines, which can be easily exploited using the optical spectra obtained by LAMOST and encompassing wavelengths between 3690 \AA\, and 9100 \AA.
We describe our measurement methods in Sect. \ref{method}. We explore the relationship between different activity indicators in Sect. \ref{results} and establish links with the amplitude of oscillations in Sect. \ref{oscillations}. We discuss in Sect. \ref{binarity} the impact of close binarity and tidal interactions versus fast rotation on the chromospheric activity indices of RGs. We provide in Sect. \ref{classification} criteria to classify the active red giants (single or binary), based on their rotation period and magnetic activity indices.
Section \ref{conclusions} is devoted to conclusions.


\section{Method and data}\label{method}

The target sample of this work is composed of the 4465 RGs with measured photometric index $S\ind{ph}$ from \citet[][see their Fig. 1]{Gaulme_2020}, and includes 3130 RGs for which \cite{Gehan_2022} also measured the $S$-index. The characteristics of our sample are summarized in Table \ref{table:0}. Our measurements are publicly available in the database of the Centre de Donn\'ees de Strasbourg (CDS).

\subsection{Change of nomenclature}\label{Sph}

To be consistent with the works of \citet[][]{Mathur_2014} and subsequent papers \citep[such as][]{Ceillier_2017,Mathur_2019}, we modify the symbol used to describe the photometric modulation as performed by \citet{Gaulme_2020}, and employed by \citet{Benbakoura_2021} and \citet{Gehan_2022}. \citet[][]{Mathur_2014} define $S\ind{ph}$ as the standard deviation of a given lightcurve over five rotational periods, but do not provide measurements for stars whose lightcurves do not display rotational modulation. To circumvent this lack of definition, \citet{Gaulme_2020} extended the definition of $S\ind{ph}$ to stars without rotational modulation by computing it as the moving average of the standard deviation of the lightcurve over a period of five days. To avoid confusion with the original definition of $S\ind{ph}$ by \citet{Mathur_2014}, we change the nomenclature by replacing $S\ind{ph}$ by $S\ind{ph}'$, defined such as: 
\begin{equation}
    S\ind{ph}' = 
    \begin{cases}
      S\ind{ph} & \mbox{with detectable rotational modulation}\\
      \langle\sigma\ind{5 d}\rangle & \mbox{without detectable rotational modulation}
    \end{cases}  
\end{equation}
where $\langle\sigma\ind{5 d}\rangle$ is the mean standard deviation of the lightcurve over five days.

\subsection{Measuring the NUV excess}\label{NUV}

\begin{figure}
\includegraphics[width=9.3cm]{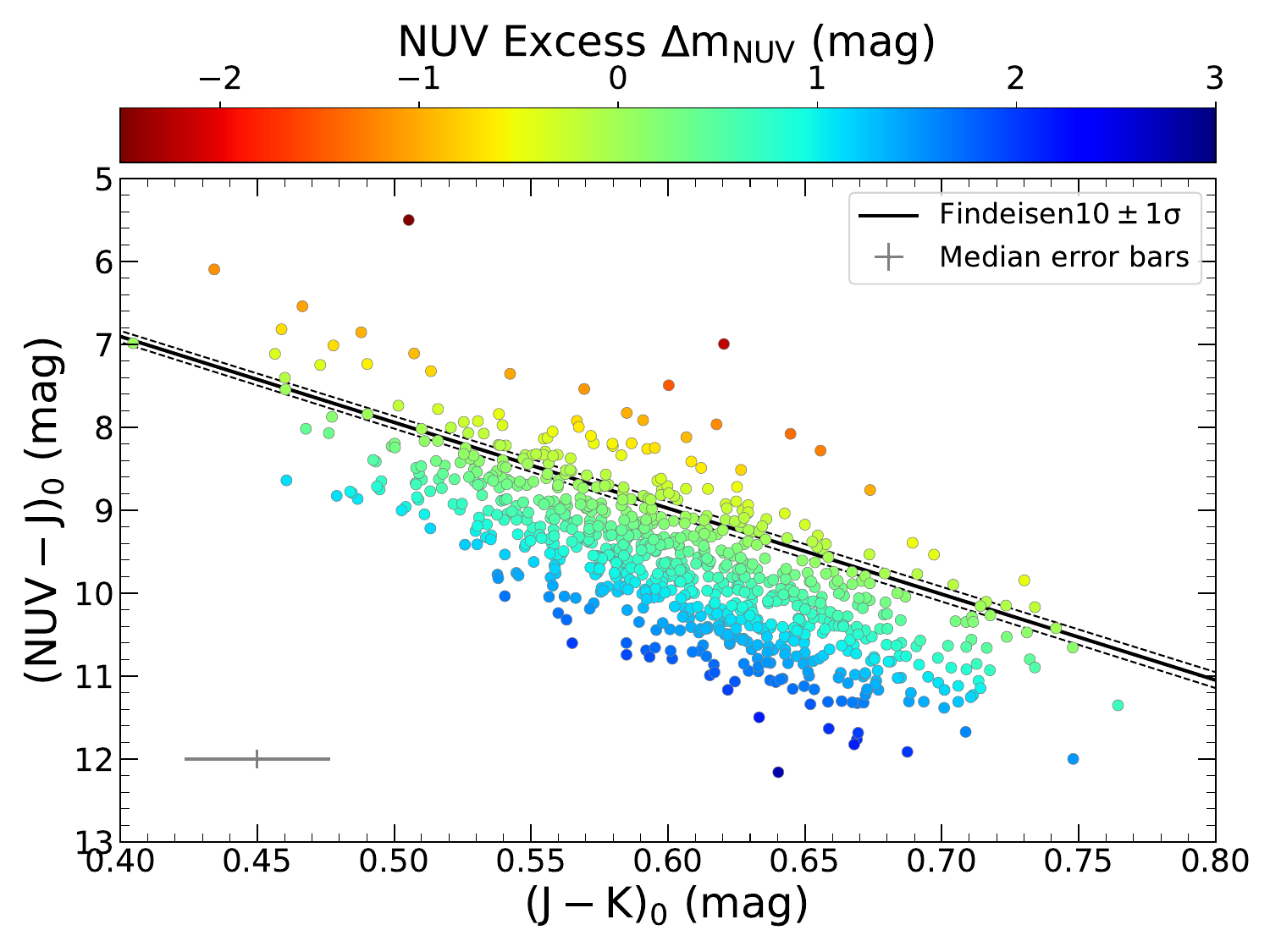}
\caption{NUV versus NIR color-color diagram for the 842 RGs with \textit{GALEX} NUV magnitudes. From this, the NUV excess is calculated by subracting the measured (NUV-J)$\ind{0}$ magnitudes and the expected values given the fiducial relation from \cite{Findeisen_2010} and the measured (J-K)$\ind{0}$ colors. The targets are color-coded by their NUV excess values, with redder colors indicating stronger excess (more negative values). The grey symbol represents the median error bars.}
\label{fig-NUV-measurement}
\end{figure}

The NUV data used in this work were taken from \textit{GALEX} \citep{Martin_2005}. The \textit{GALEX} space telescope performed an all-sky imaging survey in the far-NUV (FUV, $\lambda \sim 0.15 \mu$m) and near-UV (NUV, $\lambda \sim 0.23 \mu$m) bands. Throughout this paper, we focused on the latter of these bands, given its wider availability for our sample. 

To calculate the NUV excess, we followed the approach of \cite{Dixon_2020} and \cite{Godoy-Rivera_2021}. We first cross-matched the target sample with the latest \textit{GALEX} catalogue \citep{Bianchi_2017} (see also \citealt{Olmedo_2015}) using VizieR\footnote{\url{https://vizier.unistra.fr/}}. We imposed a maximum angular separation of 2\,arcsec, and were left with 842 RGs ($\sim$ 19\% of the initial sample) with measured \textit{GALEX} NUV magnitudes (with most targets having counterparts within a fraction of 1\,arcsec). We used the same procedure to cross-match with the near-infrared (NIR) 2MASS catalog \citep{Cutri_2003, Skrutskie_2006}, and found the entirety of the sample with measured $J$ ($\lambda \sim 1.25 \mu$m) and $K$ magnitudes ($\lambda \sim 2.16 \mu$m). We de-reddened the photometry using the extinction values from the \textit{Kepler} field catalog of Godoy-Rivera et al. (submitted), and the coefficients from \citet{Dixon_2020} (see also \citealt{Yuan_2013}). Using these data, we constructed the (NUV-J)$\ind{0}$ versus (J-K)$\ind{0}$ diagram (Fig. \ref{fig-NUV-measurement}).

To be able to detect and characterize an excess of NUV emission, we need to define a reference level, i.e., the NUV magnitude of quiet stars. Such a level, so-called the fiducial relation, was introduced by \citet{Findeisen_2010} (see their Equation 2). This relation parametrizes the behaviour of the stellar locus in the NUV-NIR color-color space (black line in Fig. \ref{fig-NUV-measurement}), by tracking expected (NUV-J)$\ind{0}$ photospheric flux as a function of the (J-K)$\ind{0}$ color. For a given star, the NUV excess $\Delta m\ind{NUV}$ is calculated as the difference between the measured (NUV-J)$\ind{0}$ color and the value predicted by the fiducial relation evaluated at the corresponding (J-K)$\ind{0}$ color. The purpose of this subtraction is to remove the photospheric contribution to the flux in the NUV wavelengths. The uncertainty on the NUV excess is calculated by propagating the errors on the fiducial relation, extinction, and measured magnitudes (being dominated by the latter). By definition, a stronger NUV excess corresponds to more negative values, i.e., higher NUV fluxes. In Fig. \ref{fig-NUV-measurement} the targets are color-coded by their NUV excess values.
The sample of 842 targets for which we measured NUV excess includes 30 confirmed close and wide binaries, where we consider that a close binary system has an orbital periods $P\ind{orb} \leq$ 150 days and a wide binary system has $P\ind{orb} >$ 150 days. Our sample includes 3 close binaries identified by \cite{Gaulme_2020} as well as 4 close binaries and 23 wide binaries identified by \textit{Gaia} data release 3 \citep[DR3,][]{Gaia_2023} in the non-single star (NSS) two-body-orbit table \citep{Gaia_2023_binaries}. Other aspects of the NUV excess are discussed in Appendix \ref{appendix-1}.

\begin{figure*}
\includegraphics[width=17cm]{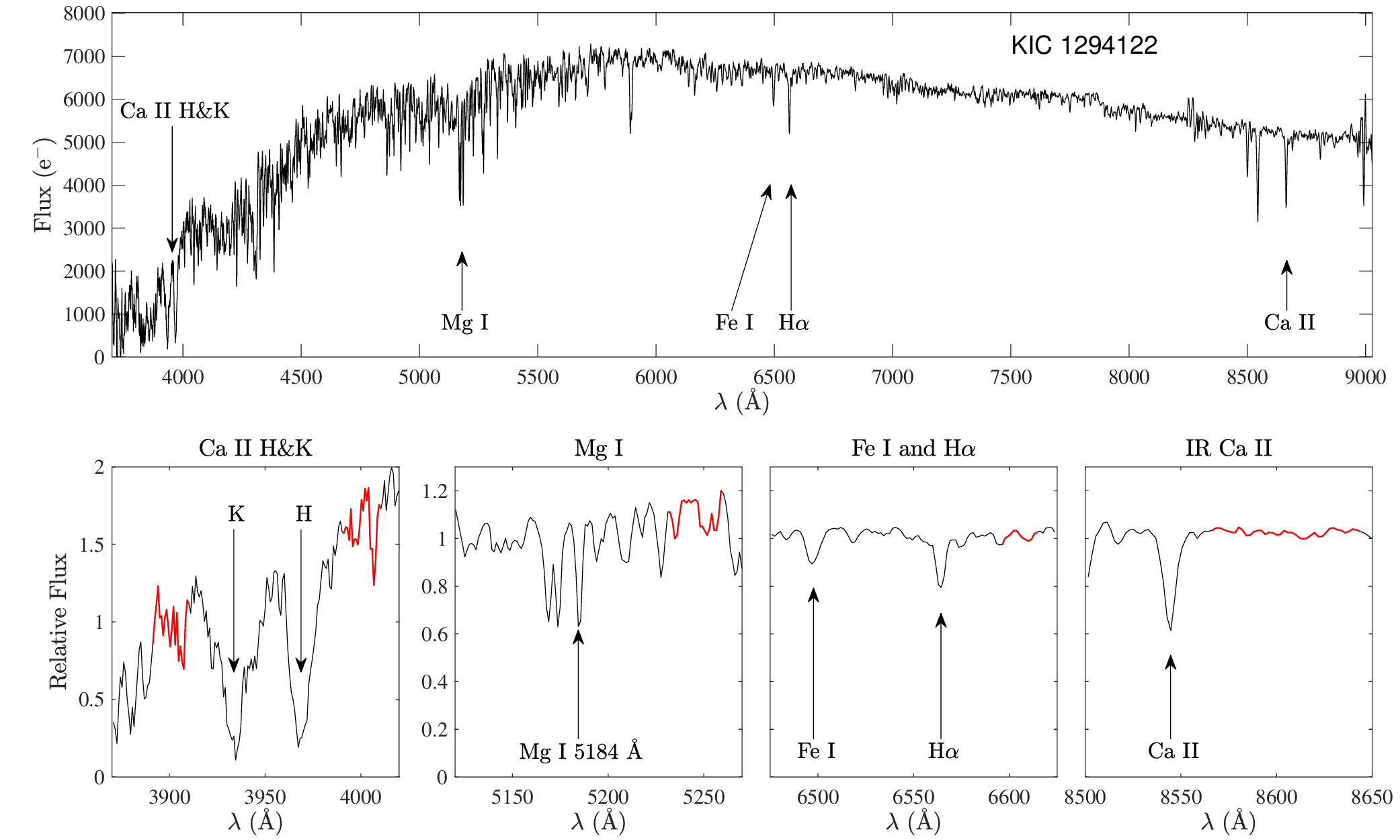}
\caption{Example of LAMOST spectrum for KIC 1294122. Top panel: Entire spectrum expressed as a function of wavelength (\AA). The lines we consider are indicated by arrows. Bottom panels: Zoom on the six absorption lines, where the flux is normalized by its average value over the 150-\AA\, range that is displayed in each of the four sub-panels. The portion of the spectrum that is used to compute the continuum level is indicated by a red line.}
\label{fig-lamost_spec}
\end{figure*}

\subsection{Measuring chromospheric emission from the H$\alpha$, \mgi and \caii infrared spectral lines}\label{H-alpha}

We follow the approach of \citet{Cincunegui_2007}, who adapted the usual method employed to measure the S-index with the chromospheric emission in the \cahk lines to the \ha line for a sample of late-type main-sequence stars. They define $S\ind{H\alpha}$ as:
\begin{equation}\label{eqt-Halpha-index}
    S\ind{H\alpha} = \frac{F\ind{H\alpha}}{F\ind{C}},
\end{equation} 
where $F\ind{H\alpha}$ is the integrated flux in the H$\alpha$ line centered at 6562.801 \AA\, using a triangular bandpass with a full width at half maximum of 1.5 \AA, and $F\ind{C}$ is the continuum level obtained by integrating the flux comprised in a square bandpass centered on 6605 \AA\, with a width of 20 \AA\, (see Fig. \ref{fig-lamost_spec}). In other words, $S\ind{H\alpha}$ is a proxy of the line depth.

In a similar way, we use the \mgi line centered at 5184 \AA\, the \caii infrared (IR) line centered at 8542 \AA\, and the \fei line centered at 6495 \AA\, to define the indices $S\ind{\mgi}$, $S\ind{\cair}$ and $S\ind{\fei}$, respectively (Eq. \ref{eqt-Halpha-index} and Fig. \ref{fig-lamost_spec}). We chose not to make use of the other absorption lines of the Mg\,\textsc{I} triplet and IR calcium triplet -- even of the renowned sodium $D_1$, $D_2$ doublet -- because they are blended with one another, which makes the pipeline fail quite often. In the same way, we do not report the Mg\,\textsc{I} at 4571\,\AA\, because it belongs to a range packed with spectral lines that are blended, although \citet{Sasso_2017} report this line to be sensitive to chromospheric activity. We used a \fei photospheric line, which is in principle insensitive to magnetic activity (see Sect. \ref{formation-height-lines}), as a control variable to look for potential systematic biases in the interpretation of our results. We used the \fei line at 6495 \AA\, because it is close to the \ha line and shows a decent signal-to-noise ratio (SNR).

In order to lead consistent comparisons between the activity indices mentioned above, we also measured $S\ind{\cahk}$ from the \cahk lines for the 3130 RGs for which \cite{Gehan_2022} measured the S-index (see Fig. \ref{fig-lamost_spec}). Here, we chose not to calibrate our $S\ind{\cahk}$ measurements to the Mount Wilson scale, since there is no reference scale to calibrate the other activity indices. Hence, our $S\ind{\cahk}$ measurements are not strictly speaking S-indices, contrarily to the measurements obtained by \cite{Gehan_2022}.

To improve the accuracy of our results, we modified the algorithm that was used by \citet{Gehan_2022}, by refining the method to compute the Doppler shift of the spectrum, and the triangular weighted average $F\ind{H\alpha}$. Indeed, \citet{Gehan_2022} determined the Doppler shift by comparing the wavelength of the bottom of the \ha line to its theoretical value, and computed the triangular moving average on the LAMOST spectrum interpolated on a finer grid obtained by linear interpolation. Regarding the location of the Doppler shift, basing our measurement only on one line limits its accuracy to the radial velocity resolution of the spectrum, which is about 69\,km\,s$^{-1}$. Regarding the use of linearly interpolated spectra, discontinuities of the flux caused by discretization introduce noise into the measurements. Our updated method includes improvements on these two aspects.

Firstly, LAMOST spectra, as most optical spectra, are provided as functions of wavelength but are evenly sampled in radial velocities. The first step consists of selecting a range to measure the Doppler shift on, and to attribute a radial-velocity axis to it. With respect to the line of interest of wavelength $\lambda_0$, the Doppler velocity axis $v\ind{D}$ is expressed as:
\begin{equation}
    v\ind{D} = \ln\left(\frac{\lambda}{\lambda_0}\right)\,c
\end{equation}
where $c$ is the speed of light. The selected range must not include telluric lines and shall exclude the noisiest parts of the spectrum. Since not all spectra were good on the complete visible range, we adapted the range used to compute the Doppler shift to the spectral lines that were considered. It means that in some cases, we may have a measurement of \ha emission but not of IR \caii. In case no zeros were present in the selected range, we computed the Doppler shift by computing the crosscorrelation of the observed spectrum with a theoretical spectrum provided by the library of PHOENIX stellar atmospheres and synthetic spectra \citep{Husser_2013}\footnote{\url{https://phoenix.astro.physik.uni-goettingen.de/}}. The location of the correlation maximum was done by picking the maximum of the crosscorrelation oversampled by a factor 1000. In the rare cases where a chunk of the selected range included zeros, we computed the Doppler shift as done by picking the minimum of the considered line on an oversampled version of the spectrum. The second step consists of oversampling the spectrum. To avoid being biased by interpolation methods, such as spline or polynomial interpolation, we oversampled the spectra by computing the fast Fourier transform of the selected range and padding it with zeros. The oversampling factor was set to 100. The triangular-weighted value of the line depth was then computed from the oversampled spectrum, as well as the continuum level.

\begin{table*}
\centering
\caption{Pearson correlation coefficients $r$ and associated p-values $p$ between different activity indicators and $S'\ind{ph}$, for inactive spotless and active spotted RGs.}
\begin{tabular}{l r r}
\hline
Activity indicator & Inactive RGs & Active RGs\\
\hline
$\Delta m\ind{NUV}$ & $r$ = -0.121; $p$ = $8.2 \times 10^{-4}$ & $r$ = 0.258; $p$ = 0.046\\
$S\ind{\cahk}$ & $r$ = -0.100; $p$ = $1.6 \times 10^{-7}$ & $r$ = 0.216; $p$ = $5.3 \times 10^{-4}$\\
$S\ind{H\alpha}$ & $r$ = 0.028; $p$ = 0.122 & $r$ = 0.291; $p$ = $1.8 \times 10^{-6}$\\
$S\ind{\mgi}$ & $r$ = 0.030; $p$ = 0.108 & $r$ = -0.081; $p$ = 0.204\\
$S\ind{\fei}$ & $r$ = -0.010; $p$ = 0.585 & $r$ = -0.158; $p$ = 0.012\\
$S\ind{\cair}$ & $r$ = -0.104; $p$ = $2.3 \times 10^{-8}$ & $r$ = 0.216; $p$ = $4.8 \times 10^{-4}$\\
\hline
\end{tabular}
\label{table-Pearson}
\end{table*}

We used LAMOST Data Release 7, which is the same we used in \cite{Gehan_2022} to measure the S-index. Firstly, we validated our  activity index measurements only for spectra presenting a high enough signal-to-noise ratio, in a similar way as in Sect. 2.2 of \cite{Gehan_2022}. Uncertainties were computed by assuming that the noise of the spectra is only due to photon noise (see details in Appendix \ref{appendix_errorbar}). When several spectra were available for a given star, we computed the activity index as the median of all the activity indices measured for each individual spectrum. 
Our sample includes 185 binaries, among which there are 24 close binaries identified by \cite{Gaulme_2020}, 9 close binaries identified by \cite{Benbakoura_2021}, as well as 22 close binaries and 130 wide binaries identified by \textit{Gaia} DR3 \citep{Gaia_2023_binaries}.


\section{Chromospheric indices of red giants from UV data and spectral lines}\label{results}

In this section and the next ones, we adopted the measurements of the photometric index $S'\ind{ph}$, the oscillation frequency at maximum amplitude $\numax$, the height of the Gaussian envelope employed to model the oscillation excess power $H\ind{max}$, and the rotation period $P\ind{rot}$ that were obtained by \cite{Gaulme_2020}.

\subsection{Formation height of spectral lines}\label{formation-height-lines}

As shown by previous works, the core of the spectral lines we use form at different heights in the solar atmosphere, and we can expect the picture to be similar for cool stars. Using radiation-MHD simulations and three-dimensional (3D) non-local thermodynamic equilibrium (LTE) radiative transfer computations, \cite{Bjorgen_2018} showed that the core of the \caii infrared line at 8542 \AA\, forms at a much lower height in the chromosphere ($\sim$ 1400 km) compared to the \cahk lines ($\sim$ 2200 km). \cite{Leenaarts_2012} showed that the \ha line also tends to form at a low height of $\sim$ 1600 km in the chromosphere. On the other hand, \fei is a minority species in stellar atmospheres with effective temperatures $\Teff$ > 4500 K \citep{Gray_2008, Mashonkina_2011}. The \fei line at 6495 \AA\, we use in this work is therefore a photospheric line, and we use it as a control variable since it is not known to be sensitive to magnetic activity \citep{Wise_2018, Quintero_Noda_2021}.

\subsection{Relation between activity indicators}\label{relation-activity-indicators}

\begin{figure*}[ht!]
\includegraphics[width=8.6cm]{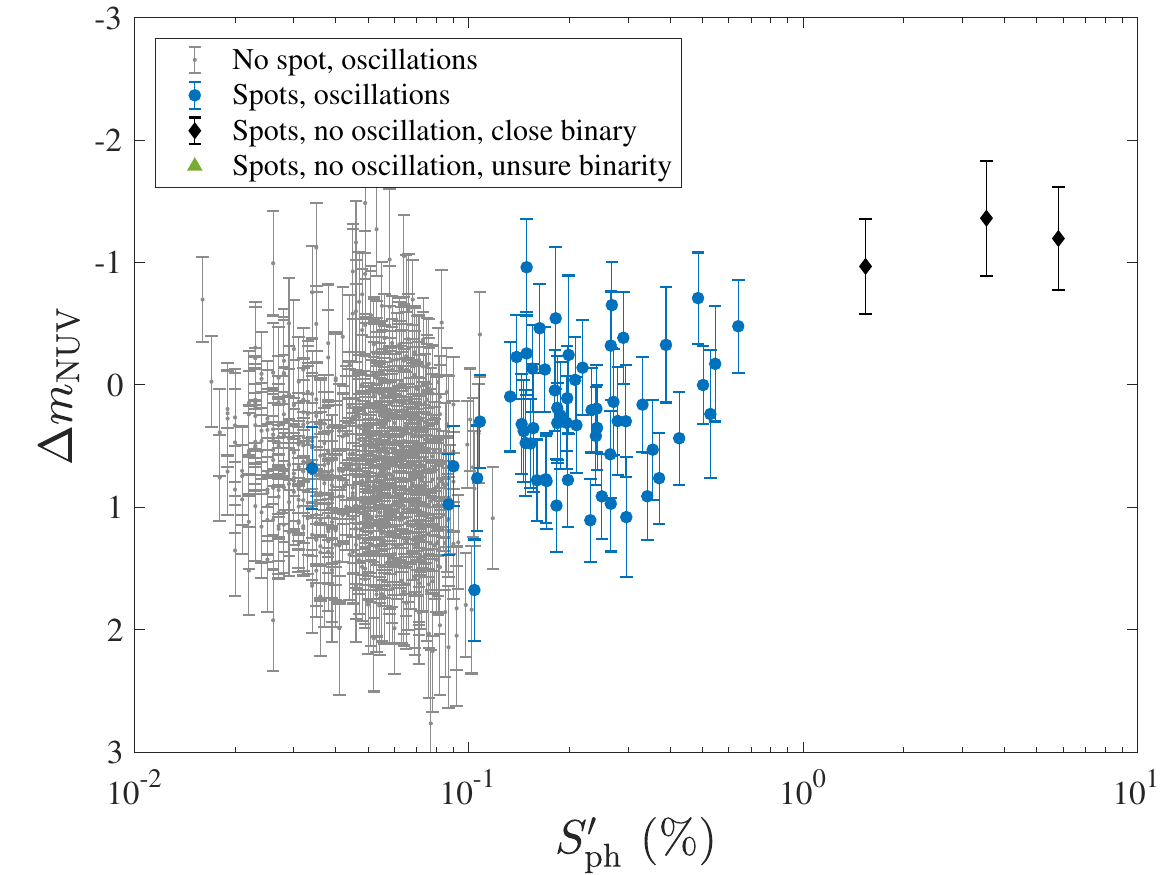}
\includegraphics[width=8.7cm]{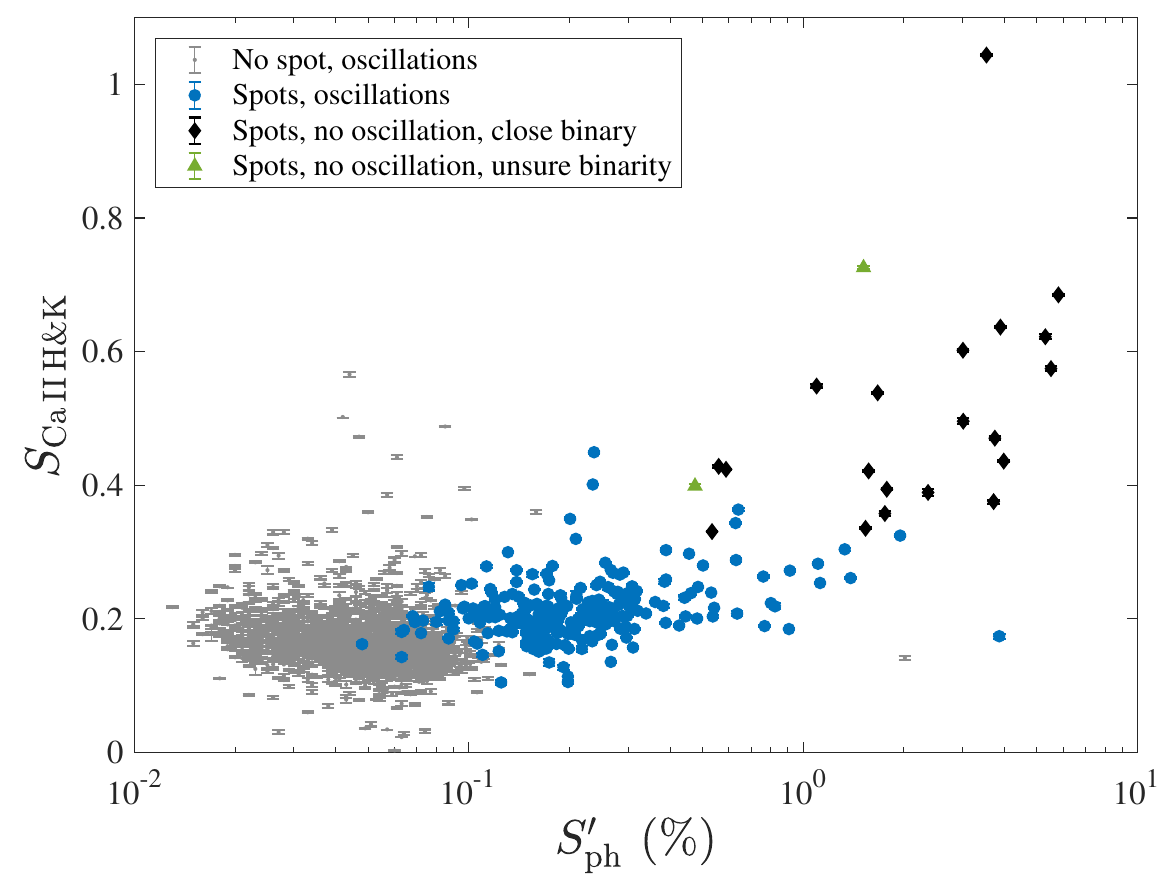}\\
\includegraphics[width=8.7cm]{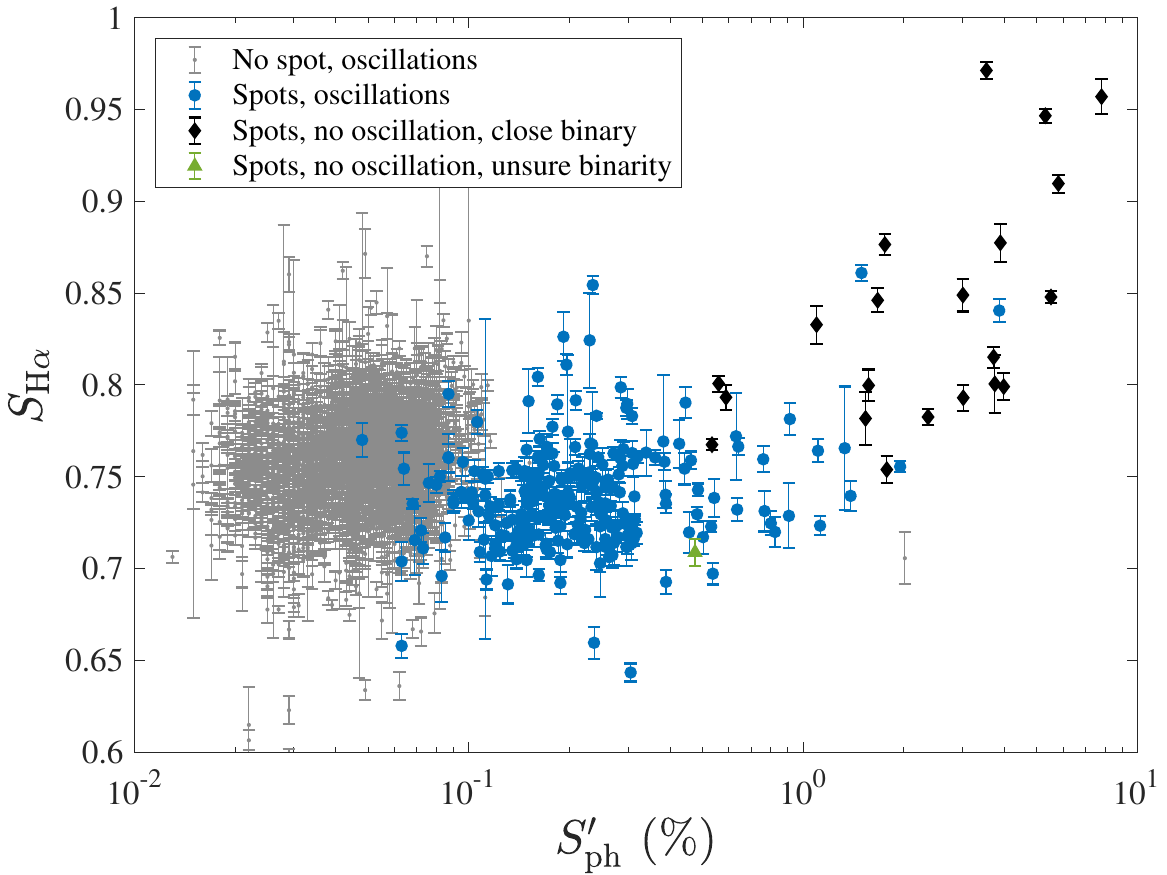}
\includegraphics[width=8.7cm]{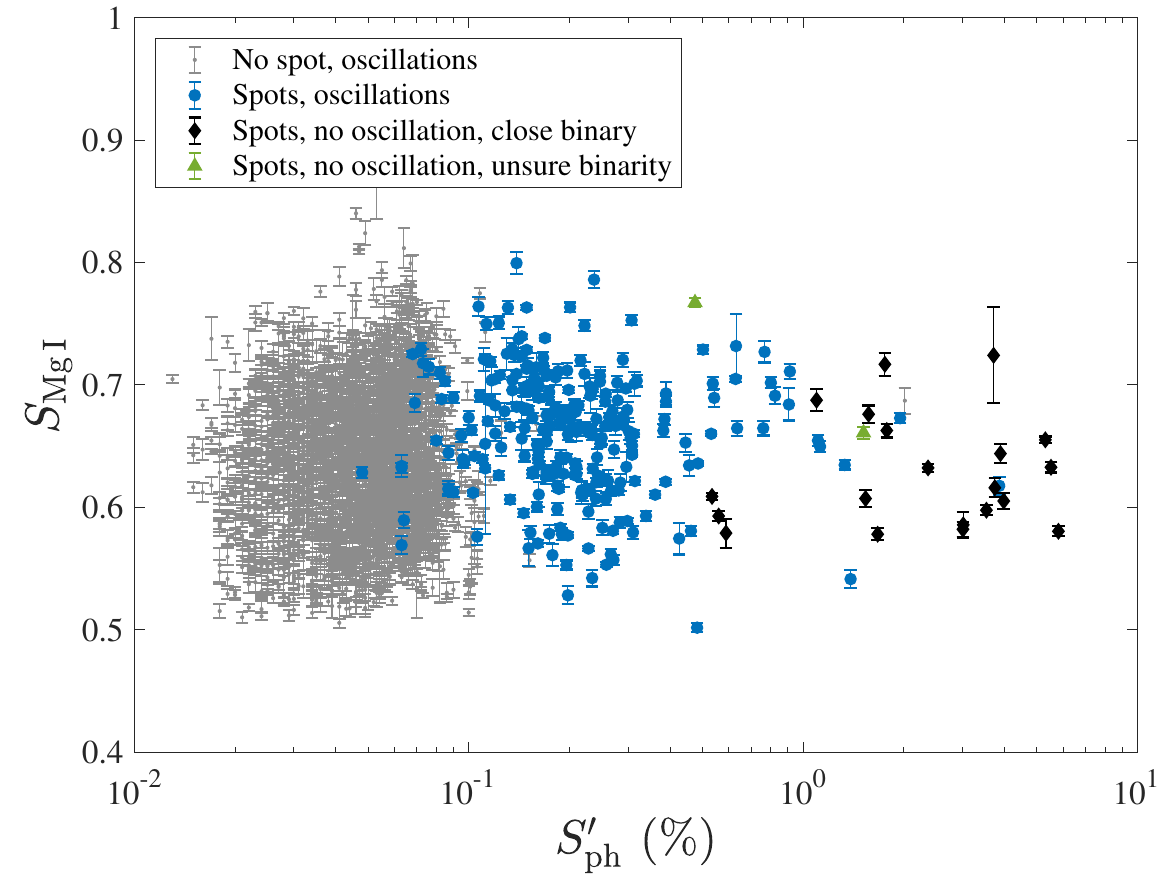}\\
\includegraphics[width=8.7cm]{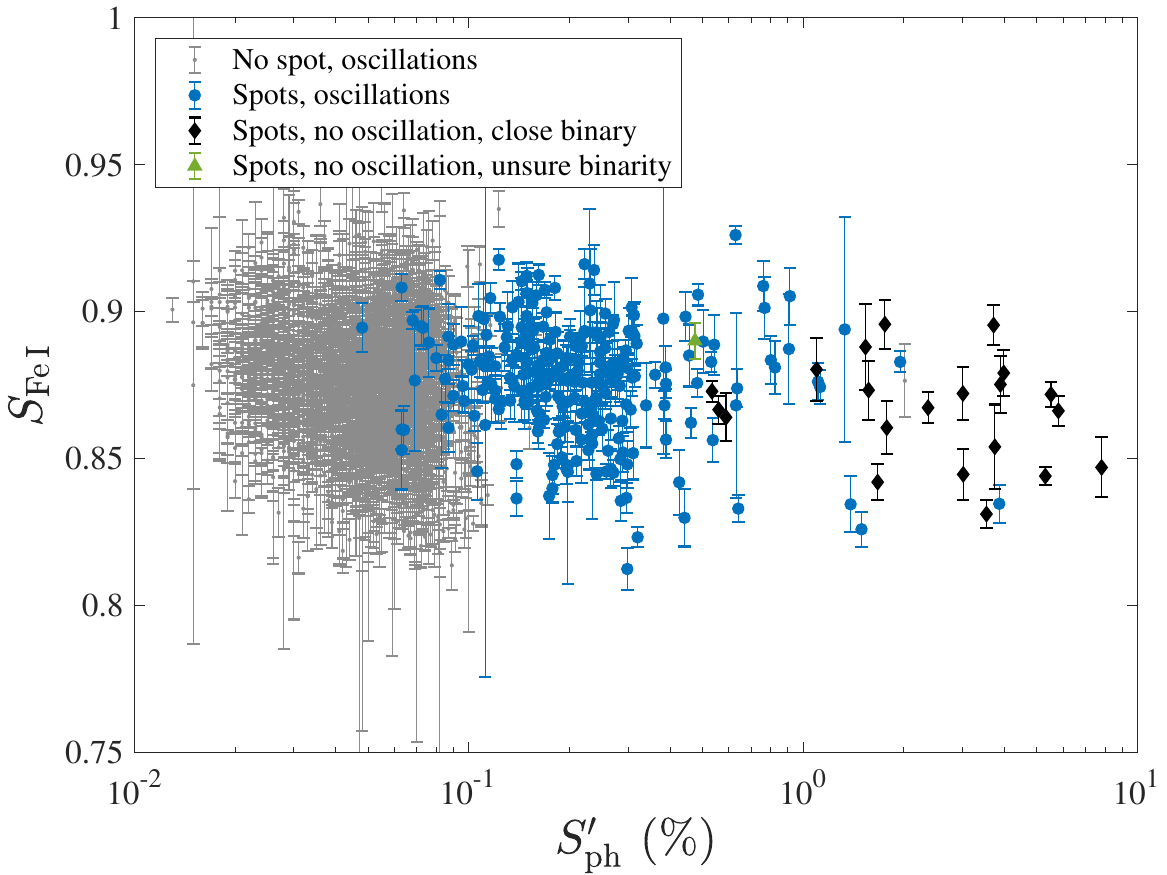}
\includegraphics[width=8.7cm]{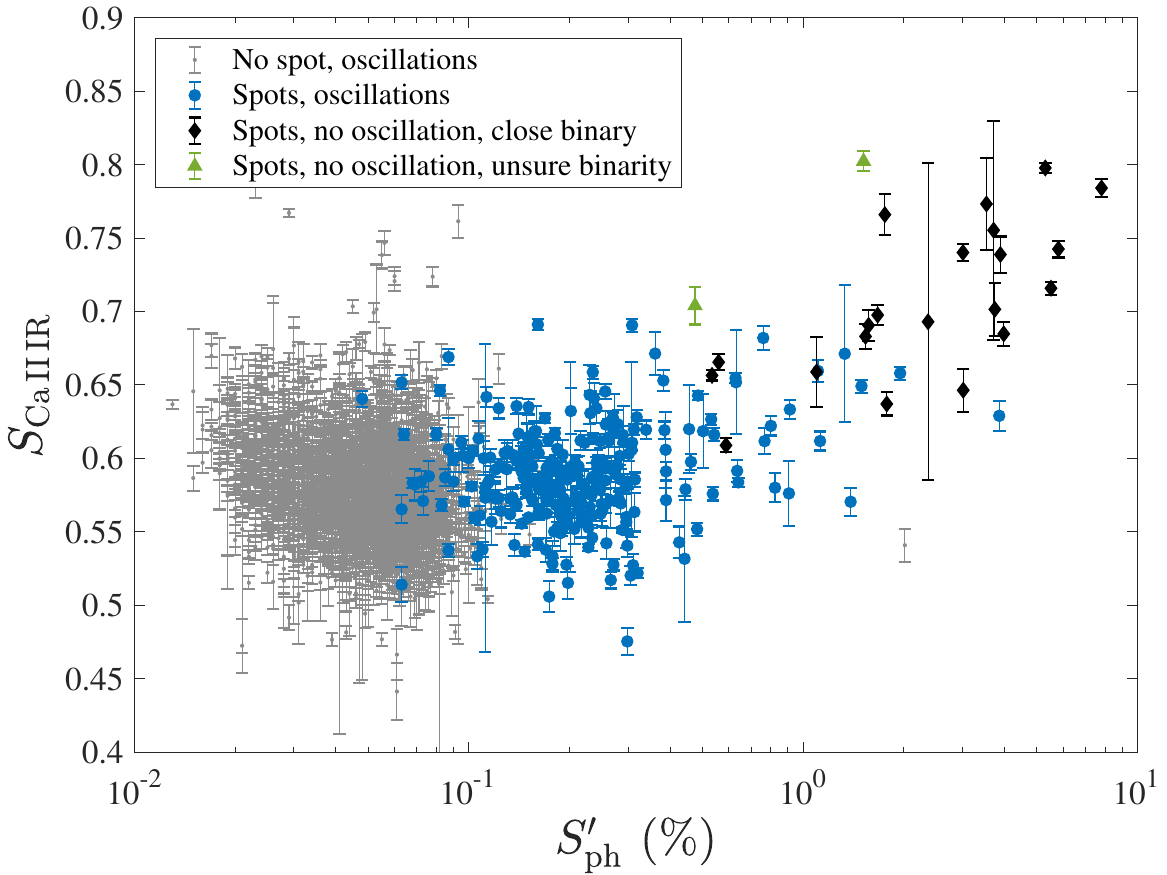}
\caption{Relation between different activity indicators and the photometric modulation $S'\ind{ph}$ as measured by \citet{Gaulme_2020} (in log scale), which is an indicator of photospheric activity. The NUV excess $\Delta m\ind{NUV}$ is a chromospheric activity indicator measured using \textit{GALEX} data. The indices $S\ind{\cahk}$, $S\ind{H\alpha}$, $S\ind{\mgi}$ and $S\ind{\cair}$ are chromospheric activity indicators measured using LAMOST data. The index $S\ind{\fei}$ measured using LAMOST data is insensitive to stellar activity and used as a control variable. Gray dots refer to regular photometrically inactive RGs (i.e. with no evidence of spot modulation) that display oscillations. Blue dots refer to photometrically active RGs (i.e. exhibiting spot modulation) with partially suppressed oscillations. Black dots refer to photometrically active and non-oscillating RGs in close binary systems, which were identified by \cite{Gaulme_2020}. Green dots refer to non-oscillating RGs with ambiguous binary versus single status; we do not have NUV excess data for these stars.}
\label{fig-activity-indicators}
\end{figure*}

In the following we compare the relation between the activity indicators $\Delta m\ind{NUV}$, $S\ind{\cahk}$, $S\ind{H\alpha}$, $S\ind{\mgi}$, $S\ind{\fei}$, $S\ind{\cair}$ and $S'\ind{ph}$ for our RG sample. These are shown in Figure \ref{fig-activity-indicators}, where our sample is split into different subsets, namely single spotless stars (in gray), single spotted stars (in blue), binary spotted stars (in black) and spotted stars with ambiguous binary versus single status (in green). The Pearson correlation coefficients and corresponding p-values are indicated in Table \ref{table-Pearson}. We recall here that more NUV-active stars have a more negative NUV-excess value $\Delta m\ind{NUV}$.

Similarly to what \citet{Gehan_2022} found for the S-index (see also upper right panel of Fig. \ref{fig-activity-indicators}), we observe an anti-correlation between $\Delta m\ind{NUV}$ and $S'\ind{ph}$ for inactive spotless RGs, as well as a marginally significant correlation between $\Delta m\ind{NUV}$ and $S'\ind{ph}$ for the photometrically active RGs with evidence of spots (upper left panel of Fig.\ref{fig-activity-indicators}).
To the extent of our knowledge, the relation between $\Delta m\ind{NUV}$ and $S'\ind{ph}$ has never been checked before, even for main-sequence stars.

Similarly, we also observe a correlation between $S'\ind{ph}$ and $S\ind{H\alpha}$ (middle left panel of Fig. \ref{fig-activity-indicators}). This is consistent with the findings of \cite{Newton_2017} and \cite{Garcia_Soto_2023} for M dwarfs, who reported that the amplitude of photometric variability is correlated with H$\alpha$ emission.  
However, the correlation that we observe emerges only when including the very active RGs that mostly belong to close binary systems (black symbols). When discarding the latter, only a marginal trend is left for the stars displaying spots and partially suppressed oscillations (blue symbols), in contrast with the clearer correlation observed with \cahk lines. 

Regarding the other spectral lines, we do not observe any significant correlation between $S'\ind{ph}$ and $S\ind{\mgi}$  (middle right panel of Fig. \ref{fig-activity-indicators}), even for close binaries. The same is observed with the $S\ind{\fei}$ line (bottom left panel of Fig. \ref{fig-activity-indicators}), which is expected since it is in principle a photospheric line that is insensitive to chromospheric activity. Finally, we observe for $S\ind{\cair}$ (bottom right panel of Fig. \ref{fig-activity-indicators}) similar trends as for $\Delta m\ind{NUV}$ and $S\ind{\cahk}$ (upper right panel of Fig. \ref{fig-activity-indicators}), although the smaller range of values measured for $S\ind{\cair}$ suggests that it is less sensitive to magnetic activity than $S\ind{\cahk}$. 

We also do not observe any significant correlation or anti-correlation between $S'\ind{ph}$ and $S\ind{\fei}$ for inactive RGs, as well as a slight anti-correlation for active RGs (bottom left panel of Fig. \ref{fig-activity-indicators}). This is expected for $S\ind{\fei}$, which is in principle insensitive to activity and that we use as a control variable.

In summary, the RGs that are in close binaries present in average larger values of $\Delta m \ind{NUV}$, $S\ind{H\alpha}$ and $S\ind{\cair}$, similarly to what was already observed for $S\ind{ph}$ \citep{Gaulme_2020} and the S-index \citep[][, see also upper right panel of Fig. \ref{fig-activity-indicators}]{Gehan_2022}. We discuss the case of close binaries in Sect. \ref{binarity}.

\subsection{Comparisons with other studies}\label{comparison-litterature}

\cite{Findeisen_2011} observed a correlation between the NUV excess and chromospheric activity from the H and K lines of \caii for G-dwarfs. Here, we consistently observe a correlation between the NUV excess and $S\ind{\cahk}$, for both inactive RGs (Pearson correlation coefficient of $0.213$ and p-value of $6.9 \times 10^{-7}$) and active RGs (Pearson correlation coefficient of $0.463$ and p-value of $1.2 \times 10^{-3}$). Since $S\ind{\cahk}$ is a proxy of the strength of surface magnetic fields \citep{Babcock_1961, Petit_2008, Auriere_2015, Brown_2022}, our result indicates that the NUV excess is a proxy for the strength of surface magnetic fields.

In addition, previous works have shown that the correlation between the flux in the \ha and the \cahk lines is still not fully understood \citep{Gomes_2022}. In the case of the Sun, the flux in the \ha and the \cahk lines is known to be correlated with the presence of active regions, and tightly follows the solar magnetic cycle \citep{Livingston_2007}. However, the picture is much less straightforward for other stars, where the correlation between the flux in the \ha and the \cahk lines presents a large dispersion \citep{Cincunegui_2007, Walkowicz_2009, Gomes_2011, Gomes_2014, Meunier_2022}. The correlation appears to be strongly positive for stars with high activity levels, while all types of correlations tend to be observed in the low-activity regime, ranging from negative to positive correlations, including correlations close to 0. Since the cores of the \cahk and \ha lines are formed at different heights in the chromosphere (see Sect. \ref{formation-height-lines}),
we can then expect these two lines to trace different activity phenomena \citep{Cincunegui_2007, Meunier_2009, Gomes_2022}. The different sensitivity of these lines to spots, plages, or filaments could explain the various correlations observed for different stars \citep{Meunier_2009}. In particular, \cite{Meunier_2009} stated that the contribution of emission in plages in \cahk and \ha should lead to a correlation very close to 1 between the flux in these two lines. However, the presence of dark filaments at the surface, mostly visible in \ha, introduces flux absorption that should decrease the correlation. Moreover, many filaments do not coincide exactly with active regions, which further contributes in decreasing the correlation between the flux in the \cahk and \ha lines. On the other hand, filaments well correlated with plages would lead to an anticorrelation.
Here, we observe an anticorrelation between $S\ind{\cahk}$ and the H$\alpha$ index for inactive RGs, (Pearson correlation coefficient of $-0.107$ and p-value of $1.7 \times 10^{-8}$). For active RGs, we do not observe any significant correlation or anti-correlation (Pearson correlation coefficient of $0.085$ and p-value of 0.182). In the case of RGs in close binaries, although the statistics is low (only 21 RGs in close binaries with measured $S\ind{H\alpha}$), the S-index and the \ha index seem to be positively correlated (Pearson correlation coefficient of $0.803$ and p-value of $1.2 \times 10^{-5}$).

For a sample of nearby M-dwarfs, \cite{Stelzer_2013} found that NUV and X-ray fluxes are correlated, as well as between \ha and X-ray fluxes, but did not check the relation between NUV and H$\alpha$ fluxes. 
Here, we observe an anti-correlation between the NUV excess and the \ha index for inactive RGs (Pearson correlation coefficient of $-0.446$ and p-value of $1.0 \times 10^{-28}$), while we observe a marginally significant anti-correlation for active RGs (Pearson correlation coefficient of $-0.308$ and p-value of $0.040$). 
Since the NUV excess is a proxy of the chromospheric activity level, and the \ha line is expected to trace activity at a lower height, with an important contribution from filaments, it is not necessarily surprising to observe an anti-correlation between $\Delta m\ind{NUV}$ and $S\ind{H\alpha}$, in the same way that we observed an anti-correlation between $S\ind{\cahk}$ and $S\ind{H\alpha}$ for inactive RGs.

\begin{figure*}[ht!]
\includegraphics[width=9cm]{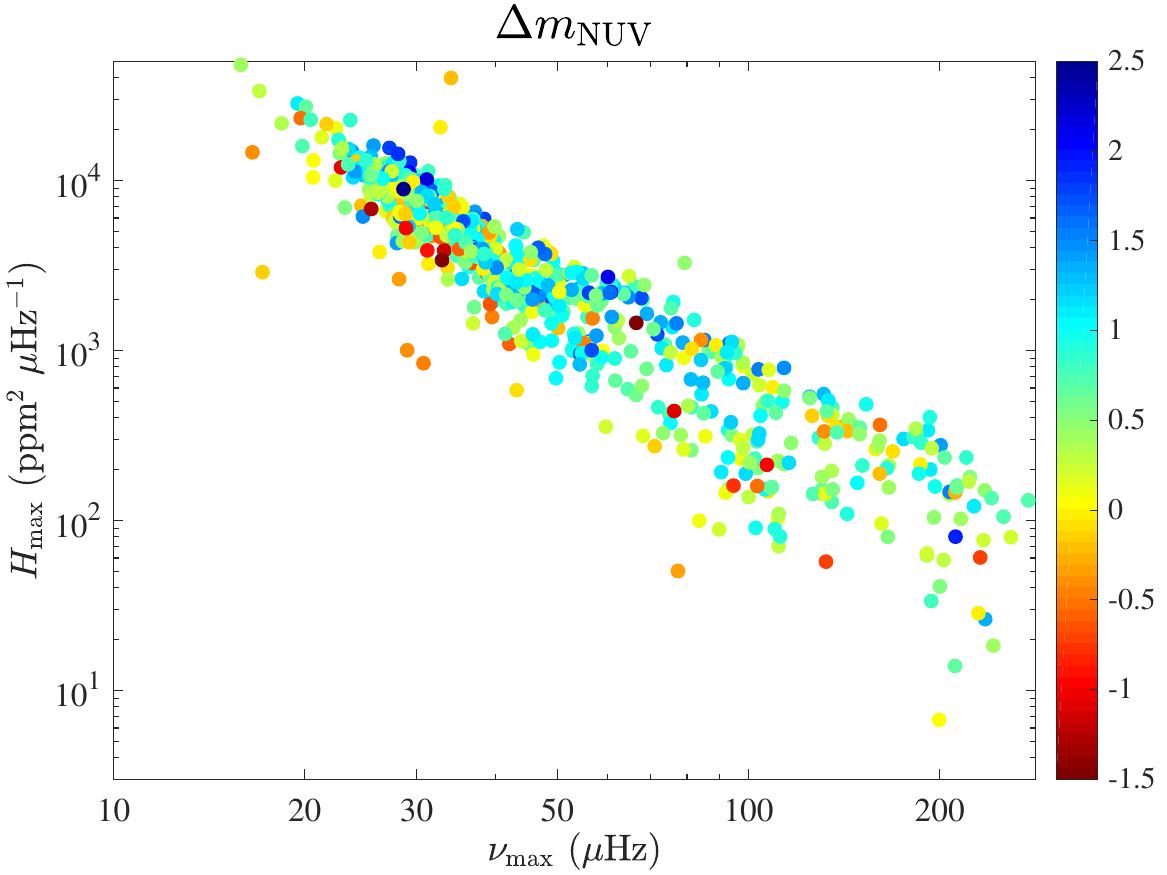}
\includegraphics[width=9cm]{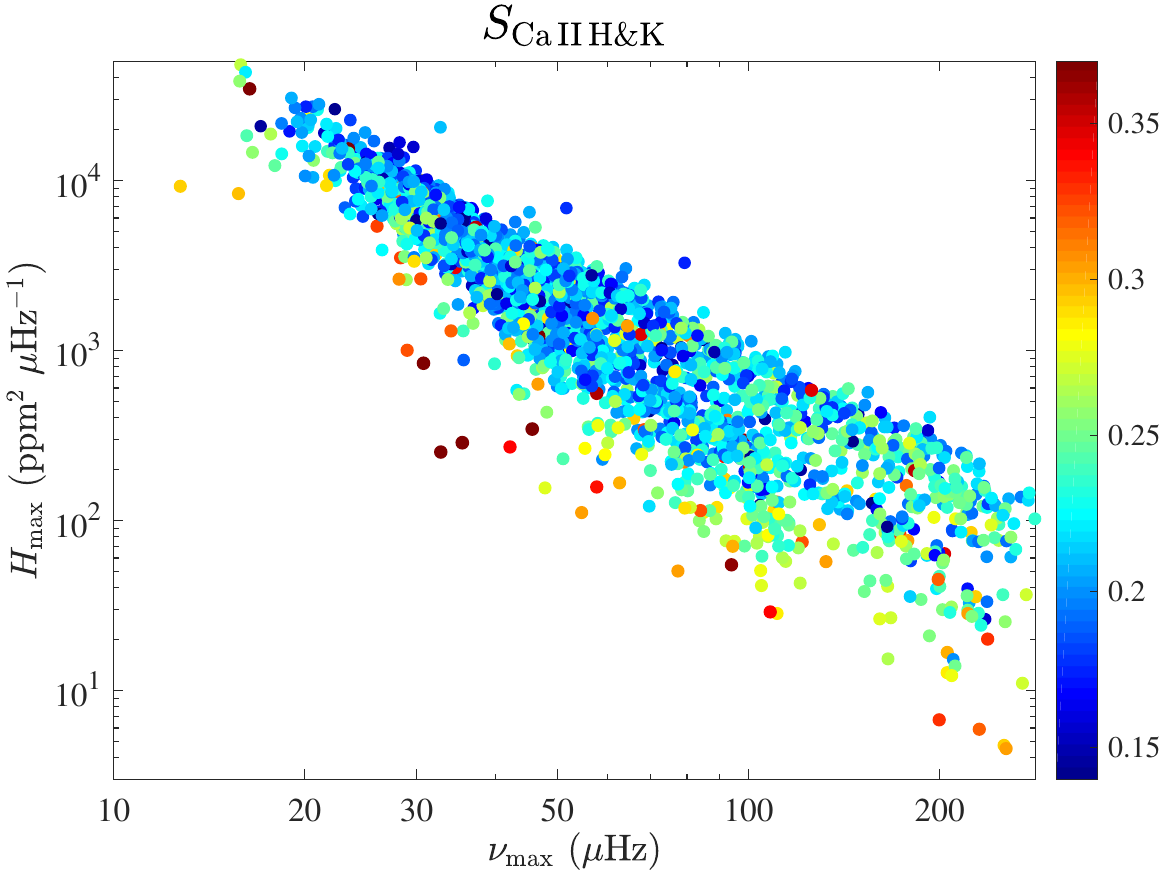}\\
\includegraphics[width=9cm]{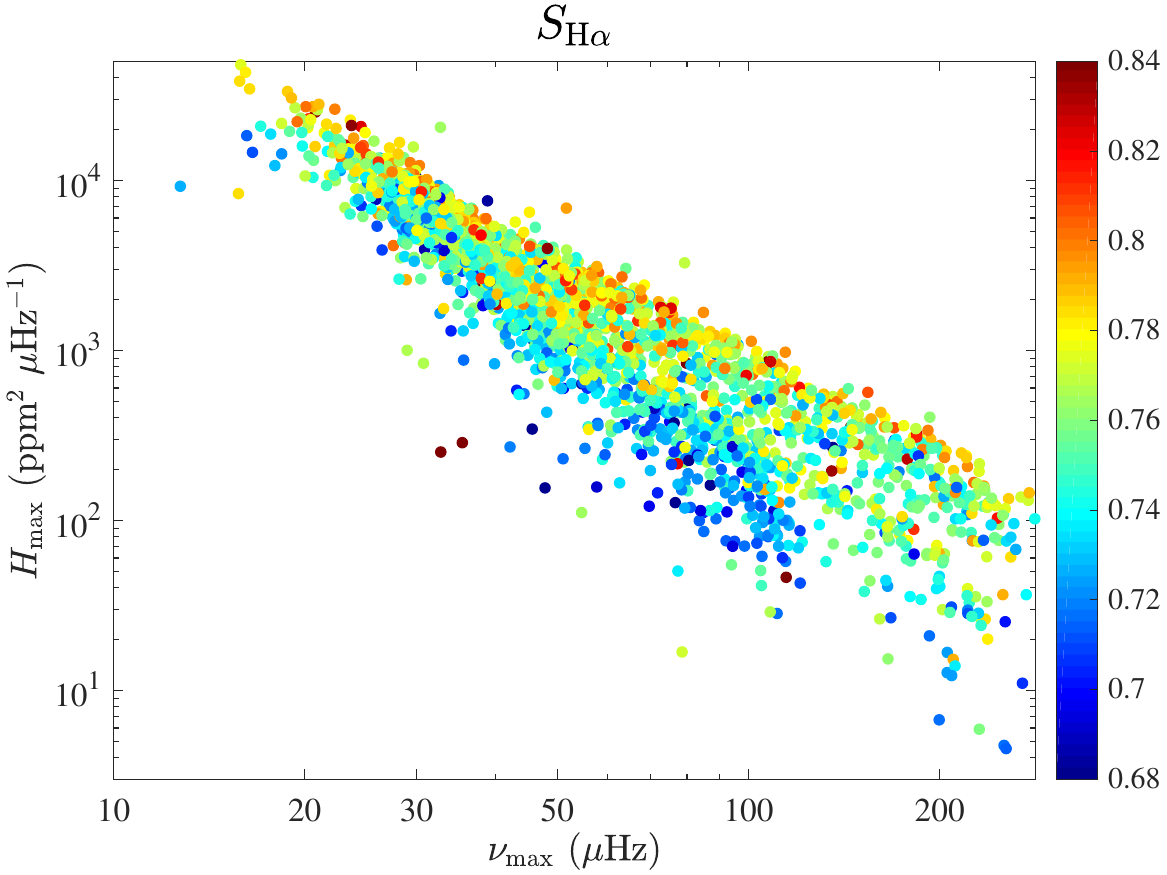}
\includegraphics[width=9cm]{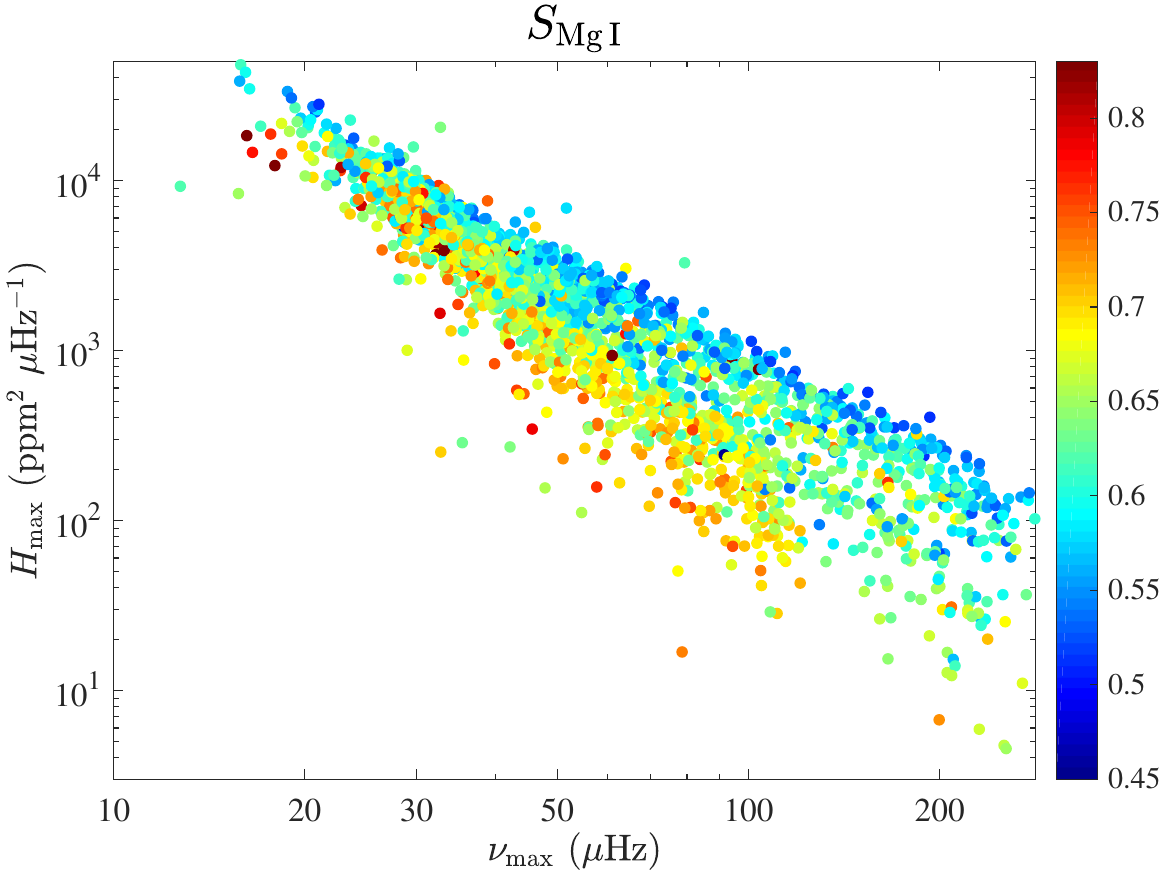}\\
\includegraphics[width=9cm]{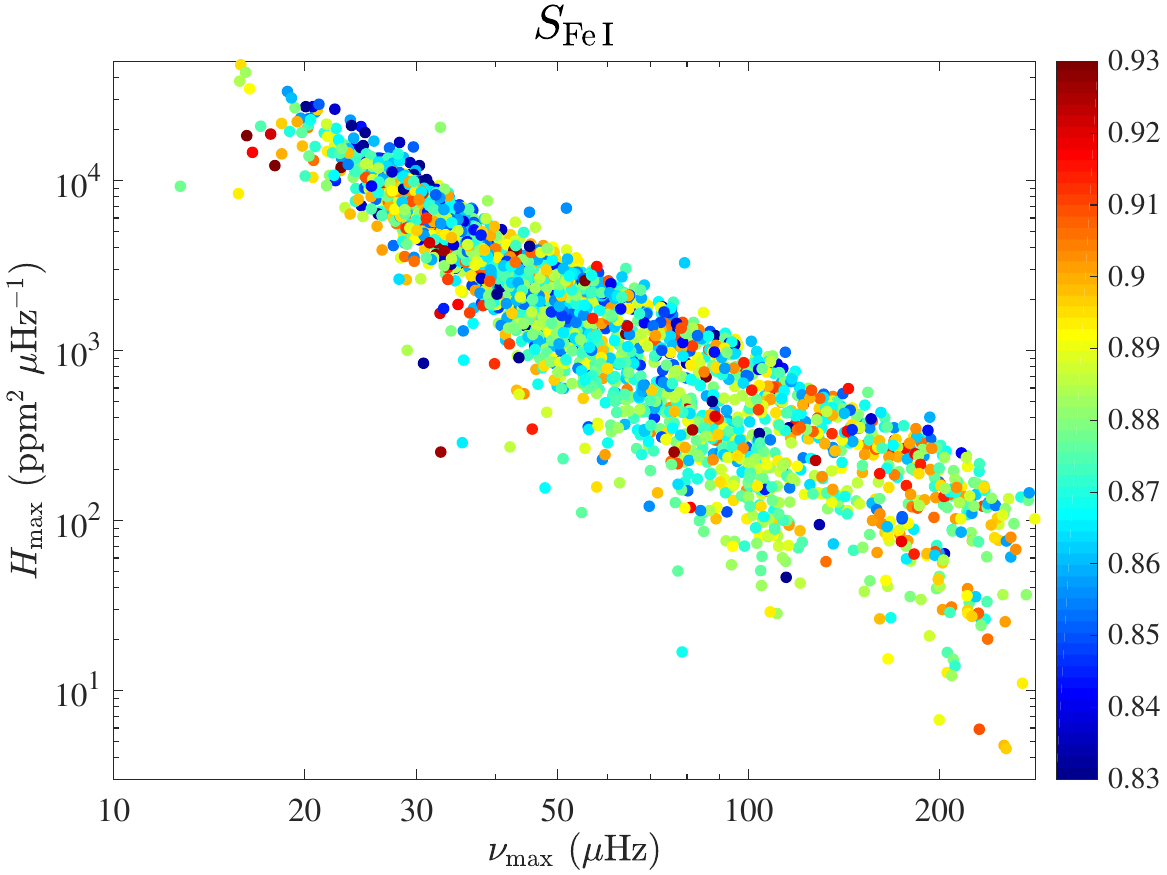}
\includegraphics[width=9cm]{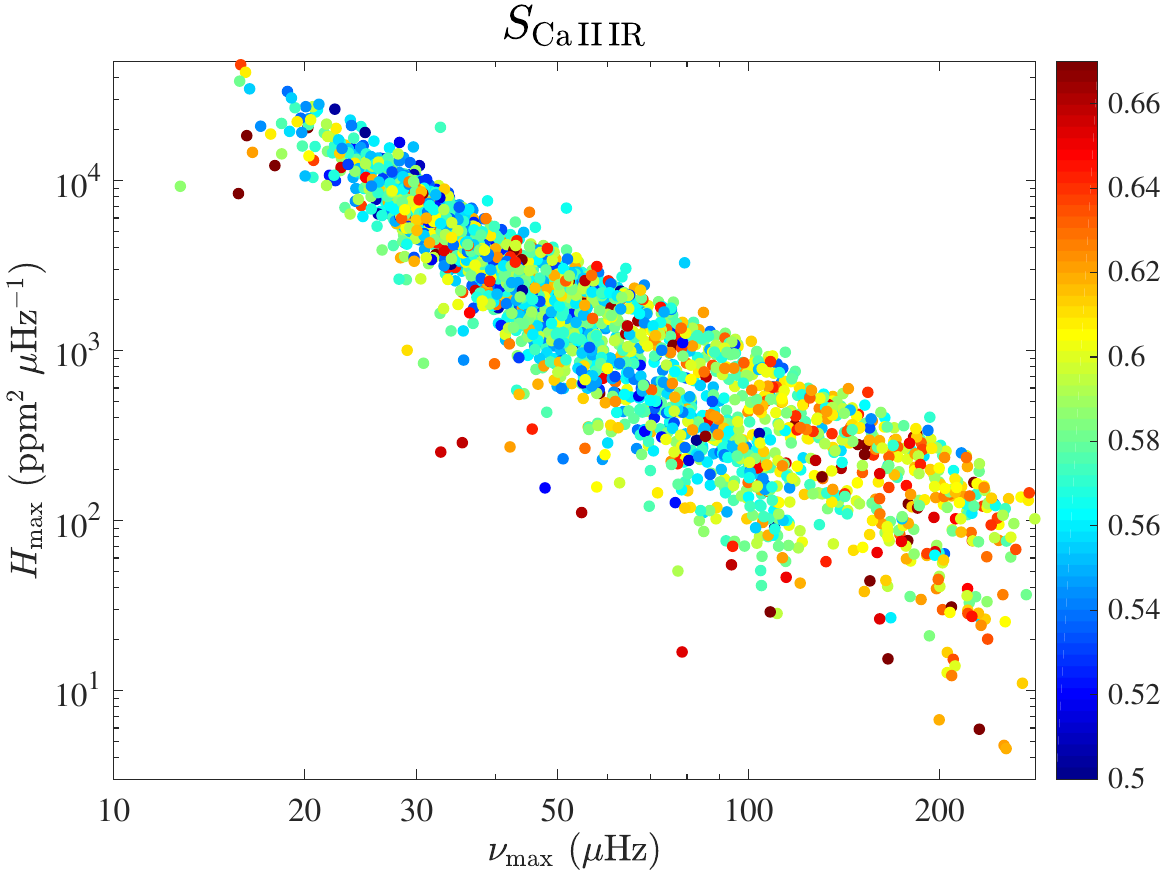}
\caption{Activity level (colorscale) as a function of oscillations frequency at maximum amplitude $\nu\ind{max}$ ($x$-axis) and height of the Gaussian envelope employed to model the oscillation excess power $H\ind{max}$ ($y$-axis), for the different activity indicators considered in this study.}
\label{fig_activity_Hmax_numax}
\end{figure*}

\section{Relation with oscillation amplitude}\label{oscillations}

Magnetic fields tend to inhibit convection, resulting in the partial or total suppression of oscillations \citep{Gaulme_2014, Gaulme_2016, Beck_2018, Benbakoura_2021}. Stars with low-amplitude oscillations or without oscillations are often associated with high S-index values \citep[][see also upper right panel of Fig. \ref{fig_activity_Hmax_numax}]{Bonanno_2014, Gehan_2022} and/or high $S\ind{ph}$ values \citep{Garcia_2010, Chaplin_2011, Gaulme_2014, Mathur_2019, Gaulme_2020, Benbakoura_2021}.

From Fig. \ref{fig_activity_Hmax_numax}, we first observe that the NUV excess is consistent with what was reported by \citet{Gehan_2022} about \cahk lines. Even though the trend is less clear than with the index of photometric variability $S'\ind{ph}$, a larger UV emission -- that is, lower negative $\Delta m\ind{NUV}$ magnitude difference -- corresponds with RG stars that display weaker oscillations. A similar trend is observed for \mgi lines even though the trend is much smoother than for \cahk and NUV. We comment on that point later in this section. The surprise comes from $S\ind{H\alpha}$: stars with relatively weak oscillations are associated with low $S\ind{H\alpha}$. In other words, oscillations are weaker when H$\alpha$ absorption lines get deeper. A similar but noisier trend is observed with \cair. As regards the control line, $S\ind{\fei}$ does not exhibit any correlation with the sample of stars that display photometric modulation and weak oscillations.

The trend of $S\ind{H\alpha}$ is then opposite to those of $\Delta m\ind{NUV}$, $S'\ind{ph}$ and $S\ind{CaII}$, and $S\ind{\mgi}$. 
As we discussed in Sect. \ref{comparison-litterature}, this different behaviour of $S\ind{H\alpha}$ could come from the larger sensitivity of the H$\alpha$ line to filaments \citep{Meunier_2009}, which appear to be dark on H$\alpha$-images of the Sun. In the case of the Sun, the number and length of filaments follow the sunspot cycle \citep{Mazumder_2021}, implying that higher solar activity levels result in an increased number and length of filaments. If the same is true for RGs, higher activity levels could result in an increase of the filaments contribution to $S\ind{H\alpha}$, hence in a decrease in $S\ind{H\alpha}$. In other terms, it would mean that in the H$\alpha$ band, the darkening caused by the presence of filaments on the stellar disk dominates the brightness brought by faculae, contrarily to the other spectral bands that we considered.

Besides the cases of stars that display rotational modulation, Fig. \ref{fig_activity_Hmax_numax} shows an interesting new finding. We note that the oscillations amplitude at a given $\numax$ is clearly correlated and anticorrelated with $S\ind{H\alpha}$ and $S\ind{\mgi}$, respectively, regardless of the detection of rotational modulation. The trend is smooth and affects the whole sample whatever $\numax$. Given that the range of effective temperature is quite narrow for RGs and that temperatures are mostly random at a given $\numax$ (see Fig. \ref{fig_activity_Hmax_numax_appendix}), a given $\numax$ means a given surface gravity $\log g$. If oscillation amplitude is driven by surface magnetic fields, the relative depth of the H$\alpha$ and \mgi lines are, to some extent, proxy of surface magnetic fields as well.

In Appendix \ref{appendix_mode_amp}, we display the oscillation amplitude versus $\numax$ where color scales are proportional to the stellar effective temperatures $\Teff$, surface gravity $\log g$, mass $M$ and radius $R$ of the sample. We note that similar figures for $M$ are displayed in \citet{Vrard_2018}.  At a given $\numax$, we see no significant trend between the oscillations amplitude, $\Teff$, and $\log g$ (see bottom panels of Fig. \ref{fig_activity_Hmax_numax_appendix}). On the contrary, the oscillations amplitude shows common features with what is observed for $S\ind{H\alpha}$ and $S\ind{\mgi}$, such as higher masses lead to lower amplitudes (see upper panels of Fig. \ref{fig_activity_Hmax_numax_appendix}). This is consistent with previous works for oscillating RGs, which have also noticed that the maximum mode amplitude decreases with $M$ \citep{Huber_2010, Huber_2011, Mosser_2011, Mosser_2012, Stello_2011, Kallinger_2014, Vrard_2018}. 
The fact that the depths of \ha and \mgi spectral lines are correlated with $M$ and $R$ is not totally surprising because surface magnetic fields must depend -- besides stellar rotation rates -- on stellar structures in the external convective envelope, which are connected to $M$ and $R$.

At last, it is worth mentioning that although the indices show an unusual behavior for $\numax$ between 25 and 35\,$\mu$Hz approximately, this mainly arise from a statistical effect. Indeed, this region includes both stars on the RGB and stars of the primary red clump (RC1), that is, low-mass stars that ignited helium fusion in their cores after passing though the helium flash and the tip of the RGB. We checked that RC1 stars do not especially show higher indices compared with more massive clump and RGB stars, but they are very numerous and highly concentrated in $\numax$, with a larger dispersion in their indices than more massive clump stars. In Fig. \ref{fig_activity_Hmax_numax}, we mainly see the RC1 stars with higher indices, while the others are masked.


\begin{table*}
\centering
\caption{Number $N$ of RGs with measured $P\ind{rot}$, as well as the Pearson correlation coefficients $r$ and p-values $p$ between different activity indicators and $P\ind{rot}$, for all RGs including the binaries as well as for single RGs only.}
\begin{tabular}{l r r}
\hline
Activity indicator & All RGs with $P\ind{rot}$ including binaries & Single RGs with $P\ind{rot}$ only\\
\hline
$\Delta m\ind{NUV}$ & $N$ = 63; $r$ = -0.369; $p$ = $2.9 \times 10^{-3}$ & $N$ = 56; $r$ = -0.366; $r$ = $5.5 \times 10^{-3}$ \\
$S\ind{\cahk}$ & $N$ = 274; $r$ = -0.428; $p$ = $1.2 \times 10^{-13}$ & $N$ = 218; $r$ = -0.271; $p$ = $4.9 \times 10^{-5}$ \\
$S\ind{H\alpha}$ & $N$ = 278; $r$ = -0.241; $p$ = $4.9 \times 10^{-5}$  & $N$ = 222; $r$ = -0.043; $p$ = 0.524 \\
$S\ind{\cair}$ & $N$ = 281; $r$ = -0.440; $p$ = $1.0 \times 10^{-14}$ & $N$ = 224; $r$ = -0.221; $p$ = $8.8 \times 10^{-4}$ \\
\hline
\end{tabular}
\label{table-Pearson-Prot}
\end{table*}

\begin{figure*}[t]
\includegraphics[width=9.7cm]{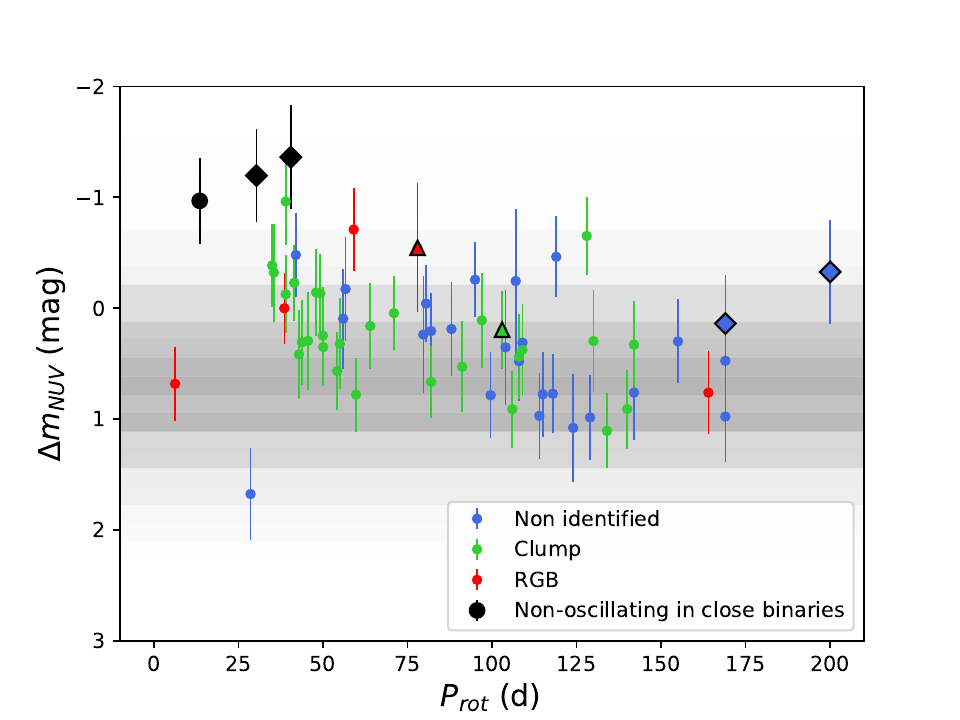}
\includegraphics[width=9.7cm]{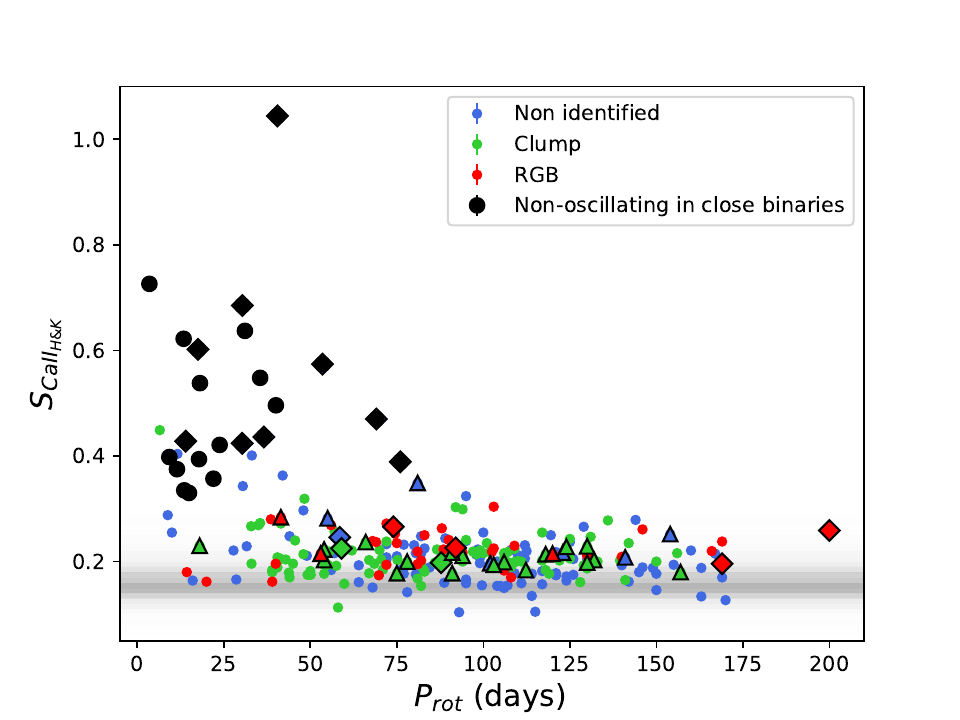}\\
\includegraphics[width=9.7cm]{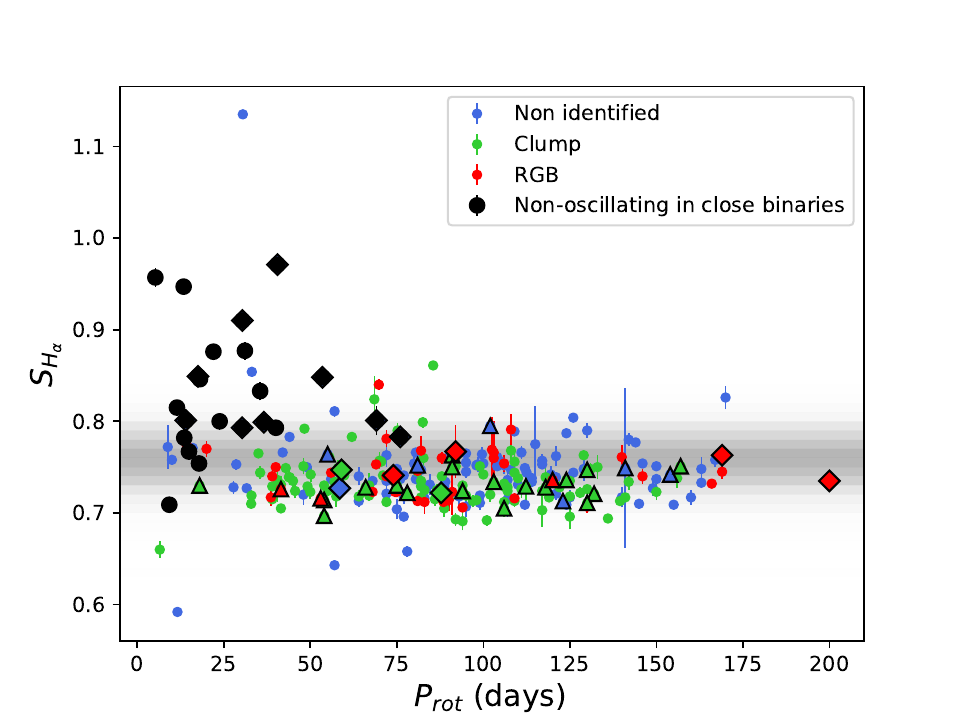}
\includegraphics[width=9.7cm]{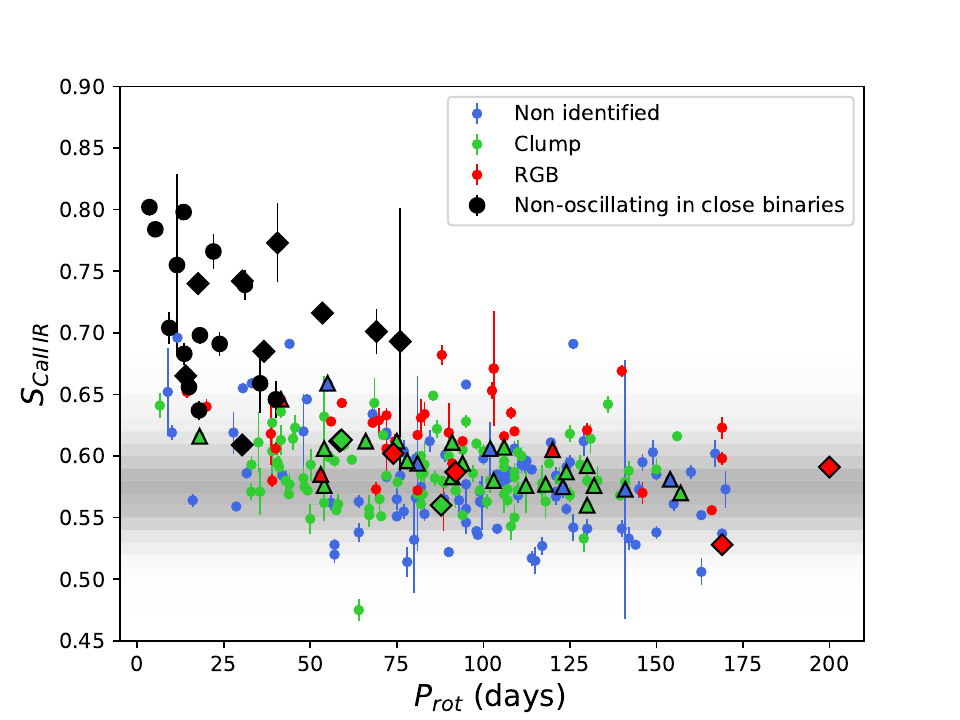}
\caption{Activity level as a function of the rotation period $P\ind{rot}$. The red giant branch (RGB) and red clump (RC) stars are represented in red and green, respectively, while RGs with an unidentified evolutionary stage are in blue. Black symbols correspond to non-oscillating RGs in close binaries from \citet{Gehan_2022}. Triangles and diamonds represent wide ($P\ind{orb}$ > 150 d) and close ($P\ind{orb} \leq$ 150 d) binaries identified by \textit{Gaia} DR3, respectively. The gray background indicates the distribution of the activity level for inactive stars, where a darker shade corresponds to a larger number of stars. \textit{Upper left:} $\Delta m\ind{NUV}$. \textit{Upper right: $S\ind{\cahk}$}. \textit{Lower left:} $S\ind{H\alpha}$. \textit{Lower right: $S\ind{\cair}$}.}
\label{fig-activity-Prot}
\end{figure*}

\section{Close binarity: impact of tidal interactions versus fast rotation}\label{binarity}

It has been observed that for a given rotation period, RGs in binary systems undergoing spin-orbit resonance or tidal locking  exhibit larger $S'\ind{ph }$ \citep{Gaulme_2020} and larger S-index \citep{Gehan_2022} than single RGs or RGs in binary systems with no special tidal configuration. Since we observe that RGs in close binaries present in average larger values of $\Delta m\ind{NUV}$, $S\ind{H\alpha}$ and $S\ind{\cair}$ (see Fig. \ref{fig-activity-indicators}), we check here the impact of fast rotation and tidal interactions on these indices. The Pearson correlation coefficients between the aforementioned indices and the rotation period $P\ind{rot}$ as well as the corresponding p-values are indicated in Table \ref{table-Pearson-Prot}.

\begin{figure*}[t]
\includegraphics[width=9.7cm]{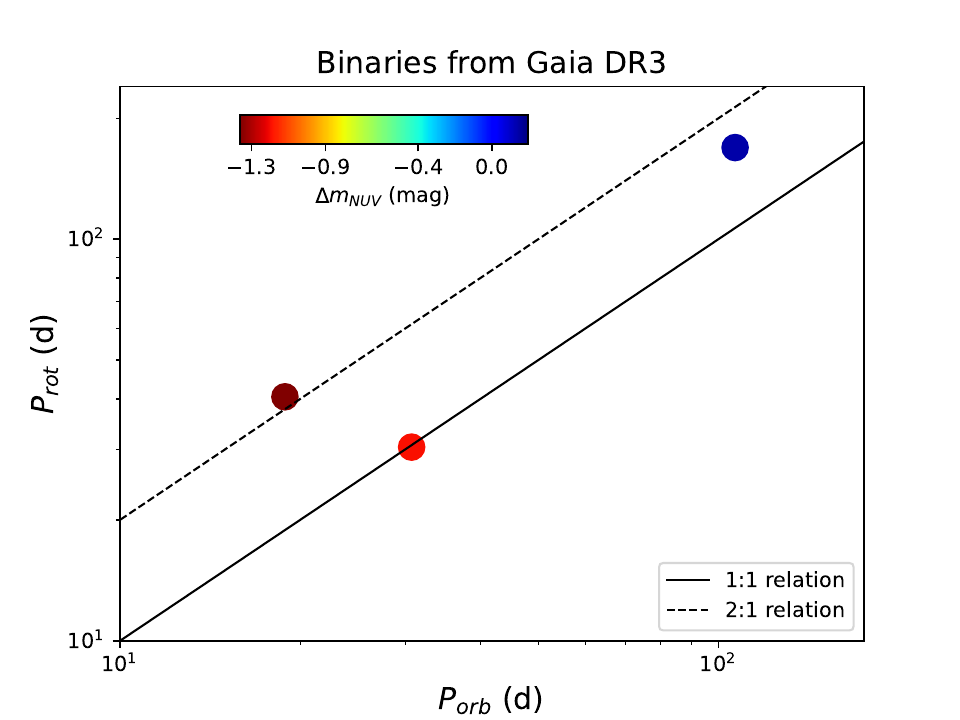}
\includegraphics[width=9.7cm]{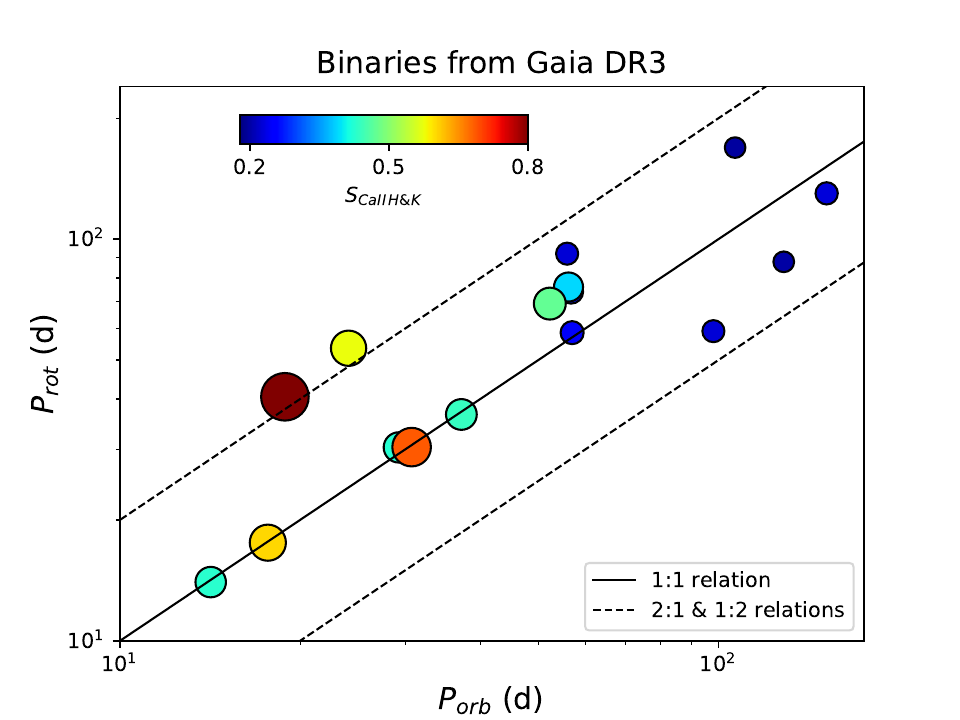}\\
\includegraphics[width=9.7cm]{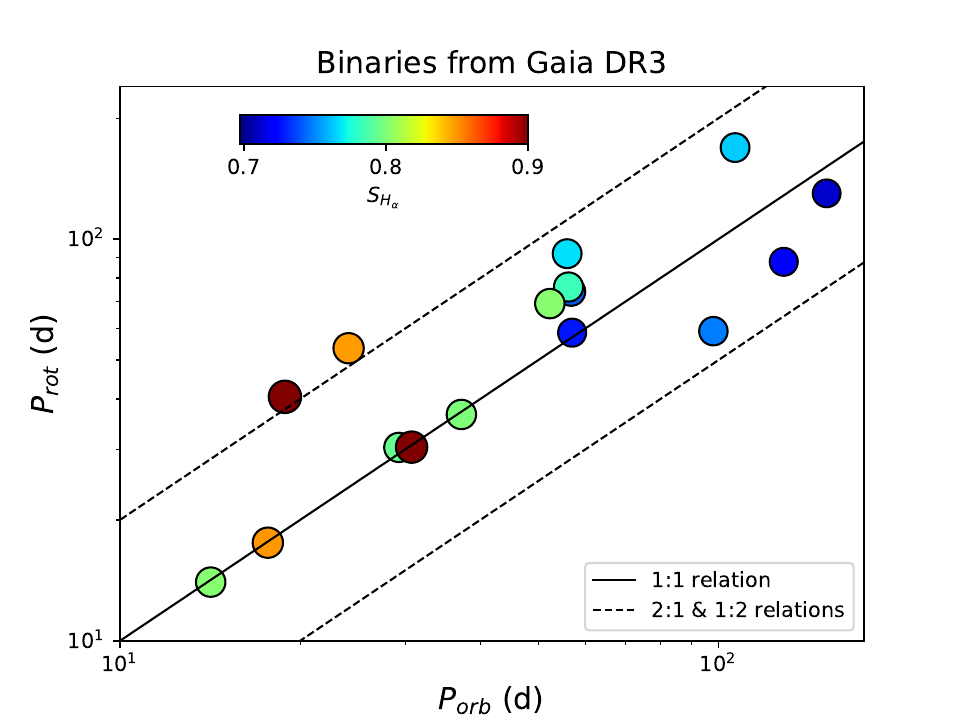}
\includegraphics[width=9.7cm]{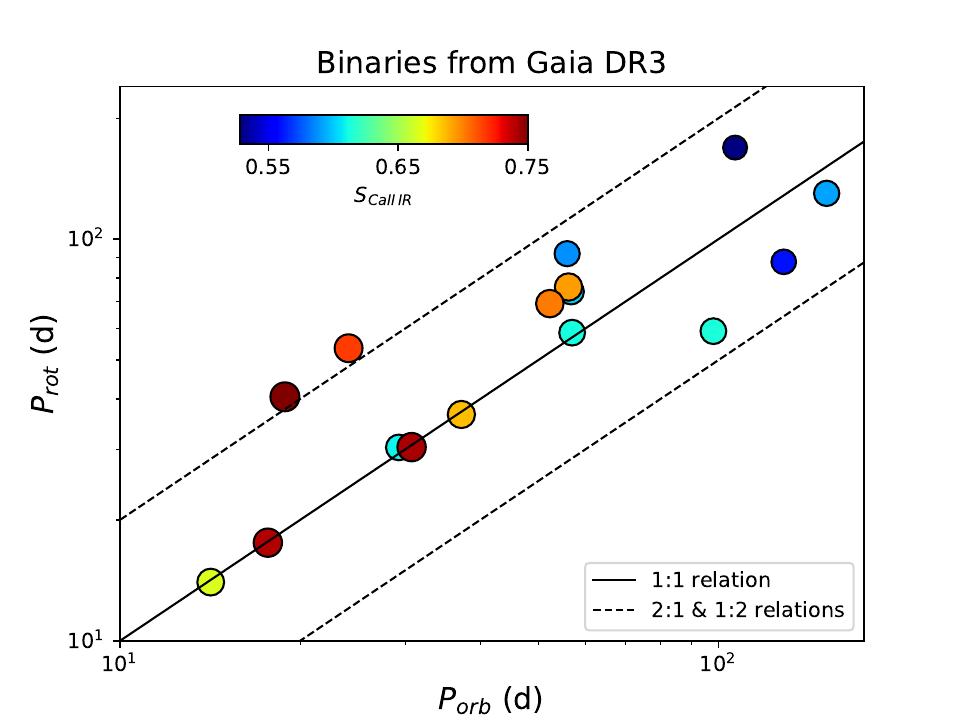}
\caption{Rotation period as a function of the orbital period for the close binaries listed by \textit{Gaia} DR3. The continuous lines represent the 1:1 relation, while the dashed lines represent the 2:1 and 1:2 relations. \textit{Upper left:} The symbols’ colors in each panel indicate the value of an activity indicator: $\Delta m\ind{NUV}$ (upper left), $S\ind{\cahk}$ (upper right), $S\ind{H\alpha}$ (lower left), $S\ind{\cair}$ (lower right).}
\label{fig-Prot-Porb}
\end{figure*}

\subsection{Impact of surface rotation}

The RGs for which \cite{Gaulme_2020} measured $P\ind{rot}$ are active in the sense that they exhibit spots on their photosphere. 
We observe that $\Delta m\ind{NUV}$, $S\ind{\cahk}$ and $S\ind{\cair}$ are anti-correlated with $P\ind{rot}$ (see Fig. \ref{fig-activity-Prot}). This trend is visible whether including the RGs in close binaries or not, although it is less pronounced when discarding the RGs in close binaries. We also find an anti-correlation between $S\ind{H\alpha}$ and $P\ind{rot}$, but there is no significant anti-correlation or correlation if we consider only the single RGs and discard the stars in close binaries. 

Such a decrease of magnetic activity as a function of the rotational period had already been observed using the S-index \citep[][see also upper right panel of Fig. \ref{fig-activity-Prot}]{Noyes_1984, Auriere_2015}. The observed trend for $\Delta m\ind{NUV}$ is consistent with the findings of \cite{Dixon_2020} for RGs and \cite{Godoy-Rivera_2021} for subgiants, as both works found that the NUV excess increases (i.e., becomes more negative) for faster rotators. We note that one single RG, KIC 11521644, stands among the fastest rotators in our sample while displaying the smallest $\Delta m\ind{NUV}$ among the RGs with measured rotation period (blue symbol on the bottom left in the left panel of Fig. \ref{fig-activity-Prot}). However, KIC 11521644 is actually likely a contaminant since we notice the presence of another close bright source on the Target Pixel File of its \textit{Kepler} light curve. In addition, we observed that its surface modulation is detectable only from the KEPSEISMIC light curve \footnote{Mathur, S., Santos, A. R. G., \& García, R. A. 2019, Kepler Light Curves Optimized For Asteroseismology (“KEPSEISMIC”), STScI/MAST, doi: \href{https://archive.stsci.edu/doi/resolve/resolve.html?doi=10.17909/t9-mrpw-gc07}{10.17909/T9-MRPW-GC07}} but not on the simple aperture photometry (SAP) light curve. Since KEPSEISMIC light curves are built on enlarged photometric apertures with respect to the SAP, it tends to confirm it is an artifact.
\cite{Newton_2017} observed a negative correlation between the \ha emission line strength and the rotation period, but they worked on a larger sample (466 stars) of less evolved stars (dwarfs), having a different spectral type than the RGs in our sample (M-type stars while our sample includes G- and K-type stars). A plausible explanation for this difference is that M dwarfs tend to be very active \citep[e.g.][]{Kiraga_2007}, which could increase the contribution of the chromospheric emission in the core of the \ha line compared to the contribution of filaments in the wings of the \ha line.
Overall, our findings provide further evidence for the rotation-activity connection in post main-sequence stars \citep{Lehtinen_2020}.

We also observe that RGs in close binaries present in average larger $\Delta m\ind{NUV}$, $S\ind{H\alpha}$ and $S\ind{\cair}$ values than single RGs, for a given rotation period. This is similar to what we observe for the photometric index \citep{Gaulme_2014, Gaulme_2020, Benbakoura_2021, Beck_2024} and the S-index \citep[][see also upper right panel of Fig. \ref{fig-activity-Prot}]{Gehan_2022}, reinforcing the idea that fast rotation alone is not enough to explain the enhanced magnetic activity observed for RGs in close binaries, and that tidal interactions are actually responsible for it.

\subsection{Impact of tidal interactions}

To investigate the role of tidal interactions in the chromospheric activity level, Fig. \ref{fig-Prot-Porb} displays the rotational versus orbital periods of the RGs in close binaries in our sample, similarly as \cite{Gaulme_2014}, \cite{Benbakoura_2021} and \cite{Beck_2024} for $S'\ind{ph}$, as well as \cite{Gehan_2022} for the S-index (see also upper right panel of Fig. \ref{fig-activity-Prot}).

We found no \textit{GALEX} NUV data for the close binaries studied by \cite{Gaulme_2014} and \cite{Benbakoura_2021} and we are left with only three RGs in close binaries with measured $\Delta m\ind{NUV}$ while having available rotation period from \cite{Gaulme_2020} and orbital period from \textit{Gaia} DR3. 
Two fast rotators exhibit large $\Delta m\ind{NUV}$ values, namely KIC 7869590 and KIC 11554998, while one slow rotator exhibits a small $\Delta m\ind{NUV}$, namely KIC 7661609. Notably, we observe that the two fast-rotating RGs belong respectively to a tidally locked system (1:1 resonance, KIC 11554998) and to a system in spin-orbit resonance (rotation period being twice the orbital period, KIC 7869590), while the system with the slowly-rotating RG does not have any special tidal configuration.
In a similar fashion, we also observe that RGs in systems that are tidally locked
or in spin-orbit resonance display the largest values of $S\ind{\cair}$ and $S\ind{H\alpha}$ compared to the systems that do not have any special tidal configuration.

\subsubsection{Comparison with previous studies}

Previous studies already noticed a possible relation between NUV emission and binarity for more evolved stars on the asymptotic giant branch \citep[stars undergoing both shell-hydrogen and shell-helium burning,][]{Sahai_2008, Sahai_2011, Ortiz_2016, Montez_2017}. For less evolved stars on the red giant branch and the red clump, \cite{Dixon_2020} observed that their binary sample presents larger $\Delta m\ind{NUV}$ values, stating however that this is likely due to fast rotation since the RGs in binaries are found to be rapidly rotating more often than single RGs. To the extent of our knowledge, the link between large $\Delta m\ind{NUV}$ values and tidal interactions was never established before.
In addition, \cite{Montes_1995} reported excess \ha emission for 51 chromospherically active binaries, including both RS Cvn  and BY Dra systems. They observed for these systems a slight decrease of the excess \ha emission with the rotation period, as well as a good correlation with the \caii K line.

Enhanced magnetic activity for RGs in systems that are either tidally locked or undergoing spin-orbit resonance has been previously observed using $S'\ind{ph}$ \citep{Gaulme_2014, Benbakoura_2021, Beck_2024} and the S-index \citep[][see also upper right panel of Fig. \ref{fig-activity-Prot}]{Gehan_2022} as activity indicators. Our results show that such systems also exhibit enhanced NUV excess, \ha index and \cair index, reinforcing the findings of \cite{Gehan_2022} that tidal interactions in close binaries drive abnormally large magnetic fields in RGs, rather than fast rotation by itself.

\subsubsection{Candidate mechanisms of tidal dynamo }

We here discuss some works that have focused on mechanisms that can in principle generate a tidal dynamo.
Tides in stars are generally decomposed into two components \citep[e.g.][] {Zahn_1977, Ogilvie_2013, Ogilvie_2014}. The first component is the equilibrium tide, which is a quasi-hydrostatic spheroidal bulge that follows the motion of the companion. This produces a large-scale flow under the form of elliptical streamlines inside the rotating star \citep{Zahn_1966, Remus_2012, Ogilvie_2013}. 
\citet{Wei_2022} found that the large-scale tidal flow can in principle induce a magnetic dynamo for solar-like close binaries with $P\ind{orb}$ up to 2-3 days. 
This is compatible with the findings of Yu et al. submitted who found an enhanced magnetic activity, above the saturation level known for single stars, for main-sequence and subgiant stars in close binaries with $P\ind{orb} \lesssim 2$ days. Although the red giants in close binaries in our sample have much larger orbital periods, the large scale tidal flow could still be responsible for the enhanced activity that we observe. Indeed, the degree of tidal interactions $T$ critically depends on $R/a$, where $a$ is the orbital distance between the two components of the system \citep{Zahn_1977, Hut_1981}. Assuming that the components have Keplerian orbits, the third Kepler law gives
\begin{equation}
a = P^{2/3}\ind{orb} \left ( \frac{G M\ind{1} M\ind{2}}{4 \pi^2} \right )^{1/3},
\end{equation}
where $M\ind{1}$ and $M\ind{2}$ are the masses of the primary and secondary components of the system, respectively.
Hence, we obtain
\begin{equation}
T \propto \frac{R}{P^{2/3}\ind{orb} (M\ind{1} M\ind{2})^{1/3}}.
\end{equation}
We can then evaluate the orbital period at which $T$ is similar for a system including a RG in comparison to a system with a MS star, under the rough assumption that $M\ind{1}$ and $M\ind{2}$ are similar for both systems, such as
\begin{equation}
 P\ind{orb, RG} \sim P\ind{orb, MS} \left ( \frac{R\ind{RG}}{R\ind{MS}} \right )^{3/2}.
\end{equation}
Taking as reference values $R\ind{RG}/R\ind{MS}$ = 10 and $P\ind{orb, MS}$ = 2-3 days, we obtain $P\ind{orb, RG} \sim$ 63-95 days. This estimate is compatible with our observations of RGs in close binaries with enhanced activity compared to single RGs (see Fig. \ref{fig-Prot-Porb}). This result indicates that the mechanism responsible for the enhanced activity observed by Yu et al. submitted for MS and subgiant stars in close binaries is likely to be the same as for the enhanced activity observed by \cite{Gehan_2022} and the present work for RGs in close binaries, with the tidal flow as a potential candidate.

The other component of stellar tides is the dynamical tide \citep{Zahn_1975}, which represents the tidal excitation of internal waves \citep[e.g.][]{Ogilvie_2007, Ogilvie_2013, Mathis_2015}. For stars with a convective envelope, such as RGs, the dynamical tide is constituted by tidal inertial waves driven by the Coriolis acceleration, which are excited under the condition that $P\ind{orb} > P\ind{rot} / 2$ \citep{Beck_2018, Astoul_2023}. \cite{Beck_2018} showed that due to the large convective envelope of RGs, the dynamical tide is expected to have a negligible contribution to tidal dissipation that drives spin-orbit evolution, confirming that the equilibrium tide is the main contributor to tidal dissipation \citep[e.g.][]{Zahn_1966, Remus_2012}.

However, the equilibrium tide can be subject to parametric instabilities \citep{Barker_2014}. One of them is the elliptical instability, which affects all rotating fluids with elliptically-deformed streamlines \citep{Kerswell_2002, Lacaze_2004, Le_Bars_2010, Cebron_2010, Cebron_2012}. The elliptical instability also drives inertial waves through a parametric resonance and is thought to be important for tidal dissipation in close binary stars \citep{Rieutord_2004}. Interestingly, we note that all the RGs in close binaries in our sample that present an enhanced magnetic activity have $P\ind{orb} > P\ind{rot} / 2$ (see Fig. \ref{fig-Prot-Porb}) and therefore lie in the regime where tidal inertial waves can propagate. It has been predicted and experimentally observed that the flows driven by the elliptical instability can induce a magnetic field \citep{Lacaze_2006, Herreman_2010}. \cite{Barker_2014} have shown using simulations that the elliptical instability is able to drive a small-scale dynamo in tidally distorted stars, and \cite{Cebron_2014} have shown using magnetohydrodynamic simulations that the elliptical instability is also able to generate a large-scale magnetic field. Hence, the elliptical instability appears as another potential candidate for the enhanced magnetic activity that we observe for RGs.


\begin{figure}[t]
\includegraphics[width=9.cm]{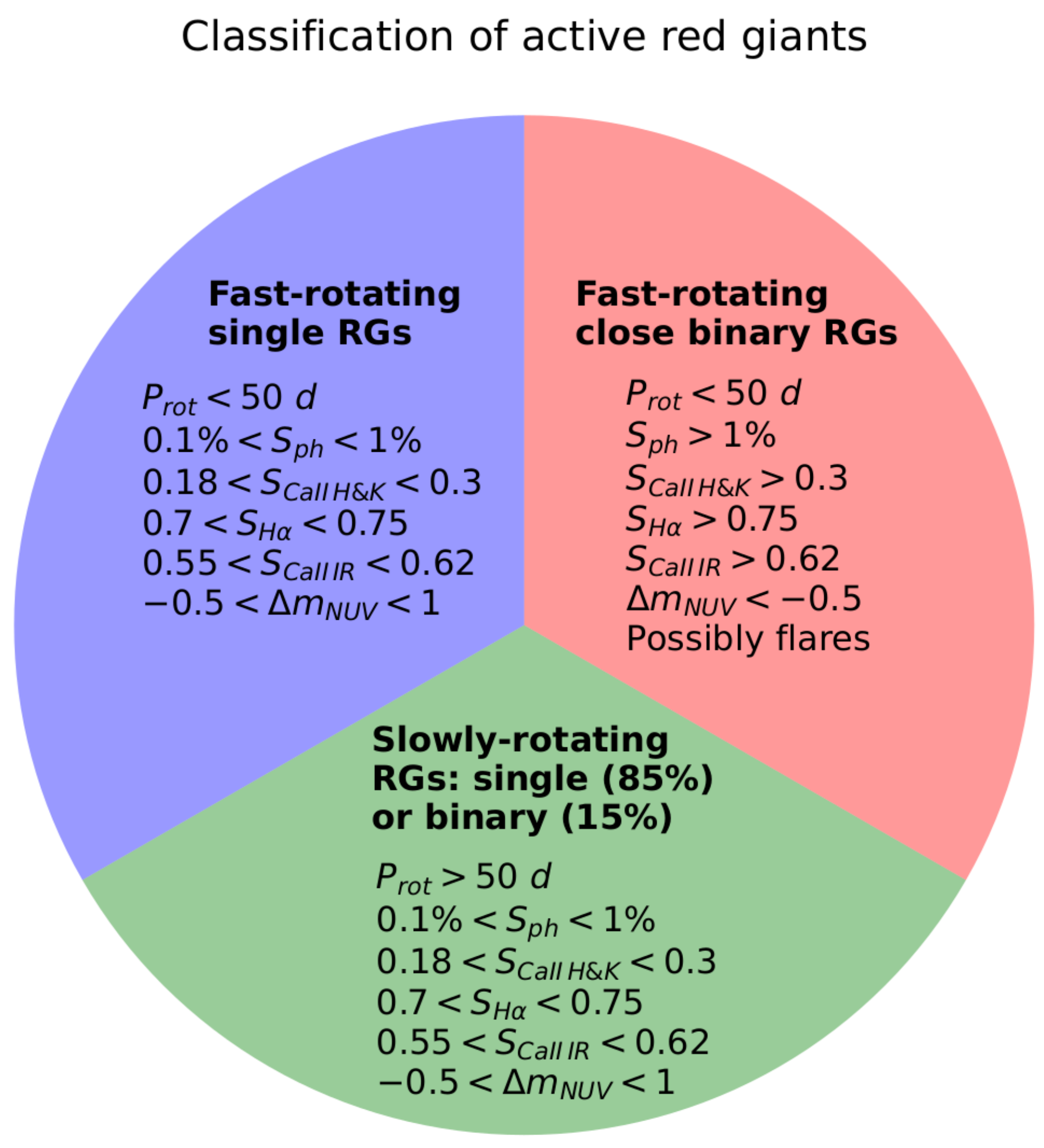}
\caption{Diagram classifying the active red giants, i.e. exhibiting rotational modulation due to the presence of spots, based on the findings of \citet[][$P\ind{rot}$ and $S\ind{ph}$]{Gaulme_2020}, \citet[][presence of flares for RGs in close binaries]{Olah_2021} and the present study ($S\ind{\cahk}$, $S\ind{H\alpha}$, $S\ind{\cair}$ and $\Delta m\ind{NUV}$).}
\label{fig-classification}
\end{figure}

\section{Classification of active red giants}\label{classification}

We here provide criteria to classify the active RGs, that have rotation period measurements resulting from the presence of spots on their photosphere producing photometric rotational modulation. Those criteria come from the findings of \cite{Gaulme_2020}, \cite{Olah_2021} and the present study and are summarized in Fig. \ref{fig-classification}.

We first focus on fast-rotating RGs ($P\ind{rot} < 50$ d). \cite{Gaulme_2020} found that RGs belonging to a close
binary system (red category in Fig. \ref{fig-classification}) displays $S\ind{ph}$ values about one
order of magnitude larger than that of single RGs (blue category in Fig. \ref{fig-classification}). This result suggested two possibilities: either tidal interactions somehow lead to stronger magnetic
fields, or the spot distribution differs between binary and single
red giants, leading to a different photometric variability.
\cite{Gehan_2022} could conclude that tidal interactions indeed lead to stronger magnetic fields, by also finding significantly larger chromospheric S-index values for close binary compared to single RGs.
In the present study, we confirm the finding of \cite{Gehan_2022} that RGs in close binaries tend to exhibit larger chromospheric $S\ind{\cahk}$ values. We also find that RGs in close binaries tend to exhibit larger chromospheric $S\ind{H\alpha}$,  $S\ind{\cair}$ and $\Delta m\ind{NUV}$ values than single RGs, reinforcing the conclusions of \cite{Gehan_2022}. In addition, \cite{Olah_2021} found that a large fraction of the 61 flaring RGs they studied are likely to belong
to close binary systems. Hence, the presence of flares is a possible additional indication of close binarity for RGs.

As regards slower rotating RGs ($P\ind{rot} \geq 50$ d, green category in Fig. \ref{fig-classification}), they tend to exhibit similar activity levels ($S\ind{ph}$, $S\ind{\cahk}$, $S\ind{H\alpha}$, $S\ind{\cair}$, $\Delta m\ind{NUV}$, low flare occurrence) than single fast-rotating RGs, whether they are in a binary or single configuration. Based on \cite{Gaulme_2020}, slower rotating RGs have a 85\% probability of being single, and a 15\% probability of being part of a binary system.


\section{Conclusions}\label{conclusions}

In the case of RGs, the amplitude of solar-like oscillations has been observed to decrease with increasing photospheric magnetic activity, as probed by measurements of the standard deviation of photometric time series \citep{Chaplin_2011, Huber_2011, Mathur_2019, Gaulme_2020}. In particular, based on a sample of about 4500 RGs observed by \textit{Kepler}, \citet{Gaulme_2020} noticed that the photometric modulation of RG light curves is about an order of magnitude larger for RG stars in close binary systems that are tidally locked than for single fast-spinning RGs rotating at the same rate. To understand whether such an increased level of photometric modulation is connected to an increased surface magnetic activity, \cite{Gehan_2022} measured the emission of \cahk lines ($S$-index) from LAMOST spectra of most of the sample used by \citet{Gaulme_2020}. They unambiguously showed that the larger is the photometric rotational modulation, the larger is the \cahk emission. In addition, as for photometric variability, there is an enhanced level of chromospheric emission for stars belonging to systems in a state of spin-orbit resonance. Similar results have subsequently been obtained with main-sequence stars observed by \textit{Gaia} and LAMOST (Yu et al., submitted).

In this paper, we looked for complementary information about active RGs by studying the depth of absorption lines that are known to  be sensitive to magnetic fields. What motivated us is the fact that the \cahk lines are at the very blue edge of LAMOST's spectral range and that the SNR is far from being optimal. The main limitation when working with LAMOST spectra is its relatively low resolution, which prevent us from working with any spectral line. Most thin and shallow lines are washed out by the resolution and blended with other lines. After a careful inspection of a sample of LAMOST spectra, we focused on a small set of deep and well defined absorption lines: H$\alpha$ at 6563\,\AA, the last of the \mgi triplet lines at 5184\,\AA, and the last of the infrared \caii triplet at 8542\,\AA. In addition, we measured the depth of a Fe line at 6495\,\AA, which actually is a blend of two iron lines, as a control variable.    
Finally, besides LAMOST spectra, we analyzed NUV data from the \textit{GALEX} survey, looking for emission around 2300\,\AA.

The outcome of our work confirms the general trend that was previously reported, but leads to a few surprises. Firstly, both H$\alpha$ and \mgi lines display a clear correlation with the oscillation amplitude. However, as noticed in Sect. \ref{oscillations}, the H$\alpha$ gets deeper as oscillations get weaker, whereas the opposite trend is observed with \mgi. We note that the observed correlation is much clearer with H$\alpha$ and \mgi than with \cahk \citep{Gehan_2022}. In particular, the correlation between line depth and oscillation amplitude is independent of the detection of photometric rotational modulation caused by spots on the photosphere. The IR \caii line appears to be less sensitive than \cahk, by showing a very noisy correlation with oscillation amplitude. The control Fe line does not exhibit any correlation with oscillation properties. We checked that there is no significant trend between the depth of the spectral lines studied in this work and the stellar parameters $\log g$ and $\Teff$. On the contrary, we found that the depth of the \ha and \mgi lines are correlated with the stellar mass $M$ and radius $R$, hence to the stellar structure. This is not surprising since the magnetic fields properties depend on the stellar structure in the external convective envelope.

Regarding the question of close binarity, \ha shows an enhanced emission for stars in close binaries, whereas no peculiar enhancement can be noticed for active RGs that display partially suppressed oscillations and a relatively low level of photometric modulation. In contrast, the \mgi line does not show any enhancement for close binaries, while \caii IR does. Similarly to chromospheric emission, we observe an enhanced level of UV emission for stars in close binary systems. In particular, we observe RGs in
systems that are tidally locked or in spin-orbit resonance display the largest values of $\Delta m \ind{NUV}$, $S\ind{H\alpha}$ and $S\ind{\cair}$ compared to the systems
that do not have any special tidal configuration. Based on previous works, a potential candidate for enhanced magnetic activity that we observe for RGs in close binaries could be the elliptical instability.

The difference of behavior that is observed between the lines that are known to be sensitive to chromospheric emission is certainly related to differences of altitude in the chromosphere where these lines probe. The fact that Mg and H$\alpha$ are anticorrelated is puzzling and understanding it is out of the scope of this paper. To go further in analyzing the spectra, we would need 3D atmospheric models of stellar atmospheres, such as CO5BOLD \citep{Freytag_2012} tuned for RG and coupled with a radiative transfer code such as OPTIM3D \citep{Chiavassa_2009}. We can imagine that combining asteroseismology, high-resolution spectroscopy, 3D modeling, with the support of RGs in eclipsing binaries to calibrate the limb darkening, will eventually lead to the possibility of determining the value of the magnetic fields at the surface of RG stars, active or quiet.

\begin{acknowledgements}
The authors thank B. Mosser, R. A. Garc\'ia, S. Mathur and E. G\"unther for useful discussions. C.G. was supported by funding from the European Research Council (ERC) under the European Union’s Horizon 2020 research and innovation programme under the ERC Synergy Grant WHOLE SUN 810218, as well as by Max Planck Society (Max Planck
Gesellschaft) grant “Preparations for PLATO Science” M.FE.A.Aero 0011.
D.G.R. acknowledges support from the Spanish Ministry of Science and Innovation (MICINN) with the grant No. PID2019-107187GB-I00. This work has made use of data collected by the NASA Galaxy Evolution Explorer (GALEX). This work has made use of data products from the Two Micron All Sky Survey, which is a joint project of the University of Massachusetts and the Infrared Processing and Analysis Center/California Institute of Technology, funded by the National Aeronautics and Space Administration and the National Science Foundation. This work has made use of data collected by Guoshoujing Telescope (the Large Sky Area Multi-Object Fiber Spectroscopic Telescope LAMOST), a National Major Scientific Project built by the Chinese Academy of Sciences. Funding for the LAMOST project has been provided by the National Development and Reform Commission. LAMOST is operated and managed by the National Astronomical Observatories, Chinese Academy of Sciences. This work has made use of data collected by the \textit{Kepler} mission, which data have been obtained from the MAST data archive at the Space Telescope Science Institute (STScI). Funding for the \textit{Kepler} mission is provided by the NASA Science Mission Directorate. STScI is operated by the Association of Universities for Research in Astronomy, Inc., under NASA contract NAS 5–26555. This work has made use of data collected by the European Space Agency (ESA) mission {\it Gaia} (\url{https://www.cosmos.esa.int/gaia}), which data have been processed by the {\it Gaia} Data Processing and Analysis Consortium (DPAC, \url{https://www.cosmos.esa.int/web/gaia/dpac/consortium}). Funding for the DPAC
has been provided by national institutions, in particular the institutions participating in the {\it Gaia} Multilateral Agreement.
\end{acknowledgements}

\bibliographystyle{aa} 
\bibliography{biblio}

\begin{thebibliography}{114}
\expandafter\ifx\csname natexlab\endcsname\relax\def\natexlab#1{#1}\fi

\bibitem[{{Airapetian} {et~al.}(2020){Airapetian}, {Barnes}, {Cohen},
  {Collinson}, {Danchi}, {Dong}, {Del Genio}, {France}, {Garcia-Sage},
  {Glocer}, {Gopalswamy}, {Grenfell}, {Gronoff}, {G{\"u}del}, {Herbst},
  {Henning}, {Jackman}, {Jin}, {Johnstone}, {Kaltenegger}, {Kay}, {Kobayashi},
  {Kuang}, {Li}, {Lynch}, {L{\"u}ftinger}, {Luhmann}, {Maehara}, {Mlynczak},
  {Notsu}, {Osten}, {Ramirez}, {Rugheimer}, {Scheucher}, {Schlieder},
  {Shibata}, {Sousa-Silva}, {Stamenkovi{\'c}}, {Strangeway}, {Usmanov},
  {Vergados}, {Verkhoglyadova}, {Vidotto}, {Voytek}, {Way}, {Zank}, \&
  {Yamashiki}}]{Airapetian_2020}
{Airapetian}, V.~S., {Barnes}, R., {Cohen}, O., {et~al.} 2020, International
  Journal of Astrobiology, 19, 136

\bibitem[{{Andrae} {et~al.}(2023){Andrae}, {Rix}, \& {Chandra}}]{Andrae_2023}
{Andrae}, R., {Rix}, H.-W., \& {Chandra}, V. 2023, \apjs, 267, 8

\bibitem[{{Astoul} \& {Barker}(2023)}]{Astoul_2023}
{Astoul}, A. \& {Barker}, A.~J. 2023, \apjl, 955, L23

\bibitem[{{Auri{\`e}re} {et~al.}(2015){Auri{\`e}re}, {Konstantinova-Antova},
  {Charbonnel}, {Wade}, {Tsvetkova}, {Petit}, {Dintrans}, {Drake}, {Decressin},
  {Lagarde}, {Donati}, {Roudier}, {Ligni{\`e}res}, {Schr{\"o}der},
  {Landstreet}, {L{\`e}bre}, {Weiss}, \& {Zahn}}]{Auriere_2015}
{Auri{\`e}re}, M., {Konstantinova-Antova}, R., {Charbonnel}, C., {et~al.} 2015,
  \aap, 574, A90

\bibitem[{{Babcock}(1961)}]{Babcock_1961}
{Babcock}, H.~W. 1961, \apj, 133, 572

\bibitem[{{Barker} \& {Lithwick}(2014)}]{Barker_2014}
{Barker}, A.~J. \& {Lithwick}, Y. 2014, \mnras, 437, 305

\bibitem[{{Beck} {et~al.}(2024){Beck}, {Grossmann}, {Steinwender}, {Schimak},
  {Muntean}, {Vrard}, {Patton}, {Merc}, {Mathur}, {Garcia}, {Pinsonneault},
  {Rowan}, {Gaulme}, {Allende Prieto}, {Arellano-C{\'o}rdova}, {Cao},
  {Corsaro}, {Creevey}, {Hambleton}, {Hanslmeier}, {Holl}, {Johnson}, {Mathis},
  {Godoy-Rivera}, {S{\'\i}mon-D{\'\i}az}, \& {Zinn}}]{Beck_2024}
{Beck}, P.~G., {Grossmann}, D.~H., {Steinwender}, L., {et~al.} 2024, \aap, 682,
  A7

\bibitem[{{Beck} {et~al.}(2018){Beck}, {Mathis}, {Gallet}, {Charbonnel},
  {Benbakoura}, {Garc{\'\i}a}, \& {do Nascimento}}]{Beck_2018}
{Beck}, P.~G., {Mathis}, S., {Gallet}, F., {et~al.} 2018, \mnras, 479, L123

\bibitem[{{Benbakoura} {et~al.}(2021){Benbakoura}, {Gaulme}, {McKeever},
  {Sekaran}, {Beck}, {Spada}, {Jackiewicz}, {Mathis}, {Mathur}, {Tkachenko}, \&
  {Garc{\'\i}a}}]{Benbakoura_2021}
{Benbakoura}, M., {Gaulme}, P., {McKeever}, J., {et~al.} 2021, \aap, 648, A113

\bibitem[{{Bianchi} {et~al.}(2005){Bianchi}, {Seibert}, {Zheng}, {Thilker},
  {Friedman}, {Wyder}, {Donas}, {Barlow}, {Byun}, {Forster}, {Heckman},
  {Jelinsky}, {Lee}, {Madore}, {Malina}, {Martin}, {Milliard}, {Morrissey},
  {Neff}, {Rich}, {Schiminovich}, {Siegmund}, {Small}, {Szalay}, \&
  {Welsh}}]{Bianchi_2005}
{Bianchi}, L., {Seibert}, M., {Zheng}, W., {et~al.} 2005, \apjl, 619, L27

\bibitem[{{Bianchi} {et~al.}(2017){Bianchi}, {Shiao}, \&
  {Thilker}}]{Bianchi_2017}
{Bianchi}, L., {Shiao}, B., \& {Thilker}, D. 2017, \apjs, 230, 24

\bibitem[{{Bj{\o}rgen} {et~al.}(2018){Bj{\o}rgen}, {Sukhorukov}, {Leenaarts},
  {Carlsson}, {de la Cruz Rodr{\'\i}guez}, {Scharmer}, \&
  {Hansteen}}]{Bjorgen_2018}
{Bj{\o}rgen}, J.~P., {Sukhorukov}, A.~V., {Leenaarts}, J., {et~al.} 2018, \aap,
  611, A62

\bibitem[{{Bonanno} {et~al.}(2014){Bonanno}, {Corsaro}, \&
  {Karoff}}]{Bonanno_2014}
{Bonanno}, A., {Corsaro}, E., \& {Karoff}, C. 2014, \aap, 571, A35

\bibitem[{{Borgniet} {et~al.}(2015){Borgniet}, {Meunier}, \&
  {Lagrange}}]{Borgniet_2015}
{Borgniet}, S., {Meunier}, N., \& {Lagrange}, A.~M. 2015, \aap, 581, A133

\bibitem[{{Borucki} {et~al.}(2010){Borucki}, {Koch}, {Basri}, {Batalha},
  {Brown}, {Caldwell}, {Caldwell}, {Christensen-Dalsgaard}, {Cochran},
  {DeVore}, {Dunham}, {Dupree}, {Gautier}, {Geary}, {Gilliland}, {Gould},
  {Howell}, {Jenkins}, {Kondo}, {Latham}, {Marcy}, {Meibom}, {Kjeldsen},
  {Lissauer}, {Monet}, {Morrison}, {Sasselov}, {Tarter}, {Boss}, {Brownlee},
  {Owen}, {Buzasi}, {Charbonneau}, {Doyle}, {Fortney}, {Ford}, {Holman},
  {Seager}, {Steffen}, {Welsh}, {Rowe}, {Anderson}, {Buchhave}, {Ciardi},
  {Walkowicz}, {Sherry}, {Horch}, {Isaacson}, {Everett}, {Fischer}, {Torres},
  {Johnson}, {Endl}, {MacQueen}, {Bryson}, {Dotson}, {Haas}, {Kolodziejczak},
  {Van Cleve}, {Chandrasekaran}, {Twicken}, {Quintana}, {Clarke}, {Allen},
  {Li}, {Wu}, {Tenenbaum}, {Verner}, {Bruhweiler}, {Barnes}, \&
  {Prsa}}]{Borucki_2010}
{Borucki}, W.~J., {Koch}, D., {Basri}, G., {et~al.} 2010, Science, 327, 977

\bibitem[{{Brown} {et~al.}(2022){Brown}, {Jeffers}, {Marsden}, {Morin}, {Boro
  Saikia}, {Petit}, {Jardine}, {See}, {Vidotto}, {Mengel}, {Dahlkemper}, \&
  {BCool Collaboration}}]{Brown_2022}
{Brown}, E.~L., {Jeffers}, S.~V., {Marsden}, S.~C., {et~al.} 2022, \mnras, 514,
  4300

\bibitem[{{Bus{\`a}} {et~al.}(2007){Bus{\`a}}, {Aznar Cuadrado}, {Terranegra},
  {Andretta}, \& {Gomez}}]{Busa_2007}
{Bus{\`a}}, I., {Aznar Cuadrado}, R., {Terranegra}, L., {Andretta}, V., \&
  {Gomez}, M.~T. 2007, \aap, 466, 1089

\bibitem[{{C{\'e}bron} \& {Hollerbach}(2014)}]{Cebron_2014}
{C{\'e}bron}, D. \& {Hollerbach}, R. 2014, \apjl, 789, L25

\bibitem[{{C{\'e}bron} {et~al.}(2010){C{\'e}bron}, {Le Bars}, {Leontini},
  {Maubert}, \& {Le Gal}}]{Cebron_2010}
{C{\'e}bron}, D., {Le Bars}, M., {Leontini}, J., {Maubert}, P., \& {Le Gal}, P.
  2010, Physics of the Earth and Planetary Interiors, 182, 119

\bibitem[{{C{\'e}bron} {et~al.}(2012){C{\'e}bron}, {Le Bars}, {Maubert}, \& {Le
  Gal}}]{Cebron_2012}
{C{\'e}bron}, D., {Le Bars}, M., {Maubert}, P., \& {Le Gal}, P. 2012,
  Geophysical and Astrophysical Fluid Dynamics, 106, 524

\bibitem[{{Ceillier} {et~al.}(2017){Ceillier}, {Tayar}, {Mathur}, {Salabert},
  {Garc{\'\i}a}, {Stello}, {Pinsonneault}, {van Saders}, {Beck}, \&
  {Bloemen}}]{Ceillier_2017}
{Ceillier}, T., {Tayar}, J., {Mathur}, S., {et~al.} 2017, \aap, 605, A111

\bibitem[{{Chaplin} {et~al.}(2011){Chaplin}, {Bedding}, {Bonanno}, {Broomhall},
  {Garc{\'\i}a}, {Hekker}, {Huber}, {Verner}, {Basu}, {Elsworth}, {Houdek},
  {Mathur}, {Mosser}, {New}, {Stevens}, {Appourchaux}, {Karoff}, {Metcalfe},
  {Molenda-{\.Z}akowicz}, {Monteiro}, {Thompson}, {Christensen-Dalsgaard},
  {Gilliland}, {Kawaler}, {Kjeldsen}, {Ballot}, {Benomar}, {Corsaro},
  {Campante}, {Gaulme}, {Hale}, {Handberg}, {Jarvis}, {R{\'e}gulo}, {Roxburgh},
  {Salabert}, {Stello}, {Mullally}, {Li}, \& {Wohler}}]{Chaplin_2011}
{Chaplin}, W.~J., {Bedding}, T.~R., {Bonanno}, A., {et~al.} 2011, \apjl, 732,
  L5

\bibitem[{{Charbonneau}(2014)}]{Charbonneau_2014}
{Charbonneau}, P. 2014, \araa, 52, 251

\bibitem[{{Chiavassa} {et~al.}(2009){Chiavassa}, {Plez}, {Josselin}, \&
  {Freytag}}]{Chiavassa_2009}
{Chiavassa}, A., {Plez}, B., {Josselin}, E., \& {Freytag}, B. 2009, \aap, 506,
  1351

\bibitem[{{Cincunegui} {et~al.}(2007){Cincunegui}, {D{\'\i}az}, \&
  {Mauas}}]{Cincunegui_2007}
{Cincunegui}, C., {D{\'\i}az}, R.~F., \& {Mauas}, P.~J.~D. 2007, \aap, 469, 309

\bibitem[{{Cutri} {et~al.}(2003){Cutri}, {Skrutskie}, {van Dyk}, {Beichman},
  {Carpenter}, {Chester}, {Cambresy}, {Evans}, {Fowler}, {Gizis}, {Howard},
  {Huchra}, {Jarrett}, {Kopan}, {Kirkpatrick}, {Light}, {Marsh}, {McCallon},
  {Schneider}, {Stiening}, {Sykes}, {Weinberg}, {Wheaton}, {Wheelock}, \&
  {Zacarias}}]{Cutri_2003}
{Cutri}, R.~M., {Skrutskie}, M.~F., {van Dyk}, S., {et~al.} 2003, VizieR Online
  Data Catalog, II/246

\bibitem[{{Dixon} {et~al.}(2020){Dixon}, {Tayar}, \& {Stassun}}]{Dixon_2020}
{Dixon}, D., {Tayar}, J., \& {Stassun}, K.~G. 2020, \aj, 160, 12

\bibitem[{{Duncan} {et~al.}(1991){Duncan}, {Vaughan}, {Wilson}, {Preston},
  {Frazer}, {Lanning}, {Misch}, {Mueller}, {Soyumer}, {Woodard}, {Baliunas},
  {Noyes}, {Hartmann}, {Porter}, {Zwaan}, {Middelkoop}, {Rutten}, \&
  {Mihalas}}]{Duncan_1991}
{Duncan}, D.~K., {Vaughan}, A.~H., {Wilson}, O.~C., {et~al.} 1991, \apjs, 76,
  383

\bibitem[{{Findeisen} \& {Hillenbrand}(2010)}]{Findeisen_2010}
{Findeisen}, K. \& {Hillenbrand}, L. 2010, \aj, 139, 1338

\bibitem[{{Findeisen} {et~al.}(2011){Findeisen}, {Hillenbrand}, \&
  {Soderblom}}]{Findeisen_2011}
{Findeisen}, K., {Hillenbrand}, L., \& {Soderblom}, D. 2011, \aj, 142, 23

\bibitem[{{Freytag} {et~al.}(2012){Freytag}, {Steffen}, {Ludwig},
  {Wedemeyer-B{\"o}hm}, {Schaffenberger}, \& {Steiner}}]{Freytag_2012}
{Freytag}, B., {Steffen}, M., {Ludwig}, H.~G., {et~al.} 2012, Journal of
  Computational Physics, 231, 919

\bibitem[{{Gaia Collaboration} {et~al.}(2023{\natexlab{a}}){Gaia
  Collaboration}, {Arenou}, {Babusiaux}, {Barstow}, {Faigler}, {Jorissen},
  {Kervella}, {Mazeh}, {Mowlavi}, {Panuzzo}, {Sahlmann}, {Shahaf}, {Sozzetti},
  {Bauchet}, {Damerdji}, {Gavras}, {Giacobbe}, {Gosset}, {Halbwachs}, {Holl},
  {Lattanzi}, {Leclerc}, {Morel}, {Pourbaix}, {Re Fiorentin}, {Sadowski},
  {S{\'e}gransan}, {Siopis}, {Teyssier}, {Zwitter}, {Planquart}, {Brown},
  {Vallenari}, {Prusti}, {de Bruijne}, {Biermann}, {Creevey}, {Ducourant},
  {Evans}, {Eyer}, {Guerra}, {Hutton}, {Jordi}, {Klioner}, {Lammers},
  {Lindegren}, {Luri}, {Mignard}, {Panem}, {Randich}, {Sartoretti}, {Soubiran},
  {Tanga}, {Walton}, {Bailer-Jones}, {Bastian}, {Drimmel}, {Jansen}, {Katz},
  {van Leeuwen}, {Bakker}, {Cacciari}, {Casta{\~n}eda}, {De Angeli},
  {Fabricius}, {Fouesneau}, {Fr{\'e}mat}, {Galluccio}, {Guerrier}, {Heiter},
  {Masana}, {Messineo}, {Nicolas}, {Nienartowicz}, {Pailler}, {Riclet}, {Roux},
  {Seabroke}, {Sordo}, {Th{\'e}venin}, {Gracia-Abril}, {Portell}, {Altmann},
  {Andrae}, {Audard}, {Bellas-Velidis}, {Benson}, {Berthier}, {Blomme},
  {Burgess}, {Busonero}, {Busso}, {C{\'a}novas}, {Carry}, {Cellino}, {Cheek},
  {Clementini}, {Davidson}, {de Teodoro}, {Nu{\~n}ez Campos}, {Delchambre},
  {Dell'Oro}, {Esquej}, {Fern{\'a}ndez-Hern{\'a}ndez}, {Fraile}, {Garabato},
  {Garc{\'\i}a-Lario}, {Haigron}, {Hambly}, {Harrison}, {Hern{\'a}ndez},
  {Hestroffer}, {Hodgkin}, {Jan{\ss}en}, {Jevardat de Fombelle}, {Jordan},
  {Krone-Martins}, {Lanzafame}, {L{\"o}ffler}, {Marchal}, {Marrese},
  {Moitinho}, {Muinonen}, {Osborne}, {Pancino}, {Pauwels}, {Recio-Blanco},
  {Reyl{\'e}}, {Riello}, {Rimoldini}, {Roegiers}, {Rybizki}, {Sarro}, {Smith},
  {Utrilla}, {van Leeuwen}, {Abbas}, {{\'A}brah{\'a}m}, {Abreu Aramburu},
  {Aerts}, {Aguado}, {Ajaj}, {Aldea-Montero}, {Altavilla}, {{\'A}lvarez},
  {Alves}, {Anders}, {Anderson}, {Anglada Varela}, {Antoja}, {Baines}, {Baker},
  {Balaguer-N{\'u}{\~n}ez}, {Balbinot}, {Balog}, {Barache}, {Barbato},
  {Barros}, {Bartolom{\'e}}, {Bassilana}, {Becciani}, {Bellazzini},
  {Berihuete}, {Bernet}, {Bertone}, {Bianchi}, {Binnenfeld}, {Blanco-Cuaresma},
  {Blazere}, {Boch}, {Bombrun}, {Bossini}, {Bouquillon}, {Bragaglia},
  {Bramante}, {Breedt}, {Bressan}, {Brouillet}, {Brugaletta}, {Bucciarelli},
  {Burlacu}, {Butkevich}, {Buzzi}, {Caffau}, {Cancelliere}, {Cantat-Gaudin},
  {Carballo}, {Carlucci}, {Carnerero}, {Carrasco}, {Casamiquela}, {Castellani},
  {Castro-Ginard}, {Chaoul}, {Charlot}, {Chemin}, {Chiaramida}, {Chiavassa},
  {Chornay}, {Comoretto}, {Contursi}, {Cooper}, {Cornez}, {Cowell}, {Crifo},
  {Cropper}, {Crosta}, {Crowley}, {Dafonte}, {Dapergolas}, {David}, {de
  Laverny}, {De Luise}, {De March}, {De Ridder}, {de Souza}, {de Torres}, {del
  Peloso}, {del Pozo}, {Delbo}, {Delgado}, {Delisle}, {Demouchy},
  {Dharmawardena}, {Diakite}, {Diener}, {Distefano}, {Dolding}, {Enke},
  {Fabre}, {Fabrizio}, {Fedorets}, {Fernique}, {Figueras}, {Fournier},
  {Fouron}, {Fragkoudi}, {Gai}, {Garcia-Gutierrez}, {Garcia-Reinaldos},
  {Garc{\'\i}a-Torres}, {Garofalo}, {Gavel}, {Gerlach}, {Geyer}, {Gilmore},
  {Girona}, {Giuffrida}, {Gomel}, {Gomez}, {Gonz{\'a}lez-N{\'u}{\~n}ez},
  {Gonz{\'a}lez-Santamar{\'\i}a}, {Gonz{\'a}lez-Vidal}, {Granvik}, {Guillout},
  {Guiraud}, {Guti{\'e}rrez-S{\'a}nchez}, {Guy}, {Hatzidimitriou}, {Hauser},
  {Haywood}, {Helmer}, {Helmi}, {Sarmiento}, {Hidalgo}, {Hilger},
  {H{\l}adczuk}, {Hobbs}, {Holland}, {Huckle}, {Jardine}, {Jasniewicz},
  {Jean-Antoine Piccolo}, {Jim{\'e}nez-Arranz}, {Juaristi Campillo}, {Julbe},
  {Karbevska}, {Khanna}, {Kordopatis}, {Korn}, {K{\'o}sp{\'a}l},
  {Kostrzewa-Rutkowska}, {Kruszy{\'n}ska}, {Kun}, {Laizeau}, {Lambert},
  {Lanza}, {Lasne}, {Le Campion}, {Lebreton}, {Lebzelter}, {Leccia},
  {Lecoeur-Taibi}, {Liao}, {Licata}, {Lindstr{\o}m}, {Lister}, {Livanou},
  {Lobel}, {Lorca}, {Loup}, {Madrero Pardo}, {Magdaleno Romeo}, {Managau},
  {Mann}, {Manteiga}, {Marchant}, {Marconi}, {Marcos}, {Marcos Santos},
  {Mar{\'\i}n Pina}, {Marinoni}, {Marocco}, {Marshall}, {Martin Polo},
  {Mart{\'\i}n-Fleitas}, {Marton}, {Mary}, {Masip}, {Massari},
  {Mastrobuono-Battisti}, {McMillan}, {Messina}, {Michalik}, {Millar}, {Mints},
  {Molina}, {Molinaro}, {Moln{\'a}r}, {Monari}, {Mongui{\'o}}, {Montegriffo},
  {Montero}, {Mor}, {Mora}, {Morbidelli}, {Morris}, {Muraveva}, {Murphy},
  {Musella}, {Nagy}, {Noval}, {Oca{\~n}a}, {Ogden}, {Ordenovic}, {Osinde},
  {Pagani}, {Pagano}, {Palaversa}, {Palicio}, {Pallas-Quintela}, {Panahi},
  {Payne-Wardenaar}, {Pe{\~n}alosa Esteller}, {Penttil{\"a}}, {Pichon},
  {Piersimoni}, {Pineau}, {Plachy}, {Plum}, {Poggio}, {Pr{\v{s}}a}, {Pulone},
  {Racero}, {Ragaini}, {Rainer}, {Raiteri}, {Ramos}, {Ramos-Lerate}, {Regibo},
  {Richards}, {Rios Diaz}, {Ripepi}, {Riva}, {Rix}, {Rixon}, {Robichon},
  {Robin}, {Robin}, {Roelens}, {Rogues}, {Rohrbasser}, {Romero-G{\'o}mez},
  {Rowell}, {Royer}, {Ruz Mieres}, {Rybicki}, {S{\'a}ez N{\'u}{\~n}ez},
  {Sagrist{\`a} Sell{\'e}s}, {Salguero}, {Samaras}, {Sanchez Gimenez}, {Sanna},
  {Santove{\~n}a}, {Sarasso}, {Schultheis}, {Sciacca}, {Segol}, {Segovia},
  {Semeux}, {Siddiqui}, {Siebert}, {Siltala}, {Silvelo}, {Slezak}, {Slezak},
  {Smart}, {Snaith}, {Solano}, {Solitro}, {Souami}, {Souchay}, {Spagna},
  {Spina}, {Spoto}, {Steele}, {Steidelm{\"u}ller}, {Stephenson}, {S{\"u}veges},
  {Surdej}, {Szabados}, {Szegedi-Elek}, {Taris}, {Taylor}, {Teixeira},
  {Tolomei}, {Tonello}, {Torra}, {Torra}, {Torralba Elipe}, {Trabucchi},
  {Tsounis}, {Turon}, {Ulla}, {Unger}, {Vaillant}, {van Dillen}, {van Reeven},
  {Vanel}, {Vecchiato}, {Viala}, {Vicente}, {Voutsinas}, {Weiler}, {Wevers},
  {Wyrzykowski}, {Yoldas}, {Yvard}, {Zhao}, {Zorec}, \&
  {Zucker}}]{Gaia_2023_binaries}
{Gaia Collaboration}, {Arenou}, F., {Babusiaux}, C., {et~al.}
  2023{\natexlab{a}}, \aap, 674, A34

\bibitem[{{Gaia Collaboration} {et~al.}(2023{\natexlab{b}}){Gaia
  Collaboration}, {Vallenari}, {Brown}, {Prusti}, {de Bruijne}, {Arenou},
  {Babusiaux}, {Biermann}, {Creevey}, {Ducourant}, {Evans}, {Eyer}, {Guerra},
  {Hutton}, {Jordi}, {Klioner}, {Lammers}, {Lindegren}, {Luri}, {Mignard},
  {Panem}, {Pourbaix}, {Randich}, {Sartoretti}, {Soubiran}, {Tanga}, {Walton},
  {Bailer-Jones}, {Bastian}, {Drimmel}, {Jansen}, {Katz}, {Lattanzi}, {van
  Leeuwen}, {Bakker}, {Cacciari}, {Casta{\~n}eda}, {De Angeli}, {Fabricius},
  {Fouesneau}, {Fr{\'e}mat}, {Galluccio}, {Guerrier}, {Heiter}, {Masana},
  {Messineo}, {Mowlavi}, {Nicolas}, {Nienartowicz}, {Pailler}, {Panuzzo},
  {Riclet}, {Roux}, {Seabroke}, {Sordo}, {Th{\'e}venin}, {Gracia-Abril},
  {Portell}, {Teyssier}, {Altmann}, {Andrae}, {Audard}, {Bellas-Velidis},
  {Benson}, {Berthier}, {Blomme}, {Burgess}, {Busonero}, {Busso},
  {C{\'a}novas}, {Carry}, {Cellino}, {Cheek}, {Clementini}, {Damerdji},
  {Davidson}, {de Teodoro}, {Nu{\~n}ez Campos}, {Delchambre}, {Dell'Oro},
  {Esquej}, {Fern{\'a}ndez-Hern{\'a}ndez}, {Fraile}, {Garabato},
  {Garc{\'\i}a-Lario}, {Gosset}, {Haigron}, {Halbwachs}, {Hambly}, {Harrison},
  {Hern{\'a}ndez}, {Hestroffer}, {Hodgkin}, {Holl}, {Jan{\ss}en}, {Jevardat de
  Fombelle}, {Jordan}, {Krone-Martins}, {Lanzafame}, {L{\"o}ffler}, {Marchal},
  {Marrese}, {Moitinho}, {Muinonen}, {Osborne}, {Pancino}, {Pauwels},
  {Recio-Blanco}, {Reyl{\'e}}, {Riello}, {Rimoldini}, {Roegiers}, {Rybizki},
  {Sarro}, {Siopis}, {Smith}, {Sozzetti}, {Utrilla}, {van Leeuwen}, {Abbas},
  {{\'A}brah{\'a}m}, {Abreu Aramburu}, {Aerts}, {Aguado}, {Ajaj},
  {Aldea-Montero}, {Altavilla}, {{\'A}lvarez}, {Alves}, {Anders}, {Anderson},
  {Anglada Varela}, {Antoja}, {Baines}, {Baker}, {Balaguer-N{\'u}{\~n}ez},
  {Balbinot}, {Balog}, {Barache}, {Barbato}, {Barros}, {Barstow},
  {Bartolom{\'e}}, {Bassilana}, {Bauchet}, {Becciani}, {Bellazzini},
  {Berihuete}, {Bernet}, {Bertone}, {Bianchi}, {Binnenfeld}, {Blanco-Cuaresma},
  {Blazere}, {Boch}, {Bombrun}, {Bossini}, {Bouquillon}, {Bragaglia},
  {Bramante}, {Breedt}, {Bressan}, {Brouillet}, {Brugaletta}, {Bucciarelli},
  {Burlacu}, {Butkevich}, {Buzzi}, {Caffau}, {Cancelliere}, {Cantat-Gaudin},
  {Carballo}, {Carlucci}, {Carnerero}, {Carrasco}, {Casamiquela}, {Castellani},
  {Castro-Ginard}, {Chaoul}, {Charlot}, {Chemin}, {Chiaramida}, {Chiavassa},
  {Chornay}, {Comoretto}, {Contursi}, {Cooper}, {Cornez}, {Cowell}, {Crifo},
  {Cropper}, {Crosta}, {Crowley}, {Dafonte}, {Dapergolas}, {David}, {David},
  {de Laverny}, {De Luise}, {De March}, {De Ridder}, {de Souza}, {de Torres},
  {del Peloso}, {del Pozo}, {Delbo}, {Delgado}, {Delisle}, {Demouchy},
  {Dharmawardena}, {Di Matteo}, {Diakite}, {Diener}, {Distefano}, {Dolding},
  {Edvardsson}, {Enke}, {Fabre}, {Fabrizio}, {Faigler}, {Fedorets}, {Fernique},
  {Fienga}, {Figueras}, {Fournier}, {Fouron}, {Fragkoudi}, {Gai},
  {Garcia-Gutierrez}, {Garcia-Reinaldos}, {Garc{\'\i}a-Torres}, {Garofalo},
  {Gavel}, {Gavras}, {Gerlach}, {Geyer}, {Giacobbe}, {Gilmore}, {Girona},
  {Giuffrida}, {Gomel}, {Gomez}, {Gonz{\'a}lez-N{\'u}{\~n}ez},
  {Gonz{\'a}lez-Santamar{\'\i}a}, {Gonz{\'a}lez-Vidal}, {Granvik}, {Guillout},
  {Guiraud}, {Guti{\'e}rrez-S{\'a}nchez}, {Guy}, {Hatzidimitriou}, {Hauser},
  {Haywood}, {Helmer}, {Helmi}, {Sarmiento}, {Hidalgo}, {Hilger},
  {H{\l}adczuk}, {Hobbs}, {Holland}, {Huckle}, {Jardine}, {Jasniewicz},
  {Jean-Antoine Piccolo}, {Jim{\'e}nez-Arranz}, {Jorissen}, {Juaristi
  Campillo}, {Julbe}, {Karbevska}, {Kervella}, {Khanna}, {Kontizas},
  {Kordopatis}, {Korn}, {K{\'o}sp{\'a}l}, {Kostrzewa-Rutkowska},
  {Kruszy{\'n}ska}, {Kun}, {Laizeau}, {Lambert}, {Lanza}, {Lasne}, {Le
  Campion}, {Lebreton}, {Lebzelter}, {Leccia}, {Leclerc}, {Lecoeur-Taibi},
  {Liao}, {Licata}, {Lindstr{\o}m}, {Lister}, {Livanou}, {Lobel}, {Lorca},
  {Loup}, {Madrero Pardo}, {Magdaleno Romeo}, {Managau}, {Mann}, {Manteiga},
  {Marchant}, {Marconi}, {Marcos}, {Marcos Santos}, {Mar{\'\i}n Pina},
  {Marinoni}, {Marocco}, {Marshall}, {Martin Polo}, {Mart{\'\i}n-Fleitas},
  {Marton}, {Mary}, {Masip}, {Massari}, {Mastrobuono-Battisti}, {Mazeh},
  {McMillan}, {Messina}, {Michalik}, {Millar}, {Mints}, {Molina}, {Molinaro},
  {Moln{\'a}r}, {Monari}, {Mongui{\'o}}, {Montegriffo}, {Montero}, {Mor},
  {Mora}, {Morbidelli}, {Morel}, {Morris}, {Muraveva}, {Murphy}, {Musella},
  {Nagy}, {Noval}, {Oca{\~n}a}, {Ogden}, {Ordenovic}, {Osinde}, {Pagani},
  {Pagano}, {Palaversa}, {Palicio}, {Pallas-Quintela}, {Panahi},
  {Payne-Wardenaar}, {Pe{\~n}alosa Esteller}, {Penttil{\"a}}, {Pichon},
  {Piersimoni}, {Pineau}, {Plachy}, {Plum}, {Poggio}, {Pr{\v{s}}a}, {Pulone},
  {Racero}, {Ragaini}, {Rainer}, {Raiteri}, {Rambaux}, {Ramos}, {Ramos-Lerate},
  {Re Fiorentin}, {Regibo}, {Richards}, {Rios Diaz}, {Ripepi}, {Riva}, {Rix},
  {Rixon}, {Robichon}, {Robin}, {Robin}, {Roelens}, {Rogues}, {Rohrbasser},
  {Romero-G{\'o}mez}, {Rowell}, {Royer}, {Ruz Mieres}, {Rybicki}, {Sadowski},
  {S{\'a}ez N{\'u}{\~n}ez}, {Sagrist{\`a} Sell{\'e}s}, {Sahlmann}, {Salguero},
  {Samaras}, {Sanchez Gimenez}, {Sanna}, {Santove{\~n}a}, {Sarasso},
  {Schultheis}, {Sciacca}, {Segol}, {Segovia}, {S{\'e}gransan}, {Semeux},
  {Shahaf}, {Siddiqui}, {Siebert}, {Siltala}, {Silvelo}, {Slezak}, {Slezak},
  {Smart}, {Snaith}, {Solano}, {Solitro}, {Souami}, {Souchay}, {Spagna},
  {Spina}, {Spoto}, {Steele}, {Steidelm{\"u}ller}, {Stephenson}, {S{\"u}veges},
  {Surdej}, {Szabados}, {Szegedi-Elek}, {Taris}, {Taylor}, {Teixeira},
  {Tolomei}, {Tonello}, {Torra}, {Torra}, {Torralba Elipe}, {Trabucchi},
  {Tsounis}, {Turon}, {Ulla}, {Unger}, {Vaillant}, {van Dillen}, {van Reeven},
  {Vanel}, {Vecchiato}, {Viala}, {Vicente}, {Voutsinas}, {Weiler}, {Wevers},
  {Wyrzykowski}, {Yoldas}, {Yvard}, {Zhao}, {Zorec}, {Zucker}, \&
  {Zwitter}}]{Gaia_2023}
{Gaia Collaboration}, {Vallenari}, A., {Brown}, A.~G.~A., {et~al.}
  2023{\natexlab{b}}, \aap, 674, A1

\bibitem[{{Gallet} \& {Bouvier}(2013)}]{Gallet_2013}
{Gallet}, F. \& {Bouvier}, J. 2013, \aap, 556, A36

\bibitem[{{Garc{\'\i}a} {et~al.}(2010){Garc{\'\i}a}, {Mathur}, {Salabert},
  {Ballot}, {R{\'e}gulo}, {Metcalfe}, \& {Baglin}}]{Garcia_2010}
{Garc{\'\i}a}, R.~A., {Mathur}, S., {Salabert}, D., {et~al.} 2010, Science,
  329, 1032

\bibitem[{{Garc{\'\i}a Soto} {et~al.}(2023){Garc{\'\i}a Soto}, {Newton},
  {Douglas}, {Burrows}, \& {Kesseli}}]{Garcia_Soto_2023}
{Garc{\'\i}a Soto}, A., {Newton}, E.~R., {Douglas}, S.~T., {Burrows}, A., \&
  {Kesseli}, A.~Y. 2023, \aj, 165, 192

\bibitem[{{Gaulme} {et~al.}(2014){Gaulme}, {Jackiewicz}, {Appourchaux}, \&
  {Mosser}}]{Gaulme_2014}
{Gaulme}, P., {Jackiewicz}, J., {Appourchaux}, T., \& {Mosser}, B. 2014, \apj,
  785, 5

\bibitem[{{Gaulme} {et~al.}(2020){Gaulme}, {Jackiewicz}, {Spada}, {Chojnowski},
  {Mosser}, {McKeever}, {Hedlund}, {Vrard}, {Benbakoura}, \&
  {Damiani}}]{Gaulme_2020}
{Gaulme}, P., {Jackiewicz}, J., {Spada}, F., {et~al.} 2020, \aap, 639, A63

\bibitem[{{Gaulme} {et~al.}(2016){Gaulme}, {McKeever}, {Jackiewicz}, {Rawls},
  {Corsaro}, {Mosser}, {Southworth}, {Mahadevan}, {Bender}, \&
  {Deshpande}}]{Gaulme_2016}
{Gaulme}, P., {McKeever}, J., {Jackiewicz}, J., {et~al.} 2016, \apj, 832, 121

\bibitem[{{Gehan} {et~al.}(2022){Gehan}, {Gaulme}, \& {Yu}}]{Gehan_2022}
{Gehan}, C., {Gaulme}, P., \& {Yu}, J. 2022, \aap, 668, A116

\bibitem[{{Godoy-Rivera} {et~al.}(2021){Godoy-Rivera}, {Tayar}, {Pinsonneault},
  {Rodr{\'\i}guez Mart{\'\i}nez}, {Stassun}, {van Saders}, {Beaton},
  {Garc{\'\i}a-Hern{\'a}ndez}, \& {Teske}}]{Godoy-Rivera_2021}
{Godoy-Rivera}, D., {Tayar}, J., {Pinsonneault}, M.~H., {et~al.} 2021, \apj,
  915, 19

\bibitem[{{Gomes da Silva} {et~al.}(2022){Gomes da Silva}, {Bensabat},
  {Monteiro}, \& {Santos}}]{Gomes_2022}
{Gomes da Silva}, J., {Bensabat}, A., {Monteiro}, T., \& {Santos}, N.~C. 2022,
  \aap, 668, A174

\bibitem[{{Gomes da Silva} {et~al.}(2021){Gomes da Silva}, {Santos},
  {Adibekyan}, {Sousa}, {Campante}, {Figueira}, {Bossini}, {Delgado-Mena},
  {Monteiro}, {de Laverny}, {Recio-Blanco}, \& {Lovis}}]{Gomes_2021}
{Gomes da Silva}, J., {Santos}, N.~C., {Adibekyan}, V., {et~al.} 2021, \aap,
  646, A77

\bibitem[{{Gomes da Silva} {et~al.}(2014){Gomes da Silva}, {Santos}, {Boisse},
  {Dumusque}, \& {Lovis}}]{Gomes_2014}
{Gomes da Silva}, J., {Santos}, N.~C., {Boisse}, I., {Dumusque}, X., \&
  {Lovis}, C. 2014, \aap, 566, A66

\bibitem[{{Gomes da Silva} {et~al.}(2011){Gomes da Silva}, {Santos}, {Bonfils},
  {Delfosse}, {Forveille}, \& {Udry}}]{Gomes_2011}
{Gomes da Silva}, J., {Santos}, N.~C., {Bonfils}, X., {et~al.} 2011, \aap, 534,
  A30

\bibitem[{{Gray}(2008)}]{Gray_2008}
{Gray}, D.~F. 2008, {The Observation and Analysis of Stellar Photospheres}

\bibitem[{{Hall}(2008)}]{Hall_2008}
{Hall}, J.~C. 2008, Living Reviews in Solar Physics, 5, 2

\bibitem[{{Harper}(2018)}]{Harper_2018}
{Harper}, G.~M. 2018, in Astronomical Society of the Pacific Conference Series,
  Vol. 517, Science with a Next Generation Very Large Array, ed. E.~{Murphy},
  265

\bibitem[{{Herreman} {et~al.}(2010){Herreman}, {Cebron}, {Le Diz{\`e}s}, \& {Le
  Gal}}]{Herreman_2010}
{Herreman}, W., {Cebron}, D., {Le Diz{\`e}s}, S., \& {Le Gal}, P. 2010, Journal
  of Fluid Mechanics, 661, 130

\bibitem[{{Huber} {et~al.}(2011){Huber}, {Bedding}, {Stello}, {Hekker},
  {Mathur}, {Mosser}, {Verner}, {Bonanno}, {Buzasi}, {Campante}, {Elsworth},
  {Hale}, {Kallinger}, {Silva Aguirre}, {Chaplin}, {De Ridder}, {Garc{\'\i}a},
  {Appourchaux}, {Frandsen}, {Houdek}, {Molenda-{\.Z}akowicz}, {Monteiro},
  {Christensen-Dalsgaard}, {Gilliland}, {Kawaler}, {Kjeldsen}, {Broomhall},
  {Corsaro}, {Salabert}, {Sanderfer}, {Seader}, \& {Smith}}]{Huber_2011}
{Huber}, D., {Bedding}, T.~R., {Stello}, D., {et~al.} 2011, \apj, 743, 143

\bibitem[{{Huber} {et~al.}(2010){Huber}, {Bedding}, {Stello}, {Mosser},
  {Mathur}, {Kallinger}, {Hekker}, {Elsworth}, {Buzasi}, {De Ridder},
  {Gilliland}, {Kjeldsen}, {Chaplin}, {Garc{\'\i}a}, {Hale}, {Preston},
  {White}, {Borucki}, {Christensen-Dalsgaard}, {Clarke}, {Jenkins}, \&
  {Koch}}]{Huber_2010}
{Huber}, D., {Bedding}, T.~R., {Stello}, D., {et~al.} 2010, \apj, 723, 1607

\bibitem[{{Husser} {et~al.}(2013){Husser}, {Wende-von Berg}, {Dreizler},
  {Homeier}, {Reiners}, {Barman}, \& {Hauschildt}}]{Husser_2013}
{Husser}, T.~O., {Wende-von Berg}, S., {Dreizler}, S., {et~al.} 2013, \aap,
  553, A6

\bibitem[{{Hut}(1981)}]{Hut_1981}
{Hut}, P. 1981, \aap, 99, 126

\bibitem[{{Kallinger} {et~al.}(2014){Kallinger}, {De Ridder}, {Hekker},
  {Mathur}, {Mosser}, {Gruberbauer}, {Garc{\'\i}a}, {Karoff}, \&
  {Ballot}}]{Kallinger_2014}
{Kallinger}, T., {De Ridder}, J., {Hekker}, S., {et~al.} 2014, \aap, 570, A41

\bibitem[{{Karoff} {et~al.}(2016){Karoff}, {Knudsen}, {De Cat}, {Bonanno},
  {Fogtmann-Schulz}, {Fu}, {Frasca}, {Inceoglu}, {Olsen}, {Zhang}, {Hou},
  {Wang}, {Shi}, \& {Zhang}}]{Karoff_2016}
{Karoff}, C., {Knudsen}, M.~F., {De Cat}, P., {et~al.} 2016, Nature
  Communications, 7, 11058

\bibitem[{{Kerswell}(2002)}]{Kerswell_2002}
{Kerswell}, R.~R. 2002, Annual Review of Fluid Mechanics, 34, 83

\bibitem[{{Kiraga} \& {Stepien}(2007)}]{Kiraga_2007}
{Kiraga}, M. \& {Stepien}, K. 2007, \actaa, 57, 149

\bibitem[{{Lacaze} {et~al.}(2006){Lacaze}, {Herreman}, {Le Bars}, {Le
  Diz{\`e}s}, \& {Le Gal}}]{Lacaze_2006}
{Lacaze}, L., {Herreman}, W., {Le Bars}, M., {Le Diz{\`e}s}, S., \& {Le Gal},
  P. 2006, Geophysical and Astrophysical Fluid Dynamics, 100, 299

\bibitem[{{Lacaze} {et~al.}(2004){Lacaze}, {Le Gal}, \& {Le
  Diz{\`e}s}}]{Lacaze_2004}
{Lacaze}, L., {Le Gal}, P., \& {Le Diz{\`e}s}, S. 2004, Journal of Fluid
  Mechanics, 505, 1

\bibitem[{{Le Bars} {et~al.}(2010){Le Bars}, {Lacaze}, {Le Diz{\`e}s}, {Le
  Gal}, \& {Rieutord}}]{Le_Bars_2010}
{Le Bars}, M., {Lacaze}, L., {Le Diz{\`e}s}, S., {Le Gal}, P., \& {Rieutord},
  M. 2010, Physics of the Earth and Planetary Interiors, 178, 48

\bibitem[{{Leenaarts} {et~al.}(2012){Leenaarts}, {Carlsson}, \& {Rouppe van der
  Voort}}]{Leenaarts_2012}
{Leenaarts}, J., {Carlsson}, M., \& {Rouppe van der Voort}, L. 2012, \apj, 749,
  136

\bibitem[{{Lehtinen} {et~al.}(2020){Lehtinen}, {Spada}, {K{\"a}pyl{\"a}},
  {Olspert}, \& {K{\"a}pyl{\"a}}}]{Lehtinen_2020}
{Lehtinen}, J.~J., {Spada}, F., {K{\"a}pyl{\"a}}, M.~J., {Olspert}, N., \&
  {K{\"a}pyl{\"a}}, P.~J. 2020, Nature Astronomy, 4, 658

\bibitem[{{Livingston} {et~al.}(2007){Livingston}, {Wallace}, {White}, \&
  {Giampapa}}]{Livingston_2007}
{Livingston}, W., {Wallace}, L., {White}, O.~R., \& {Giampapa}, M.~S. 2007,
  \apj, 657, 1137

\bibitem[{{Lu} {et~al.}(2023){Lu}, {Yuan}, {Xu}, {Zhang}, {Xiao}, {Huang},
  {Beers}, \& {Hong}}]{Lu_2023}
{Lu}, X., {Yuan}, H., {Xu}, S., {et~al.} 2023, arXiv e-prints, arXiv:2311.16901

\bibitem[{{Martin} {et~al.}(2005){Martin}, {Fanson}, {Schiminovich},
  {Morrissey}, {Friedman}, {Barlow}, {Conrow}, {Grange}, {Jelinsky},
  {Milliard}, {Siegmund}, {Bianchi}, {Byun}, {Donas}, {Forster}, {Heckman},
  {Lee}, {Madore}, {Malina}, {Neff}, {Rich}, {Small}, {Surber}, {Szalay},
  {Welsh}, \& {Wyder}}]{Martin_2005}
{Martin}, D.~C., {Fanson}, J., {Schiminovich}, D., {et~al.} 2005, \apjl, 619,
  L1

\bibitem[{{Martin} {et~al.}(2017){Martin}, {Fuhrmeister}, {Mittag}, {Schmidt},
  {Hempelmann}, {Gonz{\'a}lez-P{\'e}rez}, \& {Schmitt}}]{Martin_2017}
{Martin}, J., {Fuhrmeister}, B., {Mittag}, M., {et~al.} 2017, \aap, 605, A113

\bibitem[{{Mart{\'\i}nez-Arn{\'a}iz} {et~al.}(2011){Mart{\'\i}nez-Arn{\'a}iz},
  {L{\'o}pez-Santiago}, {Crespo-Chac{\'o}n}, \&
  {Montes}}]{Martinez-Arnaiz_2011}
{Mart{\'\i}nez-Arn{\'a}iz}, R., {L{\'o}pez-Santiago}, J., {Crespo-Chac{\'o}n},
  I., \& {Montes}, D. 2011, \mnras, 414, 2629

\bibitem[{{Mashonkina} {et~al.}(2011){Mashonkina}, {Gehren}, {Shi}, {Korn}, \&
  {Grupp}}]{Mashonkina_2011}
{Mashonkina}, L., {Gehren}, T., {Shi}, J.~R., {Korn}, A.~J., \& {Grupp}, F.
  2011, \aap, 528, A87

\bibitem[{{Mathis}(2015)}]{Mathis_2015}
{Mathis}, S. 2015, \aap, 580, L3

\bibitem[{{Mathur} {et~al.}(2014){Mathur}, {Garc{\'\i}a}, {Ballot}, {Ceillier},
  {Salabert}, {Metcalfe}, {R{\'e}gulo}, {Jim{\'e}nez}, \&
  {Bloemen}}]{Mathur_2014}
{Mathur}, S., {Garc{\'\i}a}, R.~A., {Ballot}, J., {et~al.} 2014, \aap, 562,
  A124

\bibitem[{{Mathur} {et~al.}(2019){Mathur}, {Garc{\'\i}a}, {Bugnet}, {Santos},
  {Santiago}, \& {Beck}}]{Mathur_2019}
{Mathur}, S., {Garc{\'\i}a}, R.~A., {Bugnet}, L., {et~al.} 2019, Frontiers in
  Astronomy and Space Sciences, 6, 46

\bibitem[{{Mazumder} {et~al.}(2021){Mazumder}, {Chatterjee}, {Nandy}, \&
  {Banerjee}}]{Mazumder_2021}
{Mazumder}, R., {Chatterjee}, S., {Nandy}, D., \& {Banerjee}, D. 2021, \apj,
  919, 125

\bibitem[{{Metcalfe} {et~al.}(2022){Metcalfe}, {Finley}, {Kochukhov}, {See},
  {Ayres}, {Stassun}, {van Saders}, {Clark}, {Godoy-Rivera}, {Ilyin},
  {Pinsonneault}, {Strassmeier}, \& {Petit}}]{Metcalfe_2022}
{Metcalfe}, T.~S., {Finley}, A.~J., {Kochukhov}, O., {et~al.} 2022, \apjl, 933,
  L17

\bibitem[{{Meunier} \& {Delfosse}(2009)}]{Meunier_2009}
{Meunier}, N. \& {Delfosse}, X. 2009, \aap, 501, 1103

\bibitem[{{Meunier} {et~al.}(2022){Meunier}, {Kretzschmar}, {Gravet}, {Mignon},
  \& {Delfosse}}]{Meunier_2022}
{Meunier}, N., {Kretzschmar}, M., {Gravet}, R., {Mignon}, L., \& {Delfosse}, X.
  2022, \aap, 658, A57

\bibitem[{{Montes} {et~al.}(1995){Montes}, {Fernandez-Figueroa}, {de Castro},
  \& {Cornide}}]{Montes_1995}
{Montes}, D., {Fernandez-Figueroa}, M.~J., {de Castro}, E., \& {Cornide}, M.
  1995, \aap, 294, 165

\bibitem[{{Montez} {et~al.}(2017){Montez}, {Ramstedt}, {Kastner}, {Vlemmings},
  \& {Sanchez}}]{Montez_2017}
{Montez}, Rodolfo, J., {Ramstedt}, S., {Kastner}, J.~H., {Vlemmings}, W., \&
  {Sanchez}, E. 2017, \apj, 841, 33

\bibitem[{{Mosser} {et~al.}(2011){Mosser}, {Belkacem}, {Goupil}, {Michel},
  {Elsworth}, {Barban}, {Kallinger}, {Hekker}, {De Ridder}, {Samadi}, {Baudin},
  {Pinheiro}, {Auvergne}, {Baglin}, \& {Catala}}]{Mosser_2011}
{Mosser}, B., {Belkacem}, K., {Goupil}, M.~J., {et~al.} 2011, \aap, 525, L9

\bibitem[{{Mosser} {et~al.}(2012){Mosser}, {Elsworth}, {Hekker}, {Huber},
  {Kallinger}, {Mathur}, {Belkacem}, {Goupil}, {Samadi}, {Barban}, {Bedding},
  {Chaplin}, {Garc{\'\i}a}, {Stello}, {De Ridder}, {Middour}, {Morris}, \&
  {Quintana}}]{Mosser_2012}
{Mosser}, B., {Elsworth}, Y., {Hekker}, S., {et~al.} 2012, \aap, 537, A30

\bibitem[{{Newton} {et~al.}(2017){Newton}, {Irwin}, {Charbonneau}, {Berlind},
  {Calkins}, \& {Mink}}]{Newton_2017}
{Newton}, E.~R., {Irwin}, J., {Charbonneau}, D., {et~al.} 2017, \apj, 834, 85

\bibitem[{{Noyes} {et~al.}(1984){Noyes}, {Hartmann}, {Baliunas}, {Duncan}, \&
  {Vaughan}}]{Noyes_1984}
{Noyes}, R.~W., {Hartmann}, L.~W., {Baliunas}, S.~L., {Duncan}, D.~K., \&
  {Vaughan}, A.~H. 1984, \apj, 279, 763

\bibitem[{{Ogilvie}(2013)}]{Ogilvie_2013}
{Ogilvie}, G.~I. 2013, \mnras, 429, 613

\bibitem[{{Ogilvie}(2014)}]{Ogilvie_2014}
{Ogilvie}, G.~I. 2014, \araa, 52, 171

\bibitem[{{Ogilvie} \& {Lin}(2007)}]{Ogilvie_2007}
{Ogilvie}, G.~I. \& {Lin}, D.~N.~C. 2007, \apj, 661, 1180

\bibitem[{{Ol{\'a}h} {et~al.}(2021){Ol{\'a}h}, {K{\H{o}}v{\'a}ri},
  {G{\"u}nther}, {Vida}, {Gaulme}, {Seli}, \& {P{\'a}l}}]{Olah_2021}
{Ol{\'a}h}, K., {K{\H{o}}v{\'a}ri}, Z., {G{\"u}nther}, M.~N., {et~al.} 2021,
  \aap, 647, A62

\bibitem[{{Olmedo} {et~al.}(2015){Olmedo}, {Lloyd}, {Mamajek}, {Ch{\'a}vez},
  {Bertone}, {Martin}, \& {Neill}}]{Olmedo_2015}
{Olmedo}, M., {Lloyd}, J., {Mamajek}, E.~E., {et~al.} 2015, \apj, 813, 100

\bibitem[{{Olmedo} {et~al.}(2019){Olmedo}, {Olmedo}, {Ch{\'a}vez}, {Bertone},
  {Mamajek}, \& {Lloyd}}]{Olmedo_2019}
{Olmedo}, M., {Olmedo}, D., {Ch{\'a}vez}, M., {et~al.} 2019, \memsai, 90, 680

\bibitem[{{Ortiz} \& {Guerrero}(2016)}]{Ortiz_2016}
{Ortiz}, R. \& {Guerrero}, M.~A. 2016, \mnras, 461, 3036

\bibitem[{{Petit} {et~al.}(2008){Petit}, {Dintrans}, {Solanki}, {Donati},
  {Auri{\`e}re}, {Ligni{\`e}res}, {Morin}, {Paletou}, {Ramirez Velez},
  {Catala}, \& {Fares}}]{Petit_2008}
{Petit}, P., {Dintrans}, B., {Solanki}, S.~K., {et~al.} 2008, \mnras, 388, 80

\bibitem[{{Quintero Noda} {et~al.}(2021){Quintero Noda}, {Barklem}, {Gafeira},
  {Ruiz Cobo}, {Collados}, {Carlsson}, {Mart{\'\i}nez Pillet}, {Orozco
  Su{\'a}rez}, {Uitenbroek}, \& {Katsukawa}}]{Quintero_Noda_2021}
{Quintero Noda}, C., {Barklem}, P.~S., {Gafeira}, R., {et~al.} 2021, \aap, 652,
  A161

\bibitem[{{Reinhold} {et~al.}(2019){Reinhold}, {Bell}, {Kuszlewicz}, {Hekker},
  \& {Shapiro}}]{Reinhold_2019}
{Reinhold}, T., {Bell}, K.~J., {Kuszlewicz}, J., {Hekker}, S., \& {Shapiro},
  A.~I. 2019, \aap, 621, A21

\bibitem[{{Remus} {et~al.}(2012){Remus}, {Mathis}, \& {Zahn}}]{Remus_2012}
{Remus}, F., {Mathis}, S., \& {Zahn}, J.~P. 2012, \aap, 544, A132

\bibitem[{{Rieutord}(2004)}]{Rieutord_2004}
{Rieutord}, M. 2004, in Stellar Rotation, ed. A.~{Maeder} \& P.~{Eenens}, Vol.
  215, 394

\bibitem[{{Sahai} {et~al.}(2008){Sahai}, {Findeisen}, {Gil de Paz}, \&
  {S{\'a}nchez Contreras}}]{Sahai_2008}
{Sahai}, R., {Findeisen}, K., {Gil de Paz}, A., \& {S{\'a}nchez Contreras}, C.
  2008, \apj, 689, 1274

\bibitem[{{Sahai} {et~al.}(2011){Sahai}, {Neill}, {Gil de Paz}, \& {S{\'a}nchez
  Contreras}}]{Sahai_2011}
{Sahai}, R., {Neill}, J.~D., {Gil de Paz}, A., \& {S{\'a}nchez Contreras}, C.
  2011, \apjl, 740, L39

\bibitem[{{Santos} {et~al.}(2021){Santos}, {Breton}, {Mathur}, \&
  {Garc{\'\i}a}}]{Santos_2021}
{Santos}, A.~R.~G., {Breton}, S.~N., {Mathur}, S., \& {Garc{\'\i}a}, R.~A.
  2021, \apjs, 255, 17

\bibitem[{{Sasso} {et~al.}(2017){Sasso}, {Andretta}, {Terranegra}, \&
  {Gomez}}]{Sasso_2017}
{Sasso}, C., {Andretta}, V., {Terranegra}, L., \& {Gomez}, M.~T. 2017, \aap,
  604, A50

\bibitem[{{Skrutskie} {et~al.}(2006){Skrutskie}, {Cutri}, {Stiening},
  {Weinberg}, {Schneider}, {Carpenter}, {Beichman}, {Capps}, {Chester},
  {Elias}, {Huchra}, {Liebert}, {Lonsdale}, {Monet}, {Price}, {Seitzer},
  {Jarrett}, {Kirkpatrick}, {Gizis}, {Howard}, {Evans}, {Fowler}, {Fullmer},
  {Hurt}, {Light}, {Kopan}, {Marsh}, {McCallon}, {Tam}, {Van Dyk}, \&
  {Wheelock}}]{Skrutskie_2006}
{Skrutskie}, M.~F., {Cutri}, R.~M., {Stiening}, R., {et~al.} 2006, \aj, 131,
  1163

\bibitem[{{Skumanich}(1972)}]{Skumanich_1972}
{Skumanich}, A. 1972, \apj, 171, 565

\bibitem[{{Stello} {et~al.}(2011){Stello}, {Huber}, {Kallinger}, {Basu},
  {Mosser}, {Hekker}, {Mathur}, {Garc{\'\i}a}, {Bedding}, {Kjeldsen},
  {Gilliland}, {Verner}, {Chaplin}, {Benomar}, {Meibom}, {Grundahl},
  {Elsworth}, {Molenda-{\.Z}akowicz}, {Szab{\'o}}, {Christensen-Dalsgaard},
  {Tenenbaum}, {Twicken}, \& {Uddin}}]{Stello_2011}
{Stello}, D., {Huber}, D., {Kallinger}, T., {et~al.} 2011, \apjl, 737, L10

\bibitem[{{Stelzer} {et~al.}(2013){Stelzer}, {Marino}, {Micela},
  {L{\'o}pez-Santiago}, \& {Liefke}}]{Stelzer_2013}
{Stelzer}, B., {Marino}, A., {Micela}, G., {L{\'o}pez-Santiago}, J., \&
  {Liefke}, C. 2013, \mnras, 431, 2063

\bibitem[{{Vidotto} {et~al.}(2014){Vidotto}, {Jardine}, {Morin}, {Donati},
  {Opher}, \& {Gombosi}}]{Vidotto_2014}
{Vidotto}, A.~A., {Jardine}, M., {Morin}, J., {et~al.} 2014, \mnras, 438, 1162

\bibitem[{{Vrard} {et~al.}(2018){Vrard}, {Kallinger}, {Mosser}, {Barban},
  {Baudin}, {Belkacem}, \& {Cunha}}]{Vrard_2018}
{Vrard}, M., {Kallinger}, T., {Mosser}, B., {et~al.} 2018, \aap, 616, A94

\bibitem[{{Walkowicz} \& {Hawley}(2009)}]{Walkowicz_2009}
{Walkowicz}, L.~M. \& {Hawley}, S.~L. 2009, \aj, 137, 3297

\bibitem[{{Weber} \& {Davis}(1967)}]{Weber_1967}
{Weber}, E.~J. \& {Davis}, Leverett, J. 1967, \apj, 148, 217

\bibitem[{{Wei}(2022)}]{Wei_2022}
{Wei}, X. 2022, \mnras, 513, 5474

\bibitem[{{Wilson}(1978)}]{Wilson_1978}
{Wilson}, O.~C. 1978, \apj, 226, 379

\bibitem[{{Wilson} \& {Skumanich}(1964)}]{Wilson_1964}
{Wilson}, O.~C. \& {Skumanich}, A. 1964, \apj, 140, 1401

\bibitem[{{Wise} {et~al.}(2018){Wise}, {Dodson-Robinson}, {Bevenour}, \&
  {Provini}}]{Wise_2018}
{Wise}, A.~W., {Dodson-Robinson}, S.~E., {Bevenour}, K., \& {Provini}, A. 2018,
  \aj, 156, 180

\bibitem[{{Yuan} {et~al.}(2013){Yuan}, {Liu}, \& {Xiang}}]{Yuan_2013}
{Yuan}, H.~B., {Liu}, X.~W., \& {Xiang}, M.~S. 2013, \mnras, 430, 2188

\bibitem[{{Zahn}(1966)}]{Zahn_1966}
{Zahn}, J.~P. 1966, Annales d'Astrophysique, 29, 489

\bibitem[{{Zahn}(1975)}]{Zahn_1975}
{Zahn}, J.~P. 1975, \aap, 41, 329

\bibitem[{{Zahn}(1977)}]{Zahn_1977}
{Zahn}, J.~P. 1977, \aap, 57, 383

\bibitem[{{Zhang} {et~al.}(2020){Zhang}, {Bi}, {Li}, {Jiang}, {Li}, {He}, {Yu},
  {Khanna}, {Ge}, {Liu}, {Tian}, {Wu}, \& {Zhang}}]{Zhang_2020}
{Zhang}, J., {Bi}, S., {Li}, Y., {et~al.} 2020, \apjs, 247, 9

\end{thebibliography}

\begin{appendix}
\section{Details of the NUV excess} \label{appendix-1}

\subsection{Note on the photometric systems}

We note that the \textit{GALEX} magnitudes are reported in the AB system, and the 2MASS magnitudes are reported in the Vega system. While the \textit{GALEX} magnitudes may be converted into the Vega system following the transformations of \citet[][see their Equations 2 and 3]{Bianchi_2017}, this is not needed when calculating the NUV excess. This is inherited from the definition of the NUV excess by \citet{Findeisen_2010}, who derived the fiducial relation combining the native \textit{GALEX} and 2MASS photometric systems (see their Figure 4). Regardless of the above, as the shift from one photometric system to the other is just an additive constant, the activity-rank of our targets would remain identical in either of them. 

\subsection{Impact of metallicity on the NUV excess}

Near the submission of this work, \cite{Lu_2023} posted a method for obtaining photometric metallicities based on \textit{GALEX} NUV data. Motivated by this, we tested the influence of metallicity on the NUV excess as follows. As the \citealt{Lu_2023} catalog was not yet published, we searched for metallicity information for our sample by crossmatching with the catalog of \cite{Andrae_2023}, which reported stellar parameters (T$_{\text{eff}}$, $\log{g}$, and [M/H]) derived from the \textit{Gaia} DR3 BP/RP spectra. Using these [M/H] values, we found our targets to span a range of metallicties centered around a slightly sub-solar value (median [M/H] = -0.07 dex), in agreement with the rest of the \textit{Kepler} field, (e.g., Godoy-Rivera et al. submitted). Indeed, we noted a correlation between the NUV excess and the [M/H] values, which implies that at least part of the NUV activity can be attributed to metallicity (see also \citealt{Olmedo_2019}). Nevertheless, we tested removing this effect via linear regression, and found that the majority of the NUV-active targets remain as such after accounting for this metallicity-dependence. Thus, the reported NUV excess values are robust against the effect of metallicity.

\section{Error on spectral indices}
\label{appendix_errorbar}
\begin{figure}[t]
\includegraphics[width=9.cm]{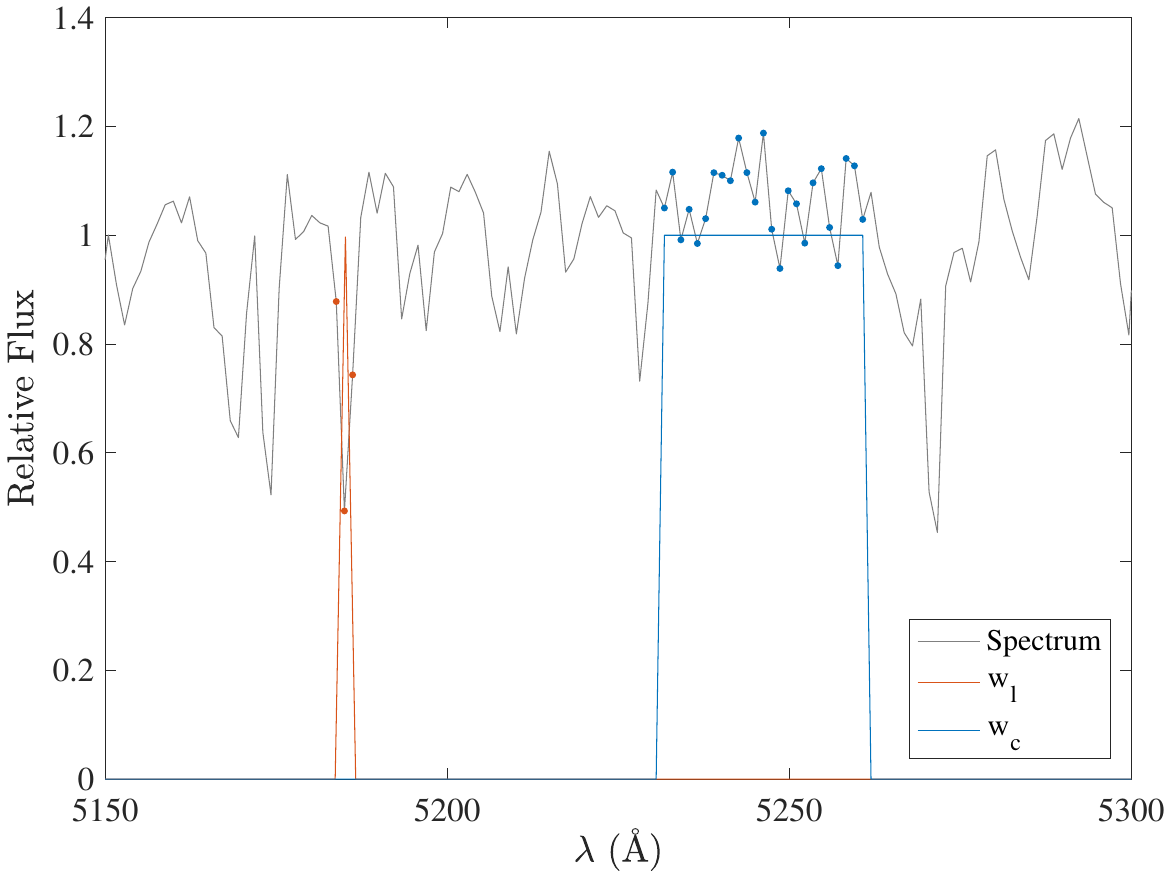}
\caption{Simulated spectrum of a RG star from a PHOENIX spectrum of a star with an effective temperature of 4800\,K, surface gravity of 2.5, and [Fe/H] of 0.0 (gray line). Resolution and sampling were downgraded to that of LAMOST. The red line correspond with the triangular weighting function $w_\ell$ of the line $\ell$ and the blue line to the flat weighting function $w_c$ employed for computing the continuum level. The $N_\ell$ red and $N\ind{c}$ blue dots indicate the points that belong to the triangular and rectangular weighting functions, respectively.}
\label{fig-weighting_fn}
\end{figure}

The spectral index $S_\ell$ of a given line $\ell$  is defined as the ratio of the triangular weighted average flux $F_\ell$ in the considered line, to the average flux in a selected range of the continuum $F\ind{c}$. We express them as a function of the discretized spectrum of flux $F_i$ per data bin of index $i$: 
\begin{eqnarray}
    F_\ell &=& \frac{\sum_{i=i_\ell}^{i_\ell+N_\ell-1} w_{\ell,i}\, F_i}{\sum_{i=i_\ell}^{i_\ell+N_\ell} w_{\ell,i}}\\
    F\ind{c} &=& \frac{\sum_{i=i\ind{c}}^{i\ind{c}+N_c-1} w_{\mathrm{c},i}\, F_i}{\sum_{i=i_\ell}^{i_\ell+N_\ell} w_{\mathrm{c},i}} \\
    &=& \frac{1}{N\ind{c}}\sum_{i=i\ind{c}}^{i\ind{c}+N_c-1} F_i
\end{eqnarray}
where $i_\ell$ and $i\ind{c}$ are the indices of the starts of the triangular and rectangular windows, $N_\ell$ and $N\ind{c}$ the number of bins inside each window, $w_{\ell,i}$ the discretized triangular weighting function, and  $w_{\mathrm{c},i}$ the discretized rectangular weighting function, such as $w_{\mathrm{c},i} = 1,\, i \in [i\ind{c}, i\ind{c}+N\ind{c}+1]$, zero elsewhere (Fig \ref{fig-weighting_fn}). By assuming that data bins of the spectrum are individual random variables of mean flux $F_i$, the variance $\sigma_{S_\ell}^2$ of the measurements of $S_\ell$ is defined by:
\begin{eqnarray}
    \sigma_{S_\ell}^2 &=& \left(\frac{\partial S_\ell}{\partial F_\ell}\right)^2\,\sigma_{F_\ell}^2\, +\, \left(\frac{\partial S_\ell}{\partial F\ind{c}}\right)^2\,\sigma_{F\ind{c}}^2 \\
    &=& S_\ell^2\, \left(\frac{\sum_{i=i_\ell}^{i_\ell+N_\ell-1} w_{\ell,i}'^2\, \sigma_i^2}{F_\ell^2} \,+\, \frac{\sum_{i=i\ind{c}}^{i\ind{c}+N_c-1} \sigma_i^2/N\ind{c}^2}{F\ind{c}^2}    \right)
\end{eqnarray}
where $w_{\ell,i}' = w_{\ell,i}/\sum{w_{\ell,i}}$. In the case of a purely Gaussian noise, the standard deviation of the measurements is $\sigma_i = \sqrt{F_i}$ for a given data bin of index $i$. We can easily derive that:
\begin{equation}
     \sigma_{S_\ell} =  S_\ell\, \sqrt{\frac{\sum_{i=i_\ell}^{i_\ell+N_\ell-1} w_{\ell,i}'^2\, \sigma_i^2}{F_\ell^2} \,+\, \frac{1}{N\ind{c}}\frac{1}{F\ind{c}} }
\end{equation}

We note that assuming the noise to be only caused by photon noise leads to underestimating the errors on the measurements as soon as other sources of noise become significant. This is certainly the case for the \cahk lines, where the detectors are notoriously less sensitive in the near UV than in the mid visible.

Then, in the case $N\ind{spec}$ spectra are available for a given target, we assume the measurements to be independent. Thus, the noise level of the mean index $\langle S_\ell\rangle$ is:
\begin{equation}
         \sigma_{\langle S_\ell\rangle} = \frac{1}{N\ind{spec}} \, \sqrt{\sum_{i=1}^{N\ind{spec}} \sigma\ind{S_{\ell,i}}^2}.
\end{equation}

\section{Mode amplitude versus physical parameters}
\label{appendix_mode_amp}

\begin{figure*}[ht!]
\includegraphics[width=9cm]{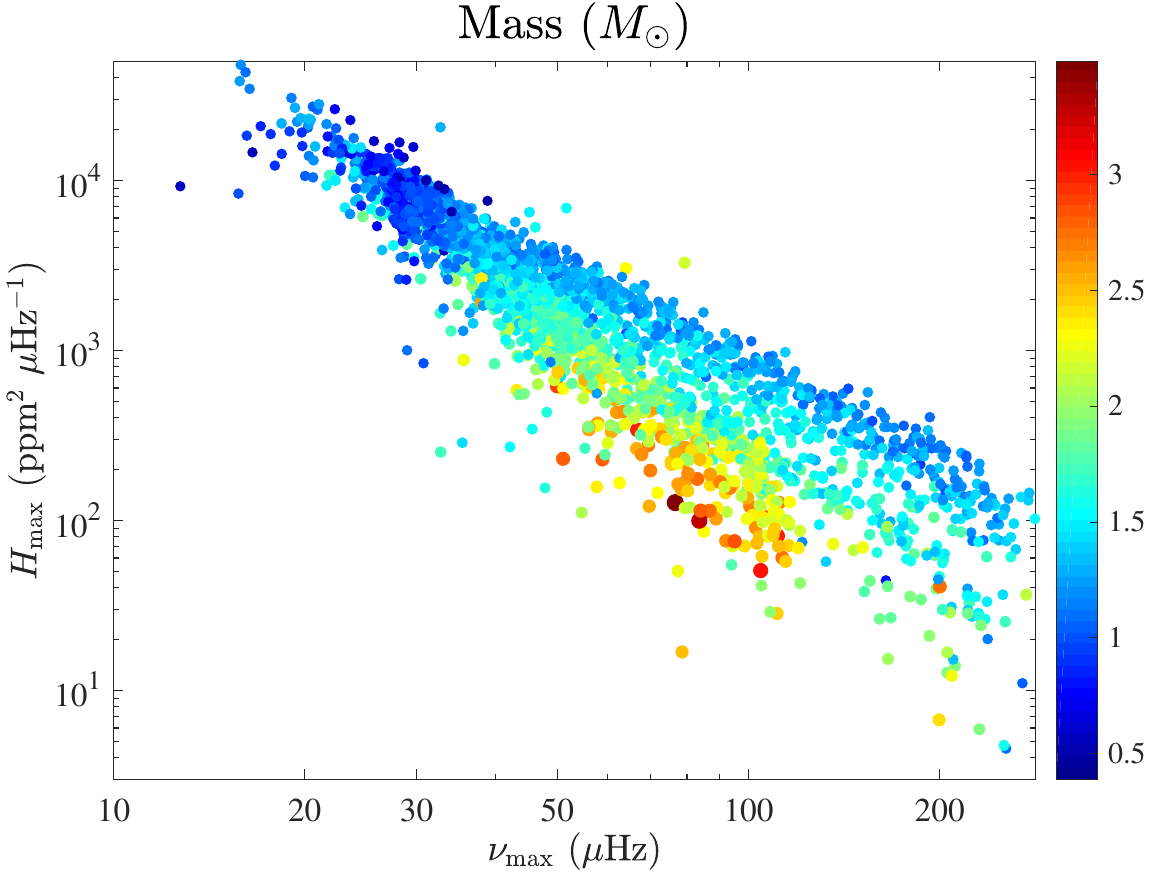}
\includegraphics[width=9cm]{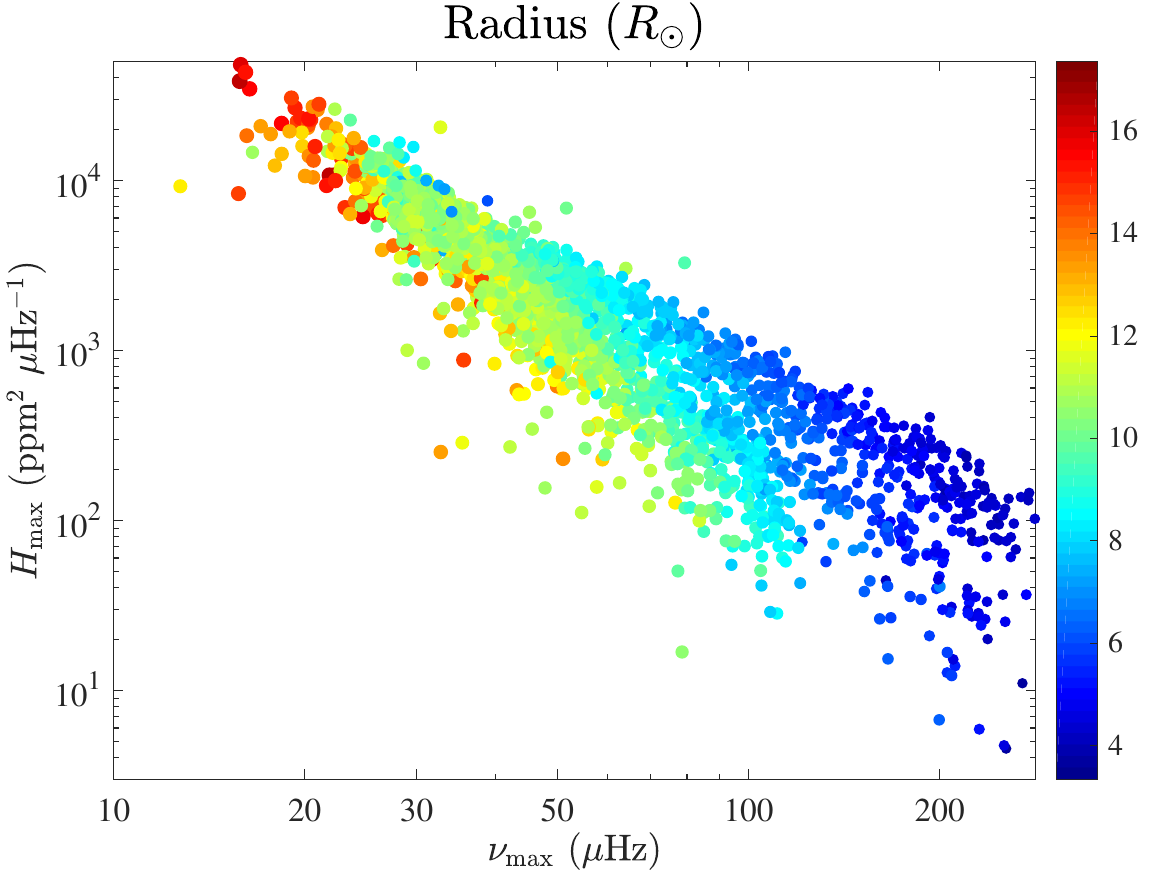}\\
\includegraphics[width=9cm]{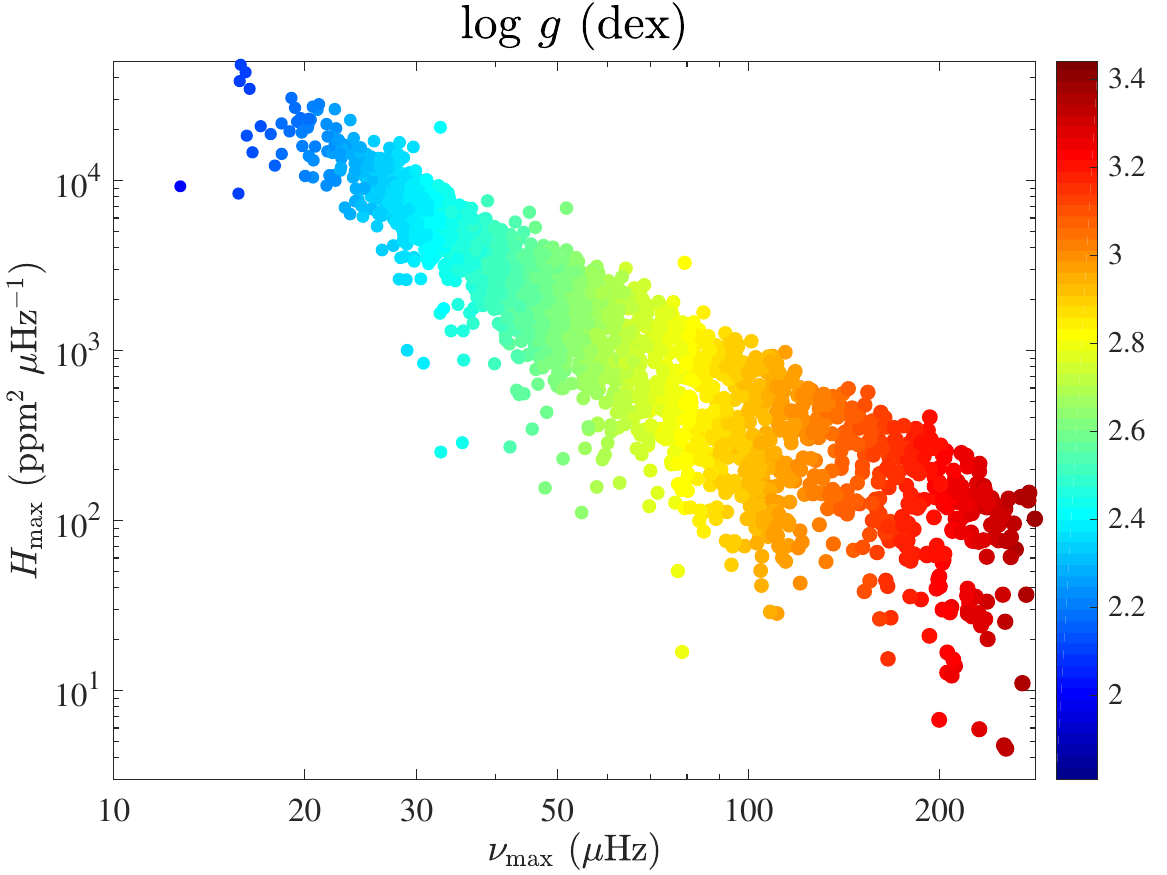}
\includegraphics[width=9cm]{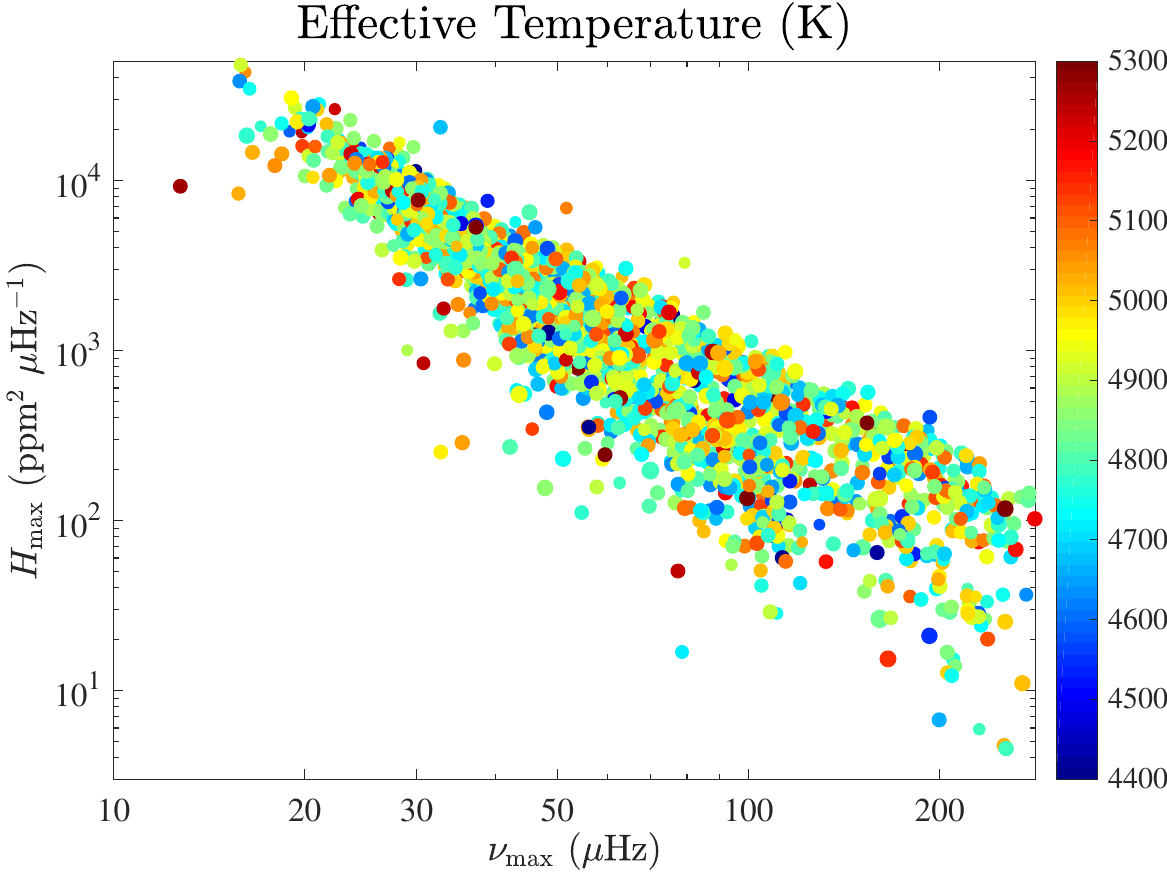}\\
\caption{Physical parameters (colorscale) as a function of oscillations frequency at maximum amplitude $\nu\ind{max}$ ($x$-axis) and height of the Gaussian envelope employed to model the oscillation excess power $H\ind{max}$ ($y$-axis). }
\label{fig_activity_Hmax_numax_appendix}
\end{figure*}

Here we display the amplitude of the Gaussian envelope of the oscillations as a function of $\nu\ind{max}$ (as in Fig. \ref{fig_activity_Hmax_numax}), where the color scales are related to the masses, radii, surface gravities, and effective temperatures of the stars. Masses, radii, and surface gravities are from \citet{Gaulme_2020}, and temperatures are from \textit{Gaia} DR2, as in \citet{Gaulme_2020} and \citet{Gehan_2022}.

\end{appendix}

\end{document}